\documentclass[a4paper,fleqn,usenatbib]{mnras}

\usepackage{newtxtext,newtxmath}

\usepackage[T1]{fontenc}
\usepackage{ae,aecompl}

\usepackage[caption=false]{subfig}

\usepackage{graphicx}	
\usepackage{amsmath}	

\usepackage{longtable}
\usepackage{lscape}

\title[Low-mass X-ray Binary Accretion Discs]{Using Optical Spectroscopy to Map the Geometry and Structure of the Irradiated Accretion Discs in Low-mass X-ray Binaries: The Pilot-Study of MAXI\,J0637$-$430}

\author[B.E. Tetarenko et al.]{
B.E. Tetarenko$^{1}$\thanks{E-mail: btetaren@umich.edu},
A.W. Shaw$^{2}$,
E.R. Manrow$^{1}$,
P.A. Charles$^{3}$,
J.M. Miller$^{1}$,
T.D. Russell$^{4}$,
\newauthor{and A.J. Tetarenko$^{5}$}
\\
$^{1}$Department of Astronomy, University of Michigan, 1085 South University Avenue, Ann Arbor, MI 48109, USA\\
$^{2}$Department of Physics, University of Nevada, Reno, NV 89557, USA\\
$^{3}$Department of Physics \& Astronomy, University of Southampton, Southampton SO17 1BJ, UK\\
$^{4}$Anton Pannekoek Institute for Astronomy, University of Amsterdam, Science Park 904, NL-1098 XH Amsterdam, The Netherlands\\
$^{5}$East Asian Observatory, 660 N. A'oh$\bar{o}$k$\bar{u}$ Place, University Park, Hilo, Hawaii 96720, USA
}

\date{Accepted XXX. Received YYY; in original form ZZZ}

\pubyear{2020}

\begin{document}
\label{firstpage}
\pagerange{\pageref{firstpage}--\pageref{lastpage}}
\maketitle

\begin{abstract}
The recurring transient outbursts in low-mass X-ray binaries (LMXBs) provide us with strong test-beds for constraining the poorly understood accretion process. While impossible to image directly, phase-resolved spectroscopy can provide a powerful diagnostic to study their highly complex, time-dependent accretion discs. 
We present an 8-month long multi-wavelength (UV, optical, X-ray) monitoring campaign of the new candidate black hole LMXB MAXI\,J0637$-$430 throughout its 2019/2020 outburst, using the {\em Neil Gehrels Swift Observatory}, as well as three quasi-simultaneous epochs of Gemini/GMOS optical spectroscopy.
We find evidence for the existence of a correlation between the X-ray irradiation heating the accretion disc and the evolution of the 
He {\sc ii} 4686 \AA\ emission line profiles detected in the optical spectra. 
Our results demonstrate a connection between the line emitting regions and  physical properties of the X-ray irradiation heating the discs during outburst cycles of LMXBs. Further, we are able to show that changes in the physical properties of the irradiation heating the disc in outburst can be imprinted within the H/He emission line profiles themselves in these systems.

\end{abstract}


%



\section{Introduction}\label{sec:intro} 

Among accreting astrophysical systems, low-mass X-ray binaries (LMXBs), containing discs of matter fed by nearby, low-mass stars flowing onto compact stellar remnants (black holes or neutron stars), provide ideal test beds for constraining the poorly understood process of accretion. Many of these systems are transient, undergoing frequently (usually timescales of $\sim$ years) recurrent outbursts. Thus, these systems offer the unique opportunity to study accretion over observable (day-month) timescales \citep{charlescoe2006,remillardmclintock2006,tetarenko2016}. 

The mechanism behind these outbursts can be broadly explained with the disc-instability model (DIM), whereby such transient behaviour is explained via the accretion disc cycling between a hot, ionized outburst state and a cool, neutral, quiescent state. This limit-cycle, triggered by the accumulation of matter in the disc as a result of mass-transfer from the companion star, heats the disc until a significant portion is ionized. The viscosity of the disc (i.e. ability to move angular momentum outwards) increases dramatically when it is in this ionized state. This results in a rapid in-fall of matter onto the compact object, and in-turn a bright outburst. 
For a recent review of the DIM see \citet{hameury2020}.

While we can image their supermassive relatives (i.e., Event Horizon Telescope (EHT); \citealt{2019EHTM87}), LMXBs are far too small and distant to be imaged directly. However, phase-resolved spectroscopy can provide an alternative diagnostic to study their highly complex, time-dependent accretion discs.
The spectral signature of an LMXB is detection of strong H and He emission lines at optical wavelengths (H$\alpha$, H$\beta$, He {\sc i}, He {\sc ii}; e.g., \citealt{dubuskimmenou2001,casares2015}) from the accretion disc \citep{charlescoe2006}. These ``accretion signatures'' typically show a double-peaked profile, due to Doppler motions in the binary \citep{crawfordkraft1956,casares2015}, with the line profile shape depending on the distribution of emission over the disc surface. As a result, these emission line profiles encode within them a projection of the disc itself along the line of sight. 
As the spatial distribution of line emission is a tracer of structure in the disc, these emission lines can, in principle, be used to effectively describe how matter in the disc behaves and evolves over time \citep[e.g.,][]{marsh2001,marsh2005}.

LMXBs harbouring stellar-mass black holes (BH-LMXBs), are of particular interest. Most of the optical light emitted by their accretion discs comes from reprocessed X-rays, arising close to the black hole, heating the outer disc \citep{vanparadijs1994,vanparadijs1996}. 
This X-ray irradiation is the dominant factor that determines the temperature over most of the disc during outburst. By illuminating the disc surface, irradiation controls the outburst decay from peak to quiescence. Consequently, the light-curve profile for an outbursting irradiated disc, as predicted by the disc-instability model (DIM), can be characterized into multiple stages \citep{dubus1999,dubus2001}, defined by how (and on what timescale) matter moves through the disc itself. 

A combination of the (i) first stage after the outburst peak, the viscous decay, and (ii) subsequent transition to the second, irradiation-controlled stage of the decay, together offer the ideal means to study both the source of the irradiation and its effect on the disc. During the viscous stage of the decay, X-ray irradiation is at its strongest and thus, the entire disc is thought to be in a hot, ionized state. At this point, the rate of mass-accretion onto the black hole ($\dot{M}_{\rm in}$) is much greater than the rate of mass-transfer from the companion to the disc ($\dot{M}_{\rm tr}$). Thus, mass in the disc is assumed to change only through viscous accretion, leading to the characteristic ``exponential'' light-curve decay profile shape. The transition to the second, irradiation-controlled decay stage occurs as a result of the temperature dropping in the outer disc regions, and subsequent formation and propagation of an inward moving cooling front at a speed determined by the strength of the decaying X-ray irradiating flux \citep{dubus2001,tetarenko2018,tetarenko2018b}. Both of these accretion regimes, to date, have not yet been well studied in LMXBs. Recently, \citet{tetarenko2018,tetarenko2018b} developed a methodology to characterize observed LMXB light-curves based on the predictions of the DIM. Together, with multi-wavelength capabilities and rapid-response times of the instruments aboard the Neil Gehrels \textit{Swift} Observatory, their analytical models allow one to track and predict the evolution of an LMXB on daily timescales during an outburst.

Despite decades of effort, the radial profile and geometry of the irradiation source heating LMXB discs are not well understood \citep{dubus1999,dubus2001,vrtilek1990,king1996,reynolds2013,rykoff2007,degenaar2014}. As the irradiation is only important in the outer regions (typically greater than hundreds of gravitational radii) of LMXB discs, H/He emission lines are an ideal means to probe it. 
In this work, we combine X-ray, optical, and ultraviolet spectral and photometric data from \textit{Swift}, modern Doppler tomography techniques (e.g., \citealt{steeghs2003}), and the H/He disc emission lines detected in Gemini Multi-Object Spectrograph (GMOS) optical spectra. With these data we attempt to map the illumination pattern of the irradiation heating a transient BH-LMXB disc during both the viscous stage and transition to the irradiation-controlled stage of an outburst decay.
Specifically, these data have allowed us to track and quantify how variations in this irradiation-heating over an outburst affect the physical properties of the disc through its emission line profiles. This is a first (and crucial) step to developing a method to use phase-resolved spectroscopy to effectively probe the geometry/structure of the gas making up an outbursting BH-LMXB accretion disc, which has not been done before.

This paper is organized as follows: Section \ref{sec:source_maxi} outlines the details of the 2019/2020 outburst of new candidate BH-LMXB MAXI\,J0637$-$430. Section \ref{sec:obsandred} discusses the multi-wavelength observational data used in this work. Section \ref{sec:results} discusses the multi-wavelength (X-ray, optical and UV) spectral and time-series analysis done on MAXI\,J0637$-$430. In Section \ref{sec:discuss} we discuss the relationship found between irradiation-heating and disc emission line profiles for this source and the implications such a finding has for LMXBs in general. Finally in Section \ref{sec:summary} we summarize this work.

\section{Galactic BH-LMXB Candidate: MAXI\,J0637$-$430}\label{sec:source_maxi} 

\subsection{Discovery and Multi-wavelength Outburst Monitoring}
MAXI\,J0637$-$430 (hereafter J0637), a newly discovered candidate BH-LMXB, was first detected in outburst by the {\em Monitor of All-sky X-ray Image} ({\em MAXI}; \citealt{matsuoka2009}) on 2019 November 02 \citep{negoro2019}.
Follow-up observations with the X-ray Telescope (XRT; \citealt{burrows2005}) aboard \textit{Swift} (2019 November 03; \citealt{kennea2019}) and the {\em Nuclear Spectroscopic Telescope Array} ({\em NuSTAR}; \citealt{harrison2013} - 2019 November 05; \citealt{tomsick2019}) showed a soft X-ray spectrum, well fit by an absorbed power-law + disc blackbody model. In addition, a bright optical counterpart with a position not coincident with any known star, was also detected by the Ultraviolet and Optical Telescope (UVOT; \citealt{roming2005}) aboard \textit{Swift} inside the X-ray error circle. Both X-ray and optical observations were highly suggestive of an outbursting X-ray binary in the soft accretion state, and furthermore the X-ray spectral shape favours a BH over a neutron star as the compact object.

Further, optical spectral observations, taken on 2019 November 03 with the Southern Astrophysical Research (SOAR) telescope \citep{strader2019}, revealed strong, broad, double-peaked H$\alpha$, and He {\sc ii} 4686 \AA\ emission at Galactic velocities, and weaker H$\beta$, H$\gamma$, and He {\sc i} 5875 \AA\ and 6678 \AA\ emission as well, providing strong evidence for J0637 being a new, candidate BH-LMXB in outburst.

J0637, first localized by \textit{Swift}, was also later detected in the radio band by the Australia Telescope Compact Array (ATCA; at frequencies of 5.5 and 9 GHz) on 2019 November 06 \citep{russelltd2019}, giving a refined position of: RA, Dec (J2000) = 06$^{\rm h}$36$^{\rm m}$23$\fs$7 $\pm$ 0\farcs2, $-$42$^{\circ}$52\arcmin04\farcs1 $\pm$ 0\farcs7. At this position, with a galactic latitude of $-20$ degrees, the interstellar reddening is low, making J0637 a great target for both blue and red time-resolved spectroscopy.

\begin{figure*}
    \center
    \includegraphics[width=0.95\linewidth,height=0.7\linewidth]{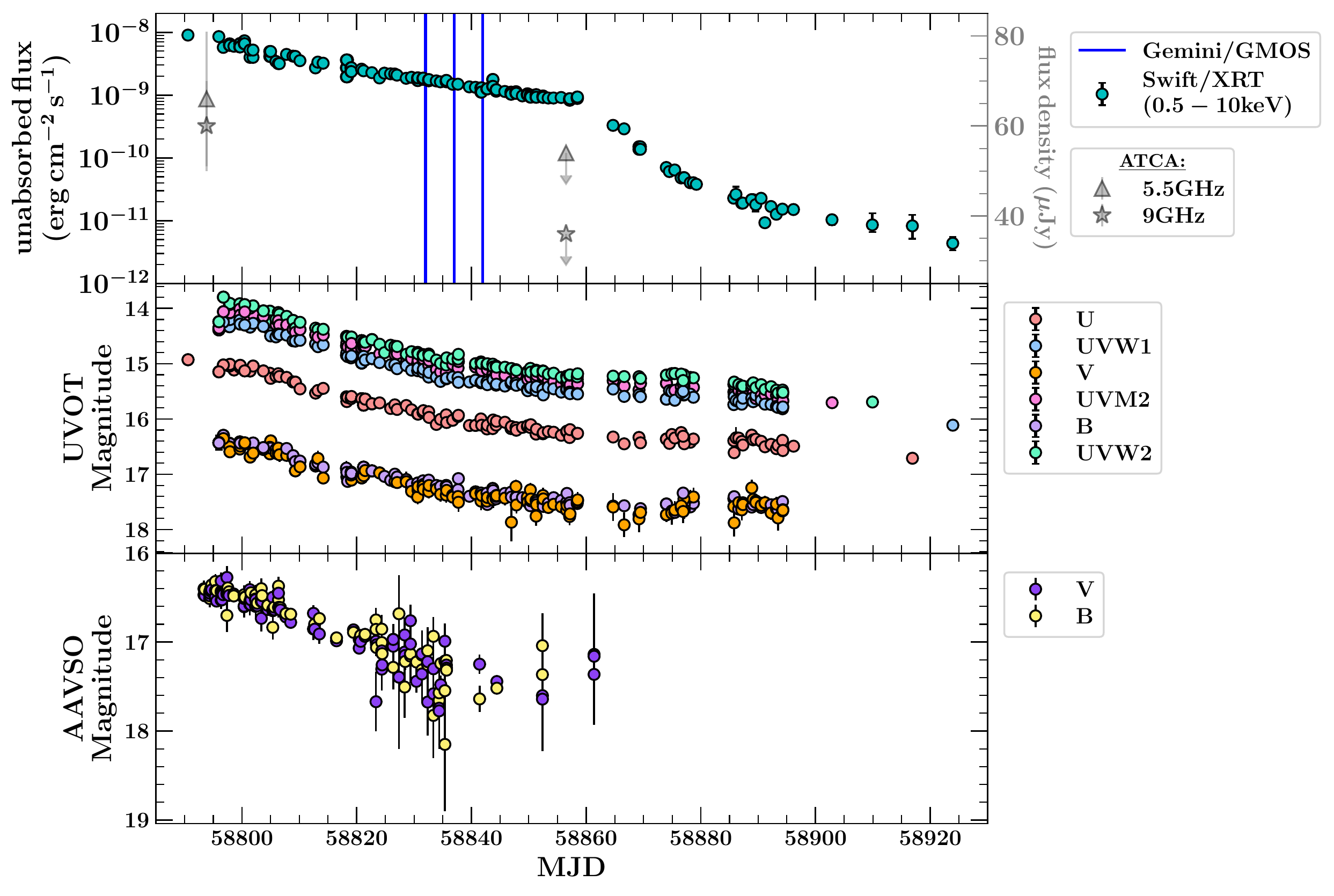}
    \caption{Multi-wavelength light-curve showing the evolution of the 2019/2020 outburst of J0637: \textit{(top)} \textit{Swift}/XRT ($0.5-10$ keV), \textit{(middle)} \textit{Swift}/UVOT, and \textit{(bottom)} AAVSO. UVOT and AAVSO magnitudes have not been de-reddened. The grey data points and grey right-side axis in the \textit{top} panel show the two epochs of radio (5.5 and 9 GHz) data taken with ATCA. The vertical solid blue lines in the \textit{top} panel show the three days in which Gemini/GMOS spectra were obtained.}
    \label{fig:lc_plot}
\end{figure*}

Radio flux densities were consistent with emission arising from a discrete ejection event (e.g., steep spectrum; \citealt{russell2019b}), expected if J0637 was an outbursting BH-LMXB in a soft or soft-intermediate accretion state. A second radio epoch was taken on 2020 January 8 (as analyzed in this work), where J0637 was not detected in either the 5.5 or 9 GHz band. See Figure \ref{fig:lc_plot} for the radio light-curve and Section \ref{sec:radio_data} for a detailed discussion.

Monitoring on approximately daily timescales with both \textit{Swift} (this work; Figure \ref{fig:lc_plot}) and the {\em Neutron star Interior Composition Explorer} ({\em NICER}; \citealt{remillard2020}) showed that J0637 remained in a soft accretion state for approximately 70 days, before beginning the transition back to the hard accretion state between 2020 January 12-14 (MJD 58860-58862). 
After the state transition occurred (approximately 2020 January 29; MJD 58877), J0637 then began its final decay toward quiescence. This decay, lasting approximately 4 months, was observed at both X-ray and optical wavelengths by \textit{Swift} XRT and UVOT (this work) and the Faulkes telescopes at the Las Cumbres Observatory (as part of the XB-NEWS Monitoring

\noindent Program; \citealt{johar2020}). J0637 was thought to likely reach a quiescent level in 2020 June \citep{tomsick2020}.

\subsection{Binary Orbital Parameters}\label{sec:orb_params}

While an optical counterpart was detected during outburst \citep{kennea2019}, little is known about the orbital parameters of J0637. Thus, in this work, following \citet{tetarenko2018b}, we sample a BH mass ($M_{\rm BH}$) and binary mass ratio ($q$) from the Galactic distributions of \citet{ozel2010} and \citet{tetarenko2016}, respectively, and assume a broad, uniform distribution in distance of $d=5-15$ kpc.

Lastly, we attempted to constrain the binary orbital period using a combination of (i) a broad-band spectral energy distribution (SED) fitting method (see Section \ref{sec:sed_fits}), with data from our continuous \textit{\textit{Swift}} (XRT+UVOT) monitoring (see Section \ref{sec:swift_data} and Figure \ref{fig:lc_plot}), and (ii) the detection of Hydrogen emission in the optical spectrum. The SED fitting method provides an upper limit of $P_{\rm orb}\leq4$ hrs. While the detection of H$\alpha$ emission in the optical spectrum (See Section \ref{sec:line_prop_details} and Figure \ref{fig:flux_cal_spec}) rules out ultra-compact systems ($P_{\rm orb}\lesssim1.5$ hrs). Thus, we assume an orbital period of $P_{\rm orb}=2-4$ hrs. This would make it one of the shortest in this class of objects.

We note that while many light-curves of LMXBs in outbursts exhibit periodicities known as ``superhumps'', resulting from a precessing accretion disc, with a period typically close to the true $P_{\rm orb}$ of the binary
system (e.g., see \citealt{zurita2008,patterson2018}), no evidence for periodicity was found in the publicly available AAVSO data of J0637 (see Section \ref{sec:superhump_search} and \citealt{hambsch2019}). The available UVOT data (Section \ref{sec:uvot_data}) lacks the cadence required to perform such an analysis.

\section{Observations and Data Reduction}\label{sec:obsandred} 

\subsection{\textit{Swift}}\label{sec:swift_data}

A total of 105 \textit{Swift} observations were obtained from the High Energy Astrophysics Science Archive Research Center (HEASARC) Archive\footnote{\url{https://heasarc.gsfc.nasa.gov/docs/archive.html}}, between 2019 November 03 and 2020 March 15 (ObsId: 00012168001-00012172094; as part of University of Michigan GTO time - co-PIs: Tetarenko and Miller), covering the 2019/2020 X-ray outburst of J0637 (Figure \ref{fig:lc_plot}). The majority of the XRT observations were taken in windowed timing (WT) mode, with a few taken in photon counting (PC) mode, when the source count rate was low. The majority of the UVOT observations were obtained in all six filters, from the UV through optical (UVW2, UVM2, UVW1, U, B, V).

\subsubsection{XRT}

Data were first processed with the {\tt xrtpipeline} task from the {\sc heasoft} v6.26 software package. In WT mode, source and background spectra were extracted using circular apertures with a radius of 20 pixels. In PC mode, source spectra were first extracted using the same size aperature as WT mode, to determine if photon pile-up was significant. If the average count rate was greater than 0.5 counts/s, spectra were re-extracted, this time using an annular region with a 20 pixel outer radius. The radius of the central portion of the point spread function (PSF) excluded here was calculated with the {\tt ximage} package\footnote{\url{http://www.swift.ac.uk/analysis/xrt/pileup.php}} in each case. The background spectra of PC mode observations were extracted using an annulus, with inner and outer radii of 50 and 70 pixels, centered on the source.

Source and background spectra were then grouped such that each energy bin contained a minimum of 5 counts. Lastly, the response matrix files were obtained from the HEASARC calibration data base (CALDB) and the {\tt xrtmkarf} task was used to generate ancillary response files. 

{\sc xspec} v12.10.1f \citep{arnaud1996} was used to perform all of the X-ray (XRT) spectral fits in the $0.5-10$ keV band. The {\tt tbabs} model, utilizing abundances from \citet{wilms2000}, and photoionization cross-sections from \citet{verner1996}, was used to account for interstellar absorption in all fits. The hydrogen column was not well constrained in the XRT spectra, and thus was fixed at $N_H=4.39\times10^{20}$ cm$^{-2}$ (corresponding to $E(B-V)=0.064$; see Section \ref{sec:uvot_data}) for all fits \citep{guverozel2009}. Note that, HI4PI\footnote{https://heasarc.gsfc.nasa.gov/cgi-bin/Tools/w3nh/w3nh.pl} find a Galactic $N_H$ in the direction of the source ($N_H=5.23\times10^{20}$ cm$^{-2}$) to be similar to our estimate. Further, using the equivalent width of the NaD (I/S) absorption line present in the Gemini/GMOS optical spectra (Figure \ref{fig:flux_cal_spec}) yields an $E(B-V)=0.068_{-0.008}^{+0.009}$ (see e.g., \citealt{munari1997}), consistent with the reddening estimate we use in this work.

\subsubsection{UVOT}\label{sec:uvot_data} 

UVOT, a 30 cm diameter telescope, operates simultaneously with XRT. Thus, we have a total of 105 UVOT observations simultaneous with the XRT exposures described above, 100 of which contain exposures in multiple UVOT filters.

Using the {\sc heasoft} task {\tt uvotsource}, aperture photometry was performed on all UVOT images, using a circular region centred on the source with a radius of 5\arcsec. The background circular region used measured 20\arcsec  in radius and was centered in a source free region. All magnitudes were computed in the Vega system, with the known flux zero-points for each filter used to convert to flux densities. Uncertainties in UVOT magnitudes/flux densities include a combination of the statistical error ($1\sigma$ confidence level) and systematic uncertainty used to account for error in the shape of the PSF. Photometric UVOT data was corrected for interstellar extinction according to \citet{fitzpatrick1999}. To deredden these data, the {\tt specutils} package in {\sc python} was used, with an $E(B-V)=0.064$ \citep{strader2019}.

UVOT spectra were extracted in each available filter using the {\tt uvot2pha} task, source and background regions as described above, and the response matrices available for each filter from the HEASARC CALDB.
If necessary, the {\tt uvotimsum} task was used first to co-add individual image extensions, creating one image per filter, before spectra were extracted.
{\sc xspec} v12.10.0c \citep{arnaud1996} was used to perform all of the broad-band (XRT+UVOT) spectral fits. The {\tt redden} model, with an $E(B-V)$ fixed at a value of $0.064$, was used to account for extinction in all fits.

\subsection{Gemini/GMOS}\label{sec:gem_data}

On 2019 December 8, we triggered our Gemini/GMOS-South Target-of-Opportunity (ToO) programme (GS-2019B-Q-233; PI Tetarenko) on J0637. We subsequently obtained long-slit optical spectra over three nights: 2019 December 15 (05:15:42.2334$-$07:14:22.1334 UT), 2019 December 20 (04:09:12.3834$-$06:07:45.1500 UT), and 2019 December 25 (03:30:53.8167$-$07:08:48.9500 UT), for a total of 42$\times$600s individual exposures (7hrs on source). See Table \ref{tab:gemtab} in Appendix \ref{sec:appC} and Figure \ref{fig:lc_plot} for details. All data were obtained with the B600 grating, a central $\lambda=560$nm, and a 1\arcsec slit, with a typical resolution of 4.5-5\AA.

Data were reduced using {\sc pyraf} v2.1.15 and the available Gemini/GMOS long-slit reduction cookbook\footnote{\url{http://ast.noao.edu/sites/default/files/GMOS\_Cookbook/}}. First, bias residual and flat-field MasterCal reference files were created with the {\tt gbias} and {\tt gflat} tasks, and then used to calibrate the science exposures. Next, a bad pixel mask was created using the provided {\sc iraf} scripts\footnote{\url{https://gmos-data-reduction-problems-and-solutions.readthedocs.io/en/latest/appendix.html}}. Then, cosmic-ray rejection was performed on individual calibrated science exposures using the {\tt gemcrspec} task. This task makes use of the LA-Cosmic program \citep{vandokkum2001}, and the provided {\sc iraf} scripts\footnote{\url{http://ast.noao.edu/sites/default/files/GMOS\_Cookbook/GettingStarted.html}}. Next, basic reductions were performed and the bad pixel mask was applied to the calibrated science exposures using the {\tt gsreduce} task. The same procedure used for the science exposures, as outlined above, was then applied to the standard star (LTT1788) exposures.

\begin{figure*}
    \centering
    \includegraphics[width=0.58\linewidth,height=0.4\linewidth]{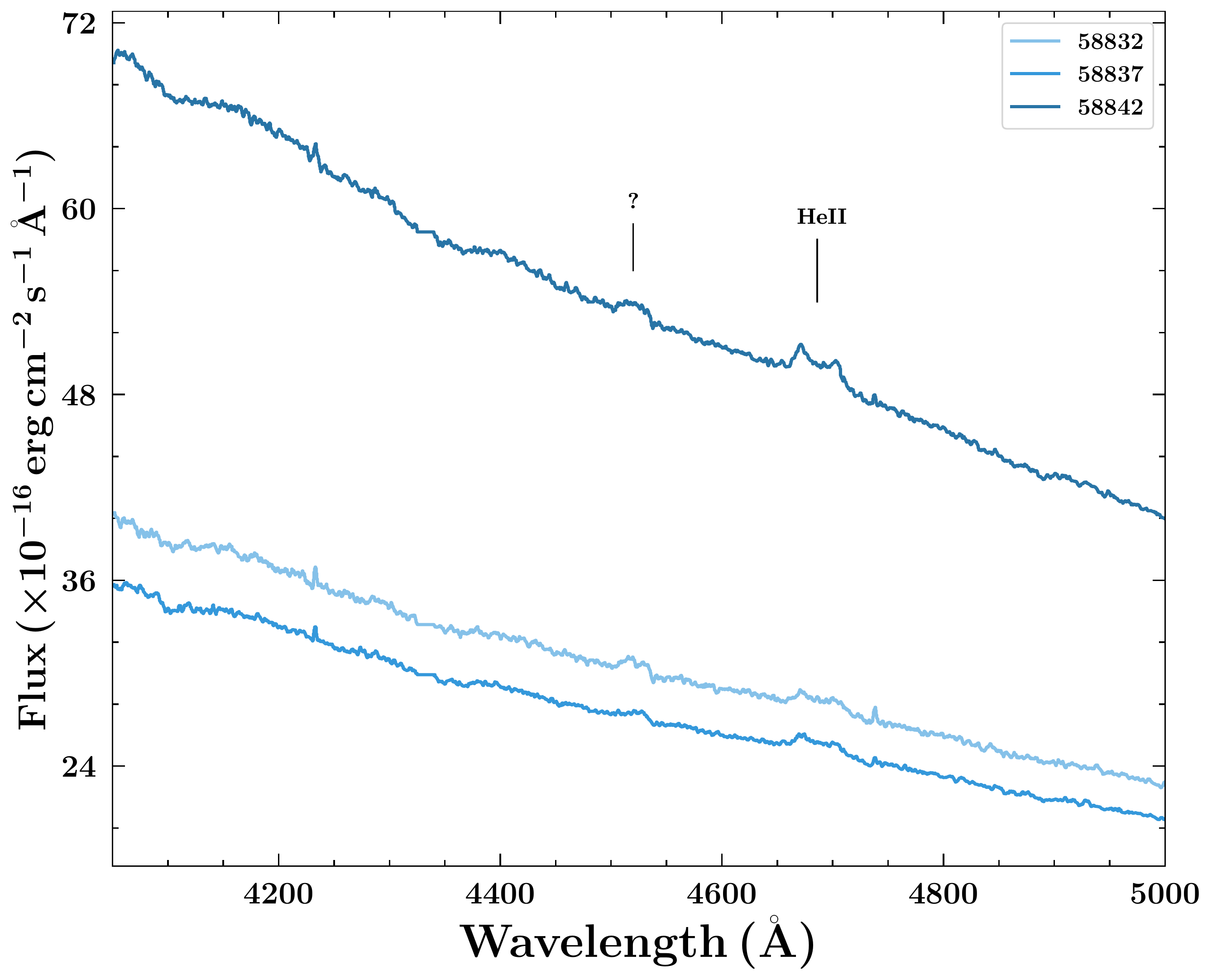}\hfill
    \includegraphics[width=0.56\linewidth,height=0.4\linewidth]{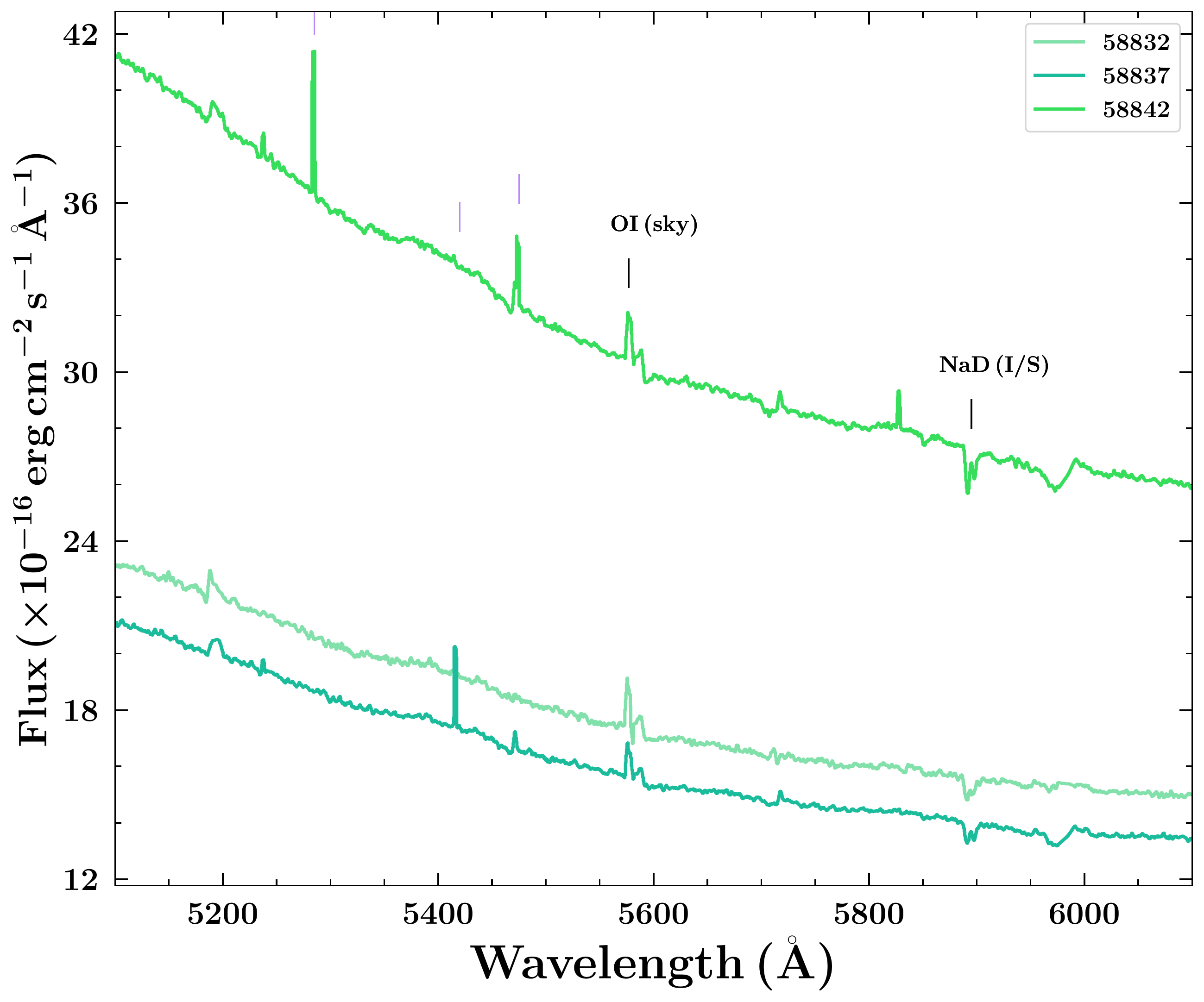}\hfill
    \includegraphics[width=0.58\linewidth,height=0.4\linewidth]{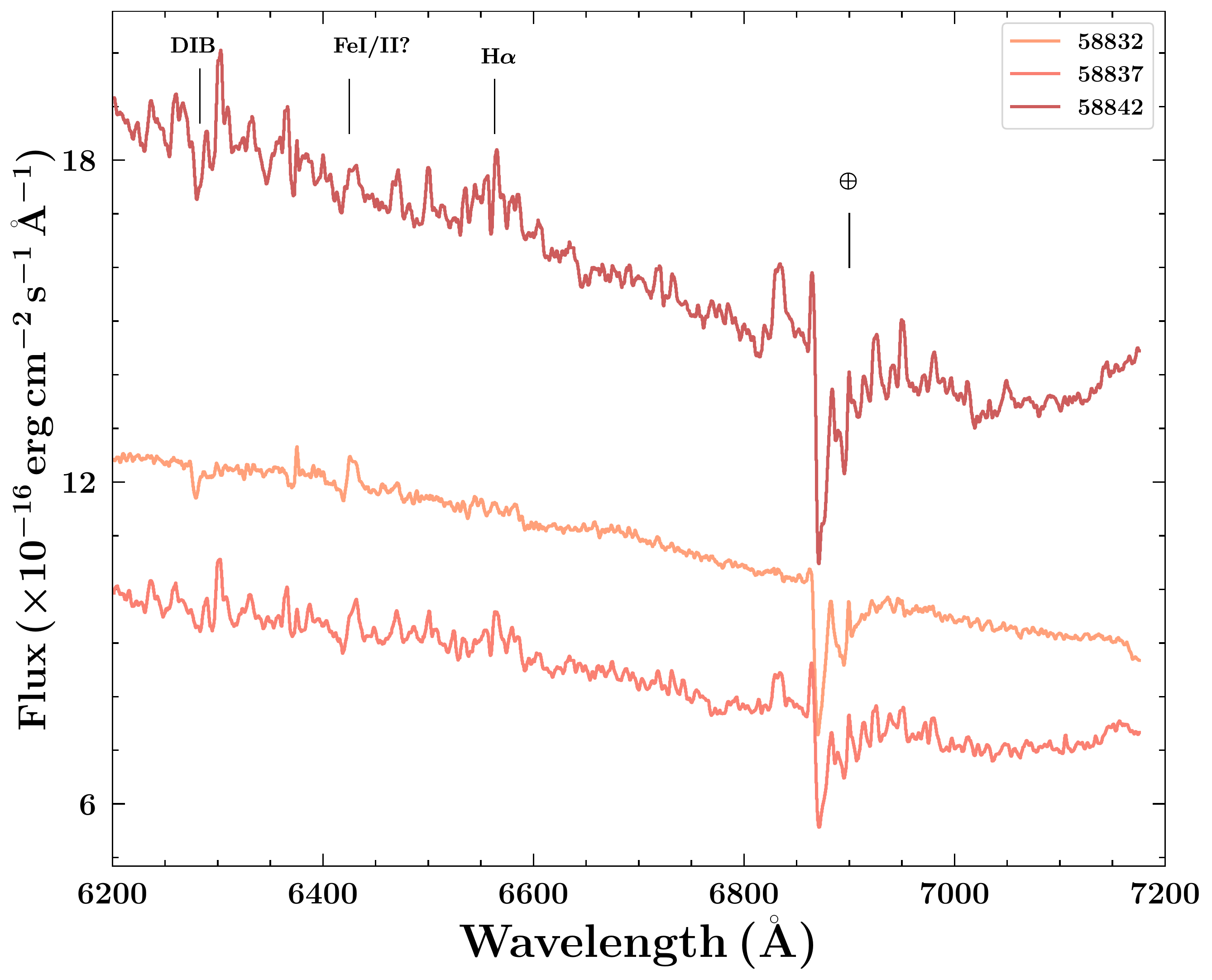}
    \caption{Flux-calibrated Gemini/GMOS optical spectra of J0637, summed over each of the three days during which data was taken: 2019 December 15 (MJD 58832), 2019 December 20 (MJD 58837), and 2019 December 25 (MJD 58842). The two strongest disc emission lines (He {\sc ii} 4686 \AA\ and H$\alpha$), interstellar absorption lines (I/S), diffuse interstellar bands (DIB), telluric absorption lines ($\oplus$), and poorly-subtracted night-sky emission lines (sky) are marked in black. Purple lines mark narrow emission features that are not real and ? signifies unknown features (see Section \ref{sec:line_prop_details} for discussion).}
    \label{fig:flux_cal_spec}
\end{figure*}

Next, arc exposures were reduced and wavelength solutions were found via the {\tt gswavelength} task. Then, the wavelength calibration was applied (via  the {\tt gstransform} task), and sky subtraction was performed (via the {\tt gsskysub} task), on the calibrated, cleaned, science and standard star exposures. Spectra were then extracted from individual, calibrated, cleaned, science and standard star exposures, using the {\tt gsextract} task. Finally, daily averaged spectra were also created from the individual extracted science spectra, on each of the three days data was taken, using the {\tt scombine} task in {\sc iraf}. Flux-calibrated spectra, summed over the three days of data taken, can be found in Figure \ref{fig:flux_cal_spec}.

\subsection{ATCA}\label{sec:radio_data} 

MAXI~J0637$-$430 was observed by the Australia Telescope Compact Array (ATCA) on 2019-11-06 (17:23 -- 19:55 UT) and 2020-01-05 (12:16 -- 14:13 UT). The observations were recorded simultaneously at central frequencies of 5.5 and 9\,GHz, with 2048\,MHz of bandwidth at each central frequency (2048 $\times$ 1\,MHz channels). For both observations PKS~1934$-$630 was used for primary flux calibration and PKS~0629$-$418 was used for secondary phase calibration. Following standard procedures within the Common Astronomy Software Application (\textsc{casa}, version 5.1.0; \citealt{mcmullin2007}\footnote{\url{https://casaguides.nrao.edu/index.php?title=Main_Page}}), the data  were first edited for radio frequency interference (RFI), before being calibrated and then imaged. For the 2019-11-06 observation, imaging used a Briggs robust parameter of 0, balancing sensitivity and resolution, providing synthesised beams of 9.3\arcsec$\times$1.1\arcsec at 5.5\,GHz and 6.0\arcsec$\times$0.7\arcsec at 9\,GHz with a position angle of 10 degrees East of North for both frequencies. For our 2020-01-05 ATCA observation, we imaged with a Briggs robust parameter of 2 to maximise the sensitivity, providing synthesised beams of 22.4\arcsec$\times$1.8\arcsec at 5.5\,GHz and 12.8\arcsec$\times$1.1\arcsec at 9\,GHz, both with positions angle of 2\,degrees West of North.

The radio counterpart to MAXI~J0637$-$430 was detected during our 2019-11-06 ATCA observations, but not on 2020-01-05. On 2019-11-06 we fitted for a point source in the image plane by applying a Gaussian with the full width at half maximum fixed to the synthesised beam parameters. Doing so, we measured flux densities of 66$\pm$15\,$\mu$Jy at 5.5 GHz and 60$\pm$10\,$\mu$Jy at 9 GHz. On 2020-01-05, we did not detect radio emission associated with the target, with $3\sigma$ radio upper-limits of 54\,$\mu$Jy/beam at 5.5\,GHz and 36\,$\mu$Jy/beam at 9\,GHz. Stacking the two bands together provides a $3\sigma$ upper-limit of 30\,$\mu$Jy/beam (at a central frequency of 7.25\,GHz).

\subsection{AAVSO}\label{sec:aavso_data} 
Publicly available observations of J0637, in the V and B bands (Vega magnitudes), were obtained from the AAVSO International Database\footnote{\url{https://www.aavso.org}}. See Figure \ref{fig:lc_plot} for the light-curves.

\subsubsection{Search For Periodicity}\label{sec:superhump_search} 

To search for periodicities in the optical/UV data we used the Lomb-Scargle periodogram implemented in the {\tt astropy} package \citep{astropy2013,astropy2018} in {\sc python} \citep{lomb1976,scargle1982,zechmeister2009}. This tool is ideal for detecting periodic signals in unevenly sampled data. We do not detect any significant ($99$\% significance level) periodicities in the data on timescales $<100$ days (i.e., length of the sampled light curves). Note that, we have also checked the equivalent width (EW) light-curves of the He {\sc ii} 4686 \AA\ emission line detected in the GMOS spectra (see \ref{sec:line_prop_details}). Again, no significant periodicities were found.

\section{Results}\label{sec:results}

\subsection{Spectral Fitting}\label{sec:sed_fits} 

\subsubsection{X-ray (XRT)}\label{sec:xrayfits} 

All X-ray spectra were adequately fit (via the $\chi^2$ statistic), in the $0.5-10$ keV band, with either an absorbed power-law ({\tt tbabs*cflux*powerlaw}), absorbed disc-blackbody ({\tt tbabs*cflux*diskbb}), or combination power-law + disc blackbody ({\tt tbabs*cflux*(diskbb+powerlaw})) model in {\sc xspec}. {\tt cflux} is convolution model used to to compute the flux of different spectral components. Used as it is here, with the normalization of either the {\tt diskbb} or {\tt powerlaw} model components fixed to a non-zero value, a flux estimate for the entire spectral model is the output.
See Figure \ref{fig:lc_plot} and Table \ref{tab:xrayfitstab} in Appendix \ref{sec:appB} for the best-fit model parameters, along with the resulting $0.5-10$ keV fluxes, for each \textit{Swift} epoch.

\subsubsection{Broad-band (XRT+UVOT)}\label{sec:broadbandspec}

Of the 100 \textit{Swift} epochs, for which there are multiple UVOT filters, we attempted to fit the broad-band (XRT+UVOT) spectrum with the {\tt redden*diskir} model in {\sc xpsec}. Adequate fits were obtained for 80 of these observations, covering the time-period: $58795-58866$. 

{\tt diskir} \citep{gierlinski2009} models an irradiated accretion disc using a total of 9 parameters\footnote{\url{https://heasarc.gsfc.nasa.gov/xanadu/xspec/manual/node165.html}}. During each epoch, we fit for the following 5 parameters: disc normalization ($N_{\rm disc}$), inner disc temperature ($T_{\rm in}$), log of the ratio between outer and inner disc radius (logrout), fraction of bolometric flux thermalized in the outer disc region ($f_{\rm out}$), and ratio of luminosity in the Compton tail to the un-illuminated disc ($L_{\rm c}/L_{\rm d}$). In addition, we fixed the remaining four parameters to typical values expected for LMXBs: power-law photon index ($\gamma=1.7$), electron temperature ($T_{\rm e}=100$ keV), fraction of the luminosity in the Compton tail
thermalized in the inner disc ($f_{\rm in}=0.1$), and the radius of the Compton illuminated disc as a fraction of the inner disc radius ($r_{\rm irr}=1.2$). See Figure \ref{fig:ex_seds} in Appendix \ref{sec:appA} for example SEDs, quasi-simultaneous (within at most 10 hrs) with the epochs where Gemini/GMOS and ATCA data were taken. See Table \ref{tab:broadfitstab} in Appendix \ref{sec:appB} for best-fit model parameters.

\begin{figure*}
    \center
    \includegraphics[width=0.7\linewidth,height=1.1\linewidth]{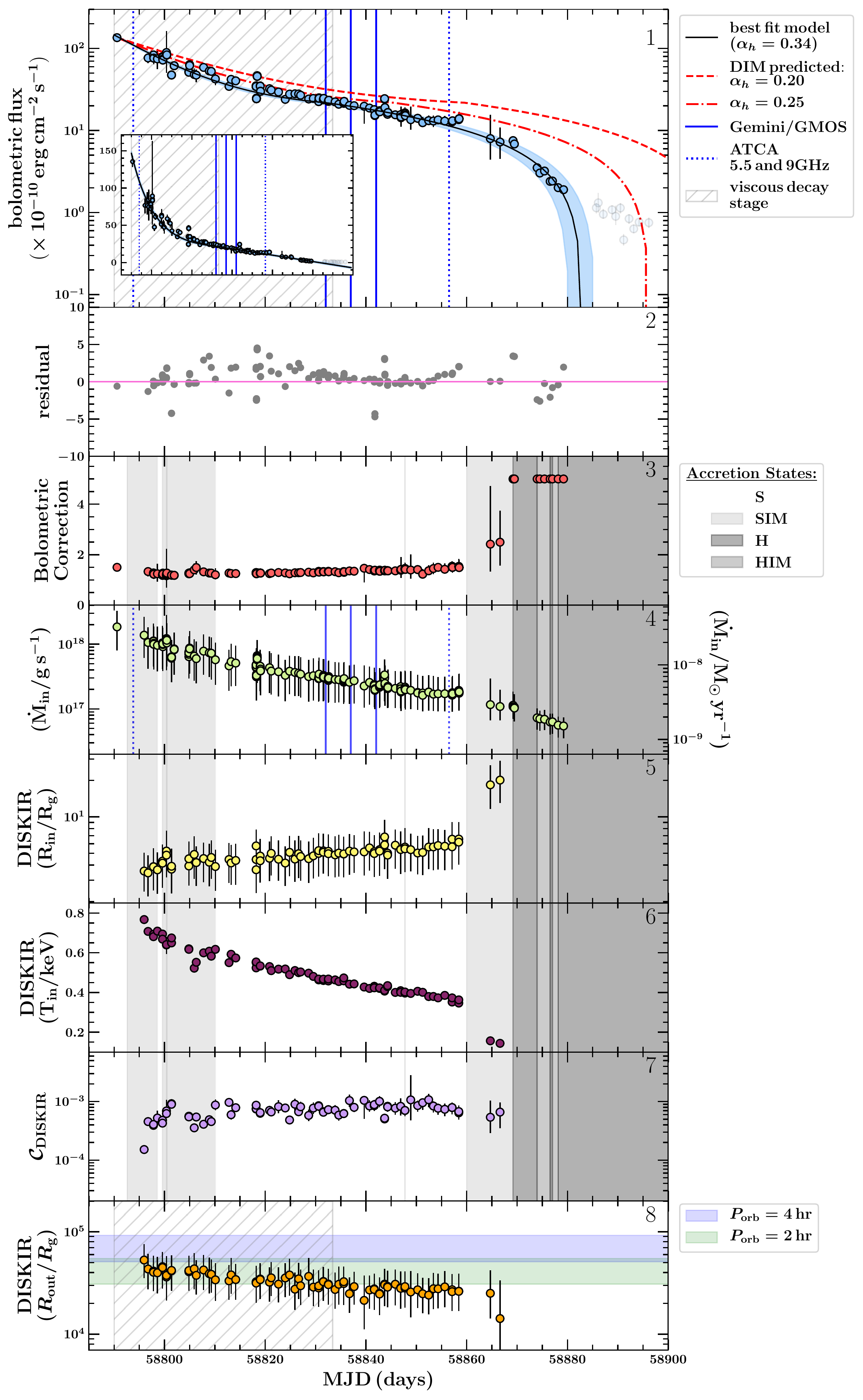}
    \caption{\textit{(top panel)} the bolometric light-curve for the 2019/2020 outburst of J0637 fit with the analytical DIM models from \citet{tetarenko2018,tetarenko2018b}. The best-fit (solid black line) and 1$\sigma$ confidence interval on the fit (blue shaded region) are shown. The timescale of the viscous stage of the decay (grey cross hatch region) corresponds to an alpha-viscosity parameter of $\alpha_h$=0.34. Also shown are model light-curve decay profiles corresponding to an $\alpha_h$=0.2 and $\alpha_h$=0.25, to demonstrate how light-curve profile shape changes with alpha-viscosity. The inset axis shows the data in linear space. The epochs in which Gemini/GMOS (solid blue lines) and ATCA (dotted blue lines) data were taken are also displayed. \textit{(second panel)} residuals on the light-curve fit. \textit{(third panel)} and \textit{(fourth panel)}: the bolometric corrections, computed from a combination of \textit{Swift} band-limited X-ray and broad-band (XRT+UVOT) spectral fitting, and central mass-accretion rate, over time, respectively. 
    \textit{(fifth panel)} through \textit{(bottom panel)}: the inner disc radius $R_{\rm in}$ (in units of $R_g$, assuming an $M_1=N(7.8,1.2) M_{\odot}$), inner disc temperature $T_{\rm in}$ (in keV), reprocessed X-ray fraction (${\cal C}_{\rm DISKIR}$), and outer disc radius $R_{\rm out}$ (also in units of $R_g$), computed from the broad-band (XRT+UVOT) spectral fitting (see Section \ref{sec:broadbandspec}).
    The estimated $R_{\rm out}$, corresponding to a $P_{\rm orb}=2$ hr (green shaded region) and $P_{\rm orb}=4$ hr (blue shaded region), calculated by sampling $M_{1}$ and $q$ from the observational Galactic distributions (see Section \ref{sec:orb_params}) are also shown in the \textit{(bottom panel)}.
    Both ${\cal C}_{\rm DISKIR}$ \textit{(seventh panel)} and $R_{\rm out}$ \textit{(bottom panel)} are computed assuming an average over all inclination angles. Lastly, the \textit{(third panel)} through \textit{(seventh panel)} also display the accretion state evolution of J0637, computed from the \textit{Swift} data (see Section \ref{sec:bcs}), in various shades of grey. See legend on the plot.}
    \label{fig:sed_fit_plot}
\end{figure*}

The best-fit {\tt diskir} parameters obtained could then be used to estimate the following physical parameters of the irradiated accretion disc in J0637: outer disc radius ($R_{\rm out}$), inner disc radius ($R_{\rm in}$), mass-accretion rate onto the BH ($\dot{M}_{\rm BH}$), and the fraction of X-ray flux intercepted and reprocessed in the outer disc ({$\cal C$}). See Figure \ref{fig:sed_fit_plot}.

The inner and outer disc radius are computed as follows, assuming a distance of $d=U(5,15)$ kpc and 
an inclination averaged over all angles:
\begin{equation}
    R_{\rm out}=R_{\rm in} 10^{\rm logrout},
\end{equation}
where,
\begin{equation}
    R_{\rm in}=1\times10^5 \left(\frac{d}{10 {\rm kpc}} \right)\left(\frac{N_{\rm disc}}{\cos{i}}\right)^{1/2}.
\end{equation}
Here the disc normalization term, $N_{\rm disc}$, is defined in the same way as the {\tt diskbb} model\footnote{\url{https://heasarc.gsfc.nasa.gov/xanadu/xspec/manual/node164.html}} in {\sc xspec}.

Using the resulting $R_{\rm out}$ estimates, computed from the broad-band spectral energy distribution (SED) fits during the viscous decay stage (when the whole disc is thought ot be in a hot, ionized state), we can estimate the orbital period ($P_{\rm orb}$) of J0637. Using typical binary orbital parameters of BH-LMXBs (see Section \ref{sec:orb_params} and \citealt{tetarenko2016}), we find a $P_{\rm orb}\lesssim4$ hrs (see Figure \ref{fig:sed_fit_plot}).

For a situation in which the outer part of an LMXB accretion disc is irradiated by a central X-ray source, the temperature profile of the (assumed steady-state) disc can be written as a combination of viscous \citep{frank2002} and irradiated \citep{dubus1999} portions such that,
\begin{equation}
T_{\rm eff}^4(R)=T_{\rm visc}^4(R)+T_{\rm irr}^4(R),
\label{eq:temp}
\end{equation}

\begin{figure*}
    \center
    \includegraphics[width=0.5\linewidth,height=0.4\linewidth]{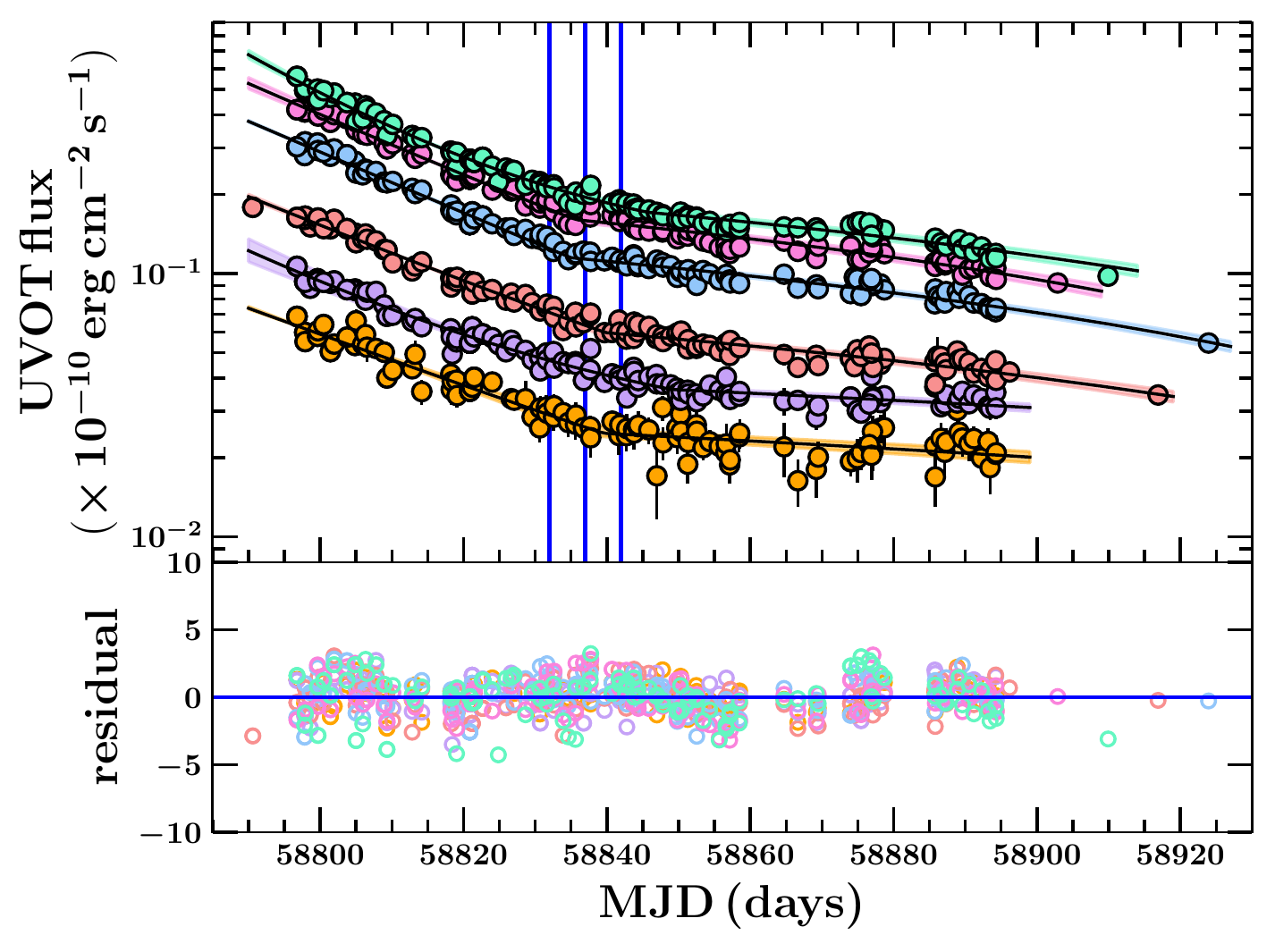}\hfill
    \includegraphics[width=0.49\linewidth,height=0.4\linewidth]{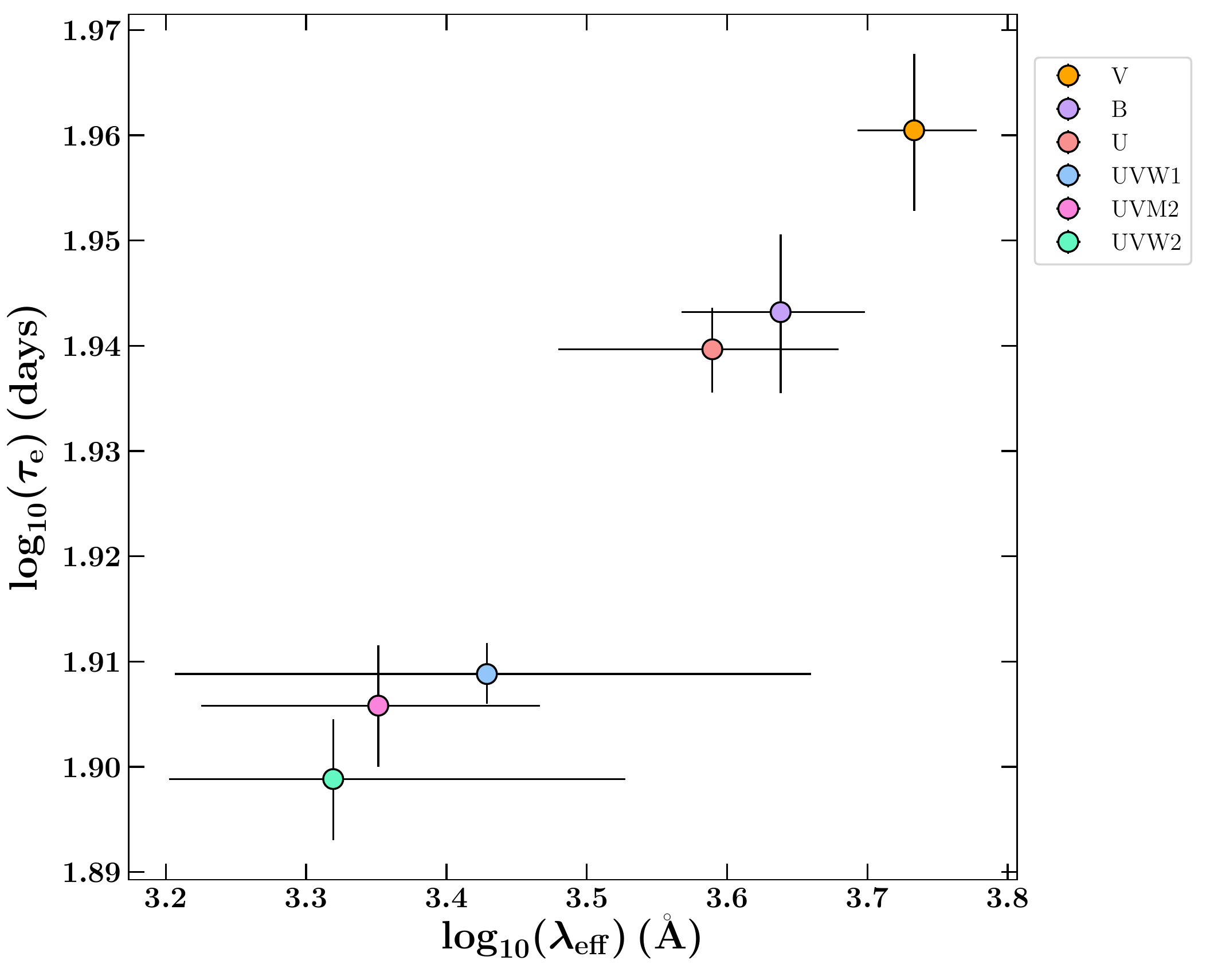}
    \caption{\textit{(left panel)} individual band-pass \textit{Swift}/UVOT light-curves, during the 2019/2020 outburst of J0637, fit with the analytical DIM models from \citet{tetarenko2018,tetarenko2018b}. The best-fit and $1\sigma$ confidence interval on the fit, for each band-pass, are displayed as solid black lines and shaded coloured regions, respectively. Also shown are the epochs in which Gemini/GMOS data were taken (solid blue lines). Residuals of each fit are displayed in the lower panel. \textit{(right panel)} correlation between exponential (viscous) decay timescale ($\tau_e$) and effective wavelength ($\lambda_{\rm eff}$) for all six UVOT band-passes. The uncertainty on wavelength is the band-pass limits of each filter.}
    \label{fig:uvot_lc_fits}
\end{figure*}

\noindent where,
\begin{equation}
T_{\rm visc}^4=\frac{3G M_1 \dot{M_{\rm BH}}}{8 \pi \sigma R^3}\left[ 1-\left(\frac{R_{\rm in}}{R}\right)^{1/2} \right],
\end{equation}
and,
\begin{equation}
T_{\rm irr}^4={\cal C} \frac{\dot{M_{\rm BH}} c^2}{4 \pi \sigma R^2}.
\label{eq:temp_irr_eq}
\end{equation}

One can write the reprocessed fraction, $\cal C$ (as defined above), in terms of {\tt diskir} parameters, as follows (see \citealt{meshcheryakov2018} for details):

\begin{equation}
    {\cal C}=\eta \sigma f_{\rm out} F_{\rm bol}^{-1}  \left(\frac{R_{\rm in}}{{\rm cm}}\right)^{2} \left(\frac{T_{\rm in}}{K}\right)^{4} \left(\frac{d}{{\rm cm}} \right)^{-2},
\end{equation}
where $F_{\rm bol}$ is the bolometric flux, obtained by integrating over the best-fit broad-band SED,and $\eta$ is the accretion efficency defined as in \citet{coriat2012},
\[ \eta=\begin{cases} 
      0.1\left(\frac{\dot{M}}{0.01 \dot{M}_{\rm edd}} \right) & L_X<0.01L_{\rm edd} \\
      0.1 & L_X\geq0.01L_{\rm edd}, 
   \end{cases}
\]
where the Eddington accretion rate is $L_{\rm edd}=0.1\dot{M}_{\rm edd}c^2$. 

Lastly, bolometric flux, $F_{\rm bol}$, can then be converted to an observed $\dot{M}_{\rm BH}$ via,
\begin{equation}
\dot{M}_{\rm BH}=\frac{F_{\rm bol}(4 \pi d^2)}{\eta c^2}.
\end{equation}

\subsubsection{Bolometric Correction}

The calculated $F_{\rm bol}$, in combination with the band-limited ($0.5-10$ keV) X-ray flux ($F_X$; see Section \ref{sec:xrayfits}), were used to compute a bolometric correction for each \textit{Swift} epoch. For (i) the first \textit{Swift} epoch (MJD 58790), in which only data from the UVOT/U filter was available, and (ii) late in the decay (post-58866), where we were unable to obtain adequate fits to the broad-band SED, standard bolometric corrections valid for LMXBs, estimated for each accretion state by \citet{migliari2006}, were assumed. See Figure \ref{fig:sed_fit_plot}.

\subsubsection{Accretion States}\label{sec:bcs}

We classify observations of J0637 into four spectral accretion states: hard (H), hard-intermediate (HIM), soft-intermediate (SIM) and soft (S), as defined in \citet{marcel2019}. Their state classification is based on two spectral signatures: the power-law fraction (PLf), defined as the ratio of the power-law flux to the total flux, and the photon index, $\gamma$. Here the PLf is computed using the best-fit to the XRT spectra (Section \ref{sec:xrayfits}), bolometric corrections (Section \ref{sec:bcs}), and $F_{\rm bol}$ (Section \ref{sec:broadbandspec}), for each \textit{Swift} epoch. See Figure \ref{fig:sed_fit_plot}.

\subsection{Light-curve Fitting}

\subsubsection{Bolometric Light-curve}\label{sec:bol_lc}
We have applied the Bayesian methodology of \citet{tetarenko2018,tetarenko2018b}, fitting the bolometric light-curve of J0637 with their analytical DIM models. We find the decay profile for the 2019/2020 outburst of J0637  to be well fit with the exponential + linear shaped decay profile, as is commonly found across the Galactic BH-LMXB population \citep{tetarenko2018}. The best-fit, as shown in Figure \ref{fig:sed_fit_plot}, gives an exponential (viscous) time-scale of $\tau_e=52.9_{-0.9}^{+0.8}$ d, a linear decay time-scale of $\tau_l=50.9_{-1.3}^{+1.4}$ d, and a transition (between viscous and irradiation-controlled stages) occurring at time (MJD) $t_{\rm break}=58832.1\pm 1.3$ and flux level of $f_t=(21.8\pm0.5)\times10^{-10} \, {\rm erg \, cm^{-2} \, s^{-1}}$. 

This best-fit profile corresponds to an: (i) alpha-viscosity parameter of $\alpha_h=0.34_{-0.05}^{+0.06}$ in the disc, and outburst-average reprocessed X-ray fraction of ${\cal C}=(9.5_{-5.6}^{+11.2})\times10^{-4}$, for an assumed $d=U(5,15)$ kpc.

Interestingly, $\tau_e$ and $\tau_l$ are comparable in the decay of J0637 as predicted by the DIM \citep{kingritter1998}. While this type of behaviour has also been seen in some short-$P_{\rm orb}$ Galactic BH-LMXBs (e.g., SwiftJ1357.2$-$0933), in the majority of the Galactic BH-LMXB population, these decay timescales have generally been found to differ significantly \citep{tetarenko2018b}.

\subsubsection{Multi-band UVOT Light-curves}

In addition, we have also fit the UVOT light-curves with these analytical DIM models, for the purpose of comparing decay timescales as measured in different wavelengths bands. Figure \ref{fig:uvot_lc_fits} shows both the best-fit models for each UVOT band-pass and the correlation 

\begin{figure*}
    \center
    \includegraphics[width=1.0\linewidth,height=0.7\linewidth]{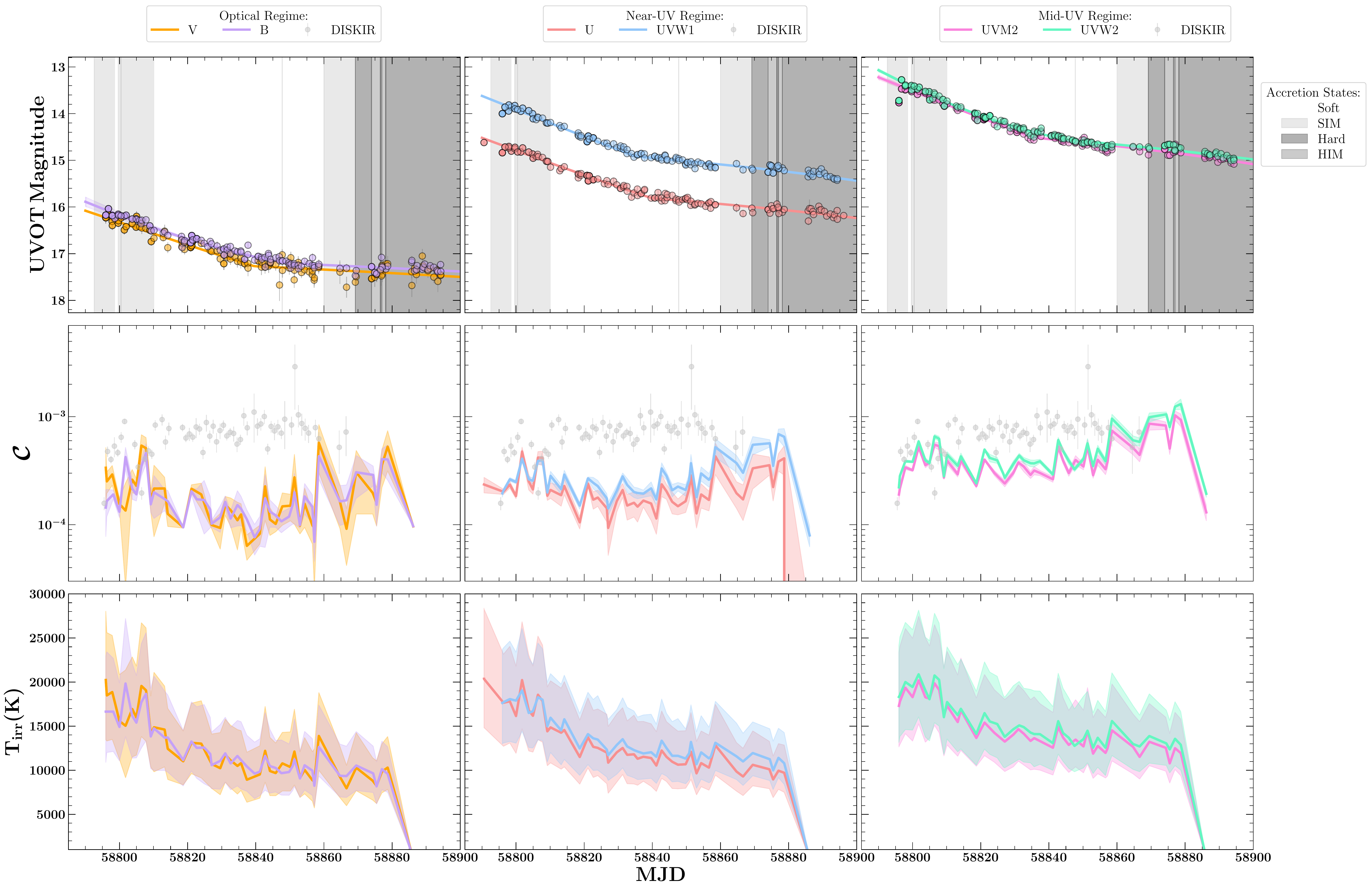}
    \caption{X-ray irradiation properties derived from directly comparing the bolometric outburst light-curve, to the optical \textit{(left)}, near-UV \textit{(middle)}, and mid-UV \textit{(right)} outburst light-curves of J0637. The \textit{(top panels)} show dereddened \textit{Swift}/UVOT magnitude as a function of time. The \textit{(middle panels)} compare
    the fraction of X-rays intercepted and reprocessed in the outer disc as a function of time ($\cal C$; solid coloured lines), calculated with the multi-wavelength light-curve method (Section \ref{sec:lc_method}), to the reprocessed fraction derived from the broad-band SED fitting (${\cal C}_{\rm DISKIR}$, grey translucent data points; Section \ref{sec:broadbandspec}). The \textit{(bottom panels)} display the irradiation temperature at the outer disc radius, $T_{\rm irr}$, calculated with the multi-wavelength light-curve method.
    Coloured shaded regions in the \textit{(middle panels)} and \textit{(bottom panels)} show the $1\sigma$ confidence interval on $\cal C$ and $T_{\rm irr}$, respectively, computed by taking into account uncertainty in X-ray flux, distance, $M_{\rm BH}$, $q$, $R_{\rm in}$, and UVOT magnitude. Shaded grey regions in the \textit{(top panels)} show the accretion state evolution of the source over time (Section \ref{sec:bcs}). See legend for details.}
    \label{fig:irr_props}
\end{figure*}

\noindent between exponential (viscous) decay timescale and effective wavelength for all six UVOT band-passes. 

Fitting the data in logspace, with a Bayesian Markov Chain Monte-Carlo algorithm (see \citealt{tetarenko2020} for full details on this method), we find a clear power-law correlation of the form: $\tau_e=N_{\rm pl}\lambda_{\rm eff}^{\beta}$, where the exponential (viscous) decay becomes faster as wavelength decreases.  The best-fit correlation yields a slope of $\beta=0.13_{-0.03}^{+0.04}$ and a normalization of $N_{\rm pl}=30.31_{-8.94}^{+9.05}$ for $\tau_e$ and $\lambda_{\rm eff}$ in units of days and angstroms, respectively.  Interestingly, such a trend is also observed in synthetic model light-curves of LMXBs (see e.g., \citealt{dubus2001}).

\subsection{Time-series Evolution of the X-ray Irradiating Source}\label{sec:lc_method}

Recently, \citet{tetarenko2020} developed a methodology that uses multi-wavelength time-series data to track how properties of the X-ray irradiation heating the discs in LMXB systems evolve over time (on daily to weekly timescales). We have applied their methodology 
to the X-ray (\textit{Swift}/XRT), and optical/UV (\textit{Swift}/UVOT) data available for J0637 (see Section \ref{sec:swift_data}). 

In doing so, we have been able to derive the time-series evolution of (i) the  fraction of X-ray intercepted and reprocessed in the outer disc $\cal C$$(t)$, and (ii) the temperature of the irradiation in the outer disc regions $T_{\rm irr}(t)$, 
throughout the 2019/2020 outburst cycle of J0637, from the light-curve data alone.

Figure \ref{fig:irr_props} displays $\cal C$$(t)$ and $T_{\rm irr}(t)$ compared to the irradiation properties derived from the broad-band (XRT+UVOT) SED fitting of the available \textit{Swift} spectral data (see Section \ref{sec:broadbandspec}). We find that (i) the mid-UV (UVM2,UVW2) UVOT bandpasses best trace the irradiated disc, and (ii) a deviation of, at most, a factor of $2-3$ between the far-UV light-curve and broad-band SED derived $\cal C$, well within the expected uncertainty of the \citet{tetarenko2020} method.

Interestingly, the outburst-averaged $\cal C$, derived from the bolometric light-curve profile (Section \ref{sec:bol_lc}), remains consistent with both irradiation properties derived from the broad-band SED fitting (Section \ref{sec:broadbandspec}) and the multi-wavelength outburst light-curves (this Section). It is also worth noting that the largest deviation between these three methods occurs during the period in which the source remained in the soft accretion state.

\subsection{Emission Line Analysis}\label{sec:line_prop_details}

The optical spectrum of J0637 is typical of what is expected for an LMXB accretion disc, with the only prominent features present being the strong double-peaked He {\sc ii} 4686 \AA\ emission lines and weak, broad features, that have a wavelength consistent with H$\alpha$ emission.
. The optical spectrum bears a resemblance to that of another short-period (candidate) BH-LMXB, Swift\,J1753.5-0127 \citep[e.g.][]{shaw2016}, most notably the relatively weak H$\alpha$ and lack of H$\beta$.  We note that the apparently strong emission features at 5285 \AA, 5420 \AA\ and 5475 \AA\ in Figure \ref{fig:flux_cal_spec} are not real. This is indicated by the fact that they are very narrow compared to the He {\sc ii} 4686 \AA\ and H$\alpha$ emission, and are not present on all 3 nights data was taken. In addition, there is also an unknown broad emission feature at $\sim4515$ \AA. While this may possibly be N {\sc iii}, there is no sign of the C {\sc iii}/N {\sc iii} Bowen blend at 4640--50 \AA. Thus, this feature remains puzzling.
See the flux-calibrated spectra, summed over the three days in which data was taken, in Figure \ref{fig:flux_cal_spec}.

\subsubsection{Deriving Line Profile Properties}
\begin{figure*}
    \centering
    \includegraphics[width=1.1\columnwidth,height=0.88\columnwidth]{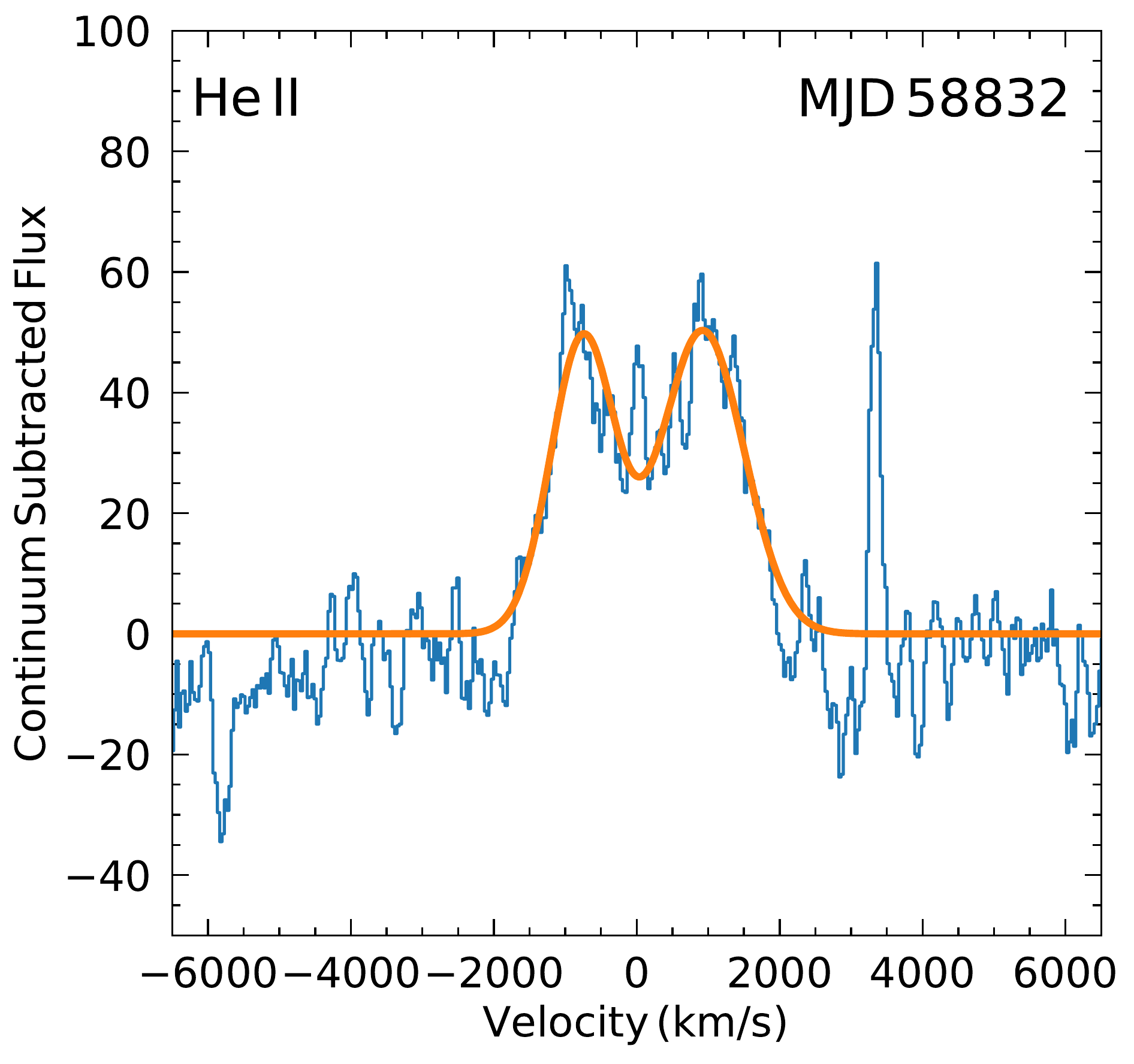}
    \includegraphics[width=1.1\columnwidth,height=0.88\columnwidth]{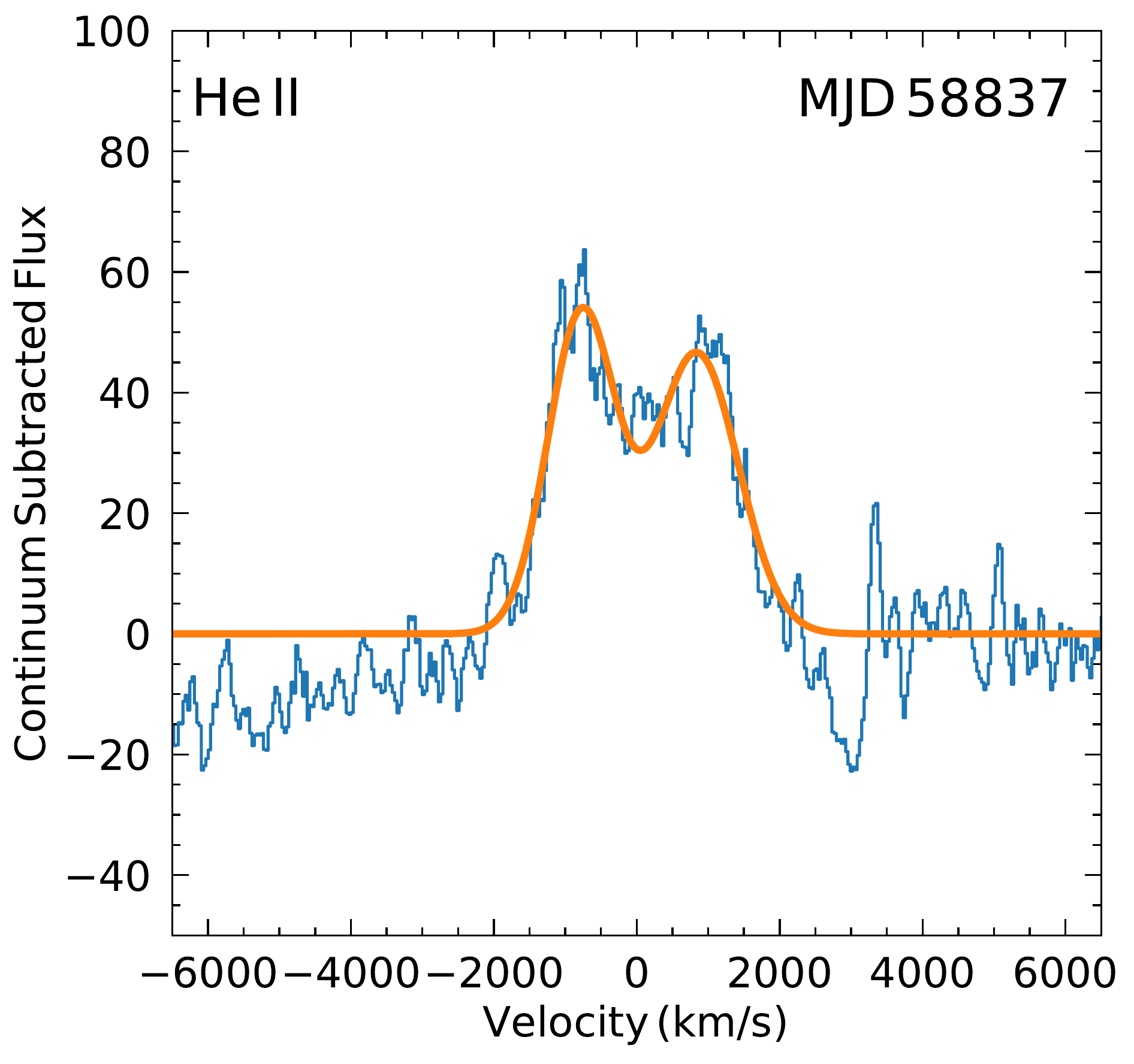}
    \includegraphics[width=1.1\columnwidth,height=0.88\columnwidth]{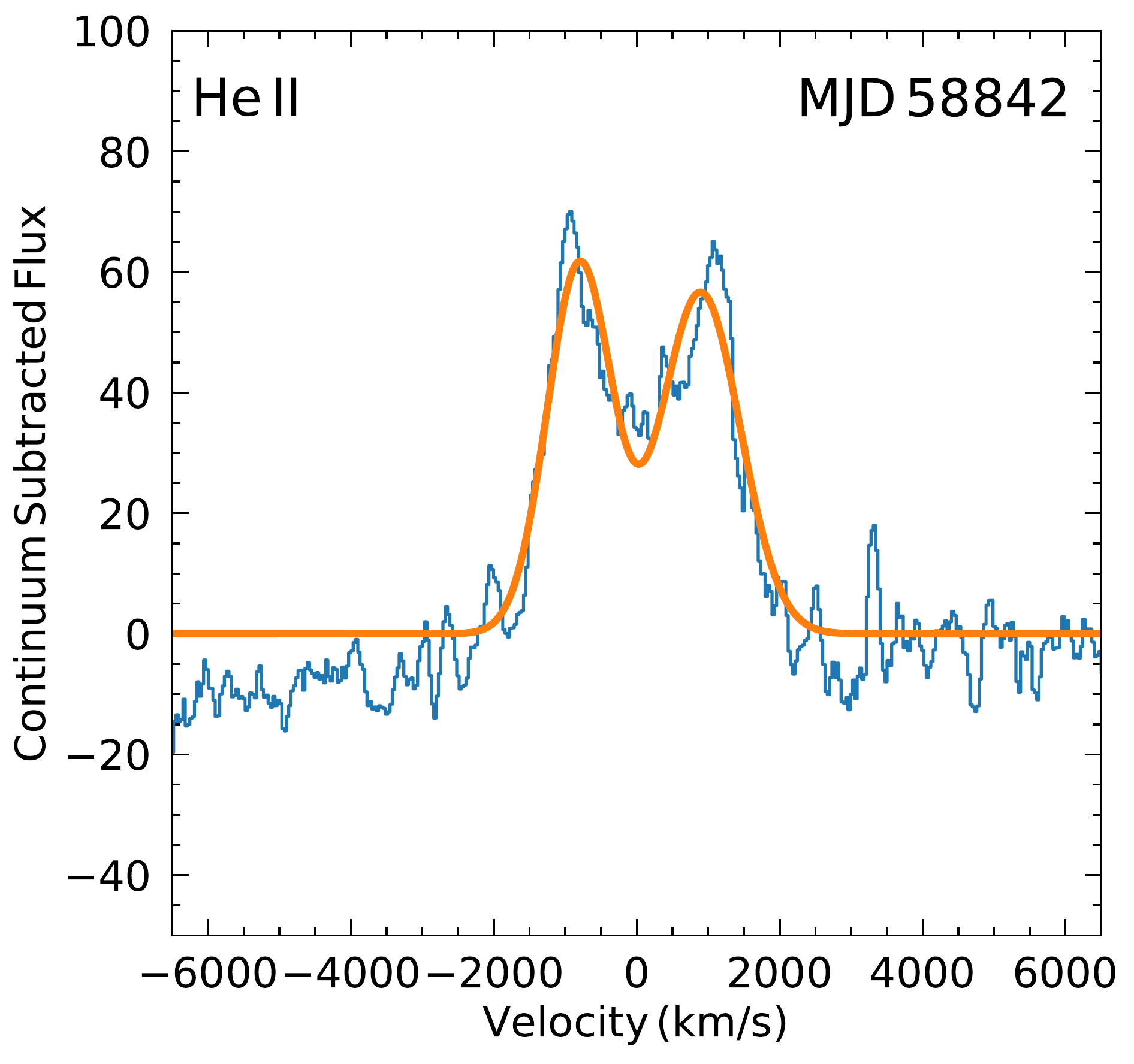} 
    \caption{Best double-peak fit to the  He {\sc ii} 4686 \AA\ emission lines found in the Gemini/GMOS daily-averaged spectra of J0637. The \textit{top}, \textit{middle}, and \textit{bottom} panels show data from MJD: 58832, 58837, and 58842, respectively. The best-fit (orange solid line) is overlaid on the data. }
    \label{fig:HeIIHalpha_fits}
\end{figure*}

For each daily-averaged spectra, we have (i) fit double gaussian line profiles to the emission lines, and (ii) computed the following emission line properties: full width half max (FWHM), equivalent width (EW), and double peak separation (DP). All analysis was done using the {\tt specutils}\footnote{\url{https://specutils.readthedocs.io/en/stable/}} package in {\sc python} and all uncertainties are purely statistical error only. See Table \ref{tab:em_line_props} and Figure \ref{fig:HeIIHalpha_fits} for results. Note that, given the complexity of the H$\alpha$ region, the accuracy of any profile parameter coming from the line fitting procedure would be suspect. For this reason, we focus our emission line analysis in this work on the He {\sc ii} 4686 \AA\ line only.

\subsubsection{Constraints on Emission Radii}\label{sec:em_radii}

We estimate the radii within the disc from which the He lines are emitted using the method of \citet{bernardini2016}. The emission radii can be defined as,
\begin{equation}
 \left(\frac{R_{\rm em}}{R_{\rm g}}\right)=\frac{1}{2}\left(c v_{\rm em} \sin i \right)^{2},
 \label{eq:emit_rad}
\end{equation}
where $i$ is the binary inclination angle and $v_{\rm em}$ is the velocity (in km/s) at the radius $R_{\rm em}$ (in units of $R_{\rm g}$). Here $v_{\rm em}$ can be estimated by computing the half-width at zero intensity (HWZI) of the emission line in question. $R_{\rm em}$ computed using this method can be thought of as a conservative upper limit on the radii in which the He emission is coming from in the disc.

\begin{figure}
    \center
    \includegraphics[width=0.99\linewidth,height=0.99\linewidth]{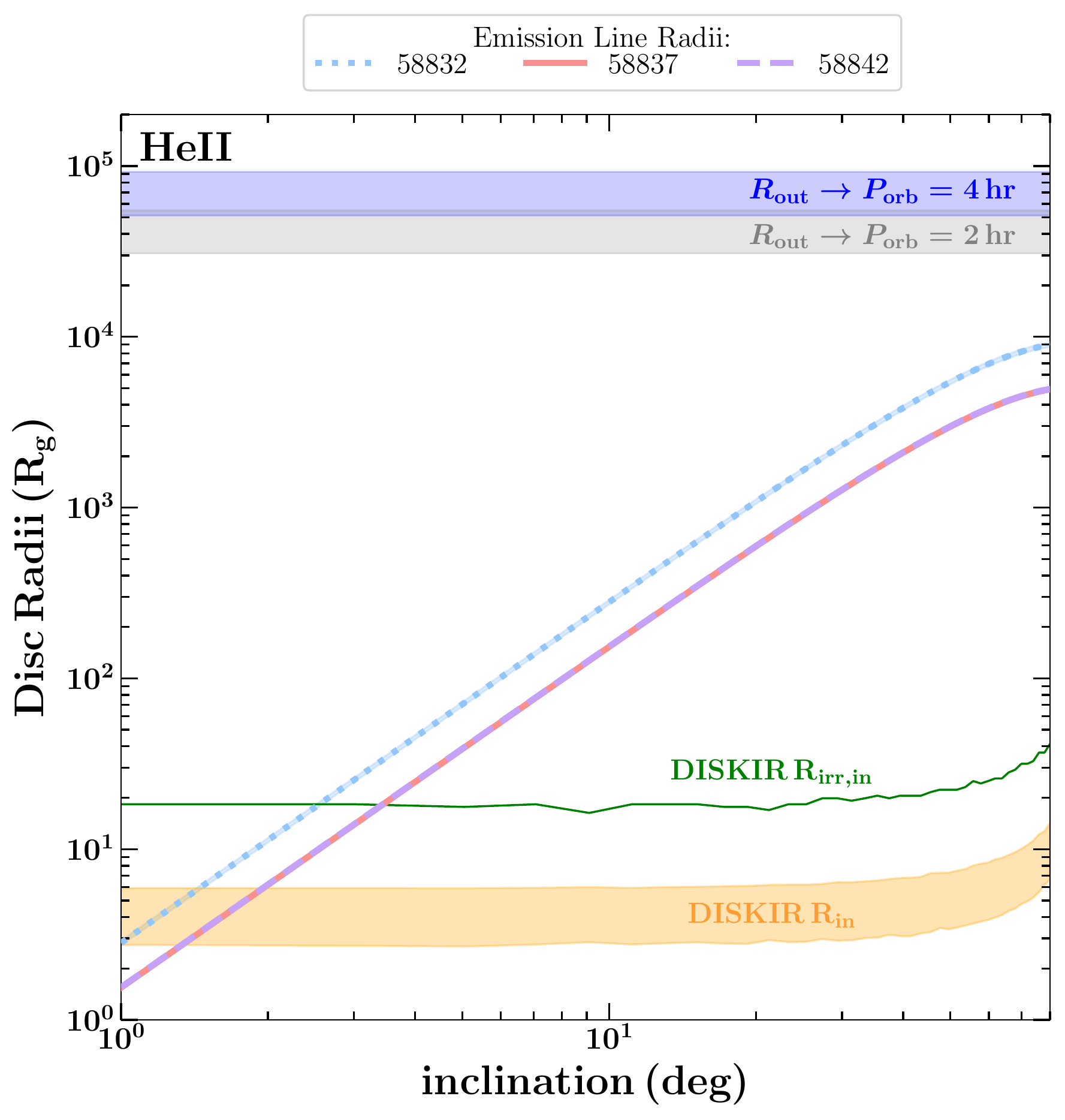}\hfill
    \caption{Radii within the disc at which the  He {\sc ii} 4686 \AA\ emission lines are emitted, as a function of binary inclination, for the three days in which Gemini/GMOS data was taken. Emission radii are compared to the inner disc radius ($R_{\rm in,DISKIR}$; orange shaded region) and inner radius of the optical/UV emitting portion of the disc ($R_{\rm irr,DISKIR}$; green shaded region) computed from the broad-band SED fits (see Section \ref{sec:broadbandspec} and \ref{sec:lc_method}). Also plotted is the estimated outer disc radius, $R_{\rm out}$, corresponding to orbital periods of 2 (grey shaded region) and 4 hrs (dark blue shaded region). This analysis rules out low inclinations for this source.}
    \label{fig:em_radii_plot}
\end{figure}

We have used the {\tt specutils} package in {\sc python} to compute the HWZI of the He {\sc ii} 4686 \AA\ emission lines present in the daily-averaged spectra (see Table \ref{tab:em_line_props}). Figure \ref{fig:em_radii_plot} displays $R_{\rm em,HeII}$ (computed using Equation \ref{eq:emit_rad}) for a wide range of inclinations, compared to the: (i) inner disc radius ($R_{\rm in,DISKIR}$) and inner radius of the optical/UV emitting portion of the disc ($R_{\rm irr,DISKIR}$) computed from the broad-band SED fits (see Section \ref{sec:broadbandspec} and \ref{sec:lc_method}), and (ii) estimated outer disc radius ($R_{\rm out}$) corresponding to different values of $P_{\rm orb}$. We observe that $R_{\rm em,HeII}$  moves inwards as the outburst evolves from late in the viscous decay to the irradiation controlled decay stage. Overall, this analysis rules out low-inclination angles for this source. This conclusion is consistent with the wide double-peak profile of the He {\sc ii} 4686 \AA\ emission. The relationship between wide emission line profiles and higher binary inclination has been noted in other BH-LMXBs as well (see e.g., \citealt{orosz1994,orosz1995,shaw2016}).

\subsection{Doppler Tomography}
To date, Doppler tomography has been used to analyze complex emission line profiles from various classes of accreting binary systems during both outburst and quiescence (see e.g., \citealt{marsh2001}).
Using emission line profiles observed through an entire orbit, this technique creates a projection of these emission lines in velocity space around the binary, effectively constructing an ``image'' of the disc on micro-arcsecond scales (see \citealt{steeghs2004,marshhorne1988,marsh2001,marsh2005}). 
 Synonymous to a CAT-scan, which uses X-ray images, taken at different angles, to reproduce an image of complex structures inside the human body, this technique uses a 2-dimensional spectral data set (i.e., time
series of line profiles) to draw a velocity-resolved blueprint (``tomogram'') of the line emission over the disc \citep{marsh2001}. 

\begin{figure*}
  \includegraphics[width=0.45\linewidth,height=.4\linewidth]{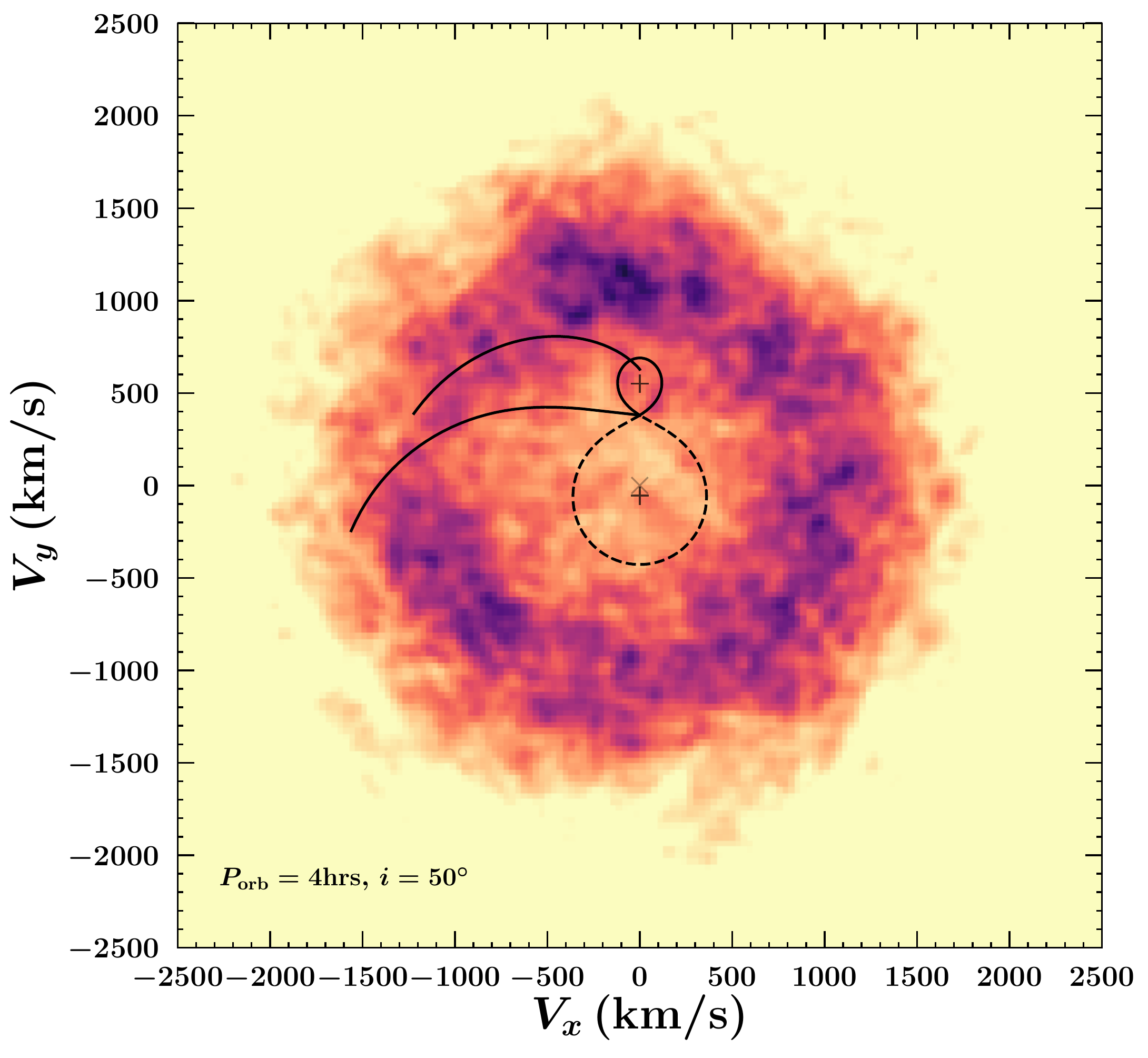}
  \includegraphics[width=0.54\linewidth,height=.46\linewidth]{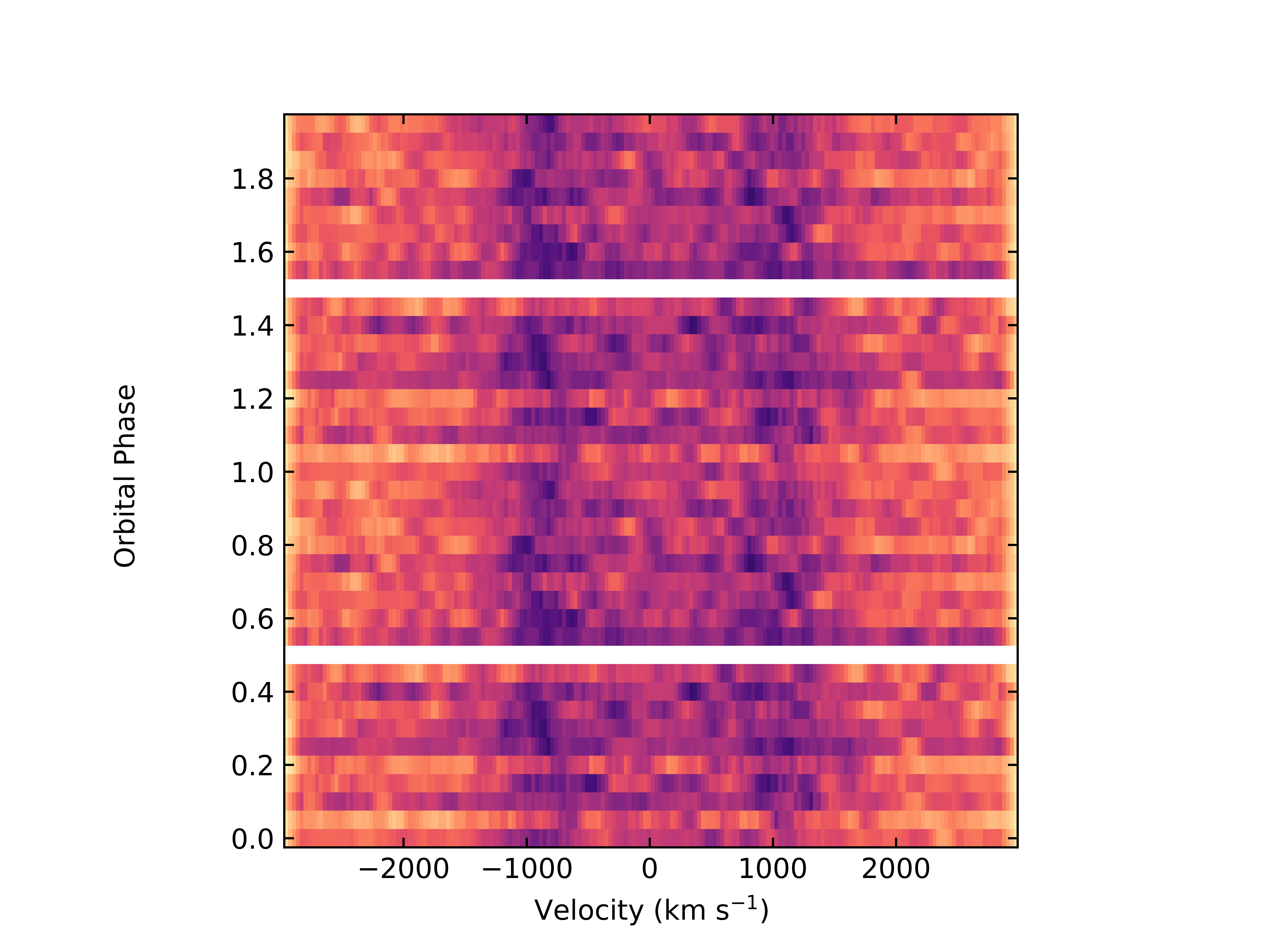}
  \caption{\textit{(left)} Doppler tomogram of He {\sc ii} 4686 \AA\ emission created from our full Gemini/GMOS data-set (7hrs on source). The tomogram is made assuming a $P_{\rm orb}=4$ hrs. The Roche lobes of the compact object (dashed black line) and companion star (solid black line) are displayed on the tomogram for an average over all binary inclination angles. \textit{(right)} trailed spectrum, phase folded on a $P_{\rm orb}=4$ hrs.}
  \label{fig:d_tomograms_heii}
 \end{figure*}

Thus, the Doppler tomography analysis technique can be used to study the structure of the disc in J0637 via orbital variation of the disc emission lines present in our Gemini/GMOS optical spectra. Accordingly, we have created doppler tomograms of the He {\sc ii} 4686 \AA\ emission (over our entire Gemini/GMOS data-set; Section \ref{sec:gem_data}) by making use of Tom Marsh's {\tt doppler}\footnote{\url{https://github.com/trmrsh/trm-doppler}} software package in {\sc python} (see \citet{marshhorne1988} for details). Figures \ref{fig:d_tomograms_heii} displays the resulting He {\sc ii} 4686 \AA\ tomograms. As the system parameters of J0637 are not yet known, we have (i) created tomograms corresponding to the estimated $P_{\rm orb}$ range ($2-4$ hrs), and (ii) over-plotted the Roche lobe of the compact object (dashed line) and companion star (solid line) using typical binary orbital parameters of BH-LMXBs (see Section \ref{sec:orb_params} and \citealt{tetarenko2016}): $M_1=7.8$, $q=0.1$. Further,  we consider both a low ($i=5^{\rm \circ}$) binary inclination angle and an average over all inclination angles when plotting the tomograms to clearly show that it is likely J0637 does not have a very low inclination.
The tomograms show He {\sc ii} 4686 \AA\ emission at large velocities. This observation is consistent with our constraints on emission radii estimated from the HWZI of each line (see Section \ref{sec:em_radii}).

\begin{table*}
	\centering
	\setlength{\tabcolsep}{2.5pt}
	\caption{Derived Emission Line Properties}
	\medskip
	\label{tab:em_line_props}
	\begin{tabular}{lccccccccccc} 
		\hline
		Time&\multicolumn{6}{c}{\underline{ Best Double-peak Fit}}&FWHM&EW&DP&HWZI&$v_{\rm em}$ \\
		&\multicolumn{3}{c}{\underline{Peak 1}}&\multicolumn{3}{c}{\underline{Peak 2}}&(km/s)&(\AA)&(\AA)&(\AA)&(km/s)\\
		(MJD)&[$\mu_1$&$\sigma_1$&$A_1$]&$[\mu_2$&$\sigma_2$&$A_2$]&&&&&\\
		\hline
		\multicolumn{12}{c}{Emission Line: He {\sc ii} 4686 \AA}\\
		\hline

58832 & ${4674.2(5)}$ & ${7.1(1)}$ & ${49(5)}$ & ${4700.4(5)}$ & ${9.0(1)}$ & ${50(5)}$ & ${2406(14)}$ & ${-1972(202)}$ & ${26.2(5)}$ & ${31.9(5)}$ & ${2041(31)}$\\[0.05cm]
58837 & ${4674.0(5)}$ & ${7.4(1)}$ & ${53(6)}$ & ${4699.1(5)}$ & ${9.1(1)}$ & ${47(6)}$ & ${2414(15)}$ & ${-2064(241)}$ & ${25.2(5)}$ & ${43.1(6)}$ & ${2759(39)}$\\[0.05cm]
58842 & ${4673.5(5)}$ & ${7.1(1)}$ & ${62(7)}$ & ${4700.0(5)}$ & ${8.6(1)}$ & ${57(6)}$ & ${2449(12)}$ & ${-2306(260)}$ & ${26.5(5)}$ & ${43.0(5)}$ & ${2752(35)}$\\[0.05cm]
		\hline
	\end{tabular}
\end{table*}

\section{Discussion}\label{sec:discuss}

\subsection{The Relationship between Irradiation-heating and Disc Emission Line Profiles}

It has become standard practice to use double-peaked emission lines in LMXB spectra to trace accretion disc dynamics (e.g., \citealt{casares2015,casares2016}). In LMXBs, the existence of these emission lines in 
outburst is thought to be the result of X-ray irradiation of the disc. Thus, we have searched for the existence of a correlation between the X-ray irradiation heating the disc in J0637 and the key properties of the H/He disc emission line profiles present in the Gemini/GMOS optical spectra. Below, we show our findings and use these correlations to demonstrate the connection between the line emitting 
regions of an LMXB disc and both the physical properties and spectrum of the X-ray irradiation heating said disc during outburst. Note that, as discussed earlier, the H$\alpha$ region of our optical spectra is quite complex, here we focus our efforts on the strongest emission line found in our optical spectra, He {\sc ii} 4686 \AA. 

Using the line properties derived from our daily-averaged Gemini/GMOS spectra (see Section \ref{sec:line_prop_details} and Table \ref{tab:em_line_props}), we find evidence for positive correlations existing between (i) the fraction of 
 X-rays intercepted and reprocessed in the outer disc ($\cal C$) and both the FWHM and EW of the He {\sc ii} 4686 \AA\ line, and (ii) the irradiation temperature in the outer disc region ($T_{\rm irr}$) and both the FWHM and EW of the He {\sc ii} 4686 \AA\ line.
 
\begin{figure}
    \center
    \includegraphics[width=1.0\linewidth,height=0.75\linewidth]{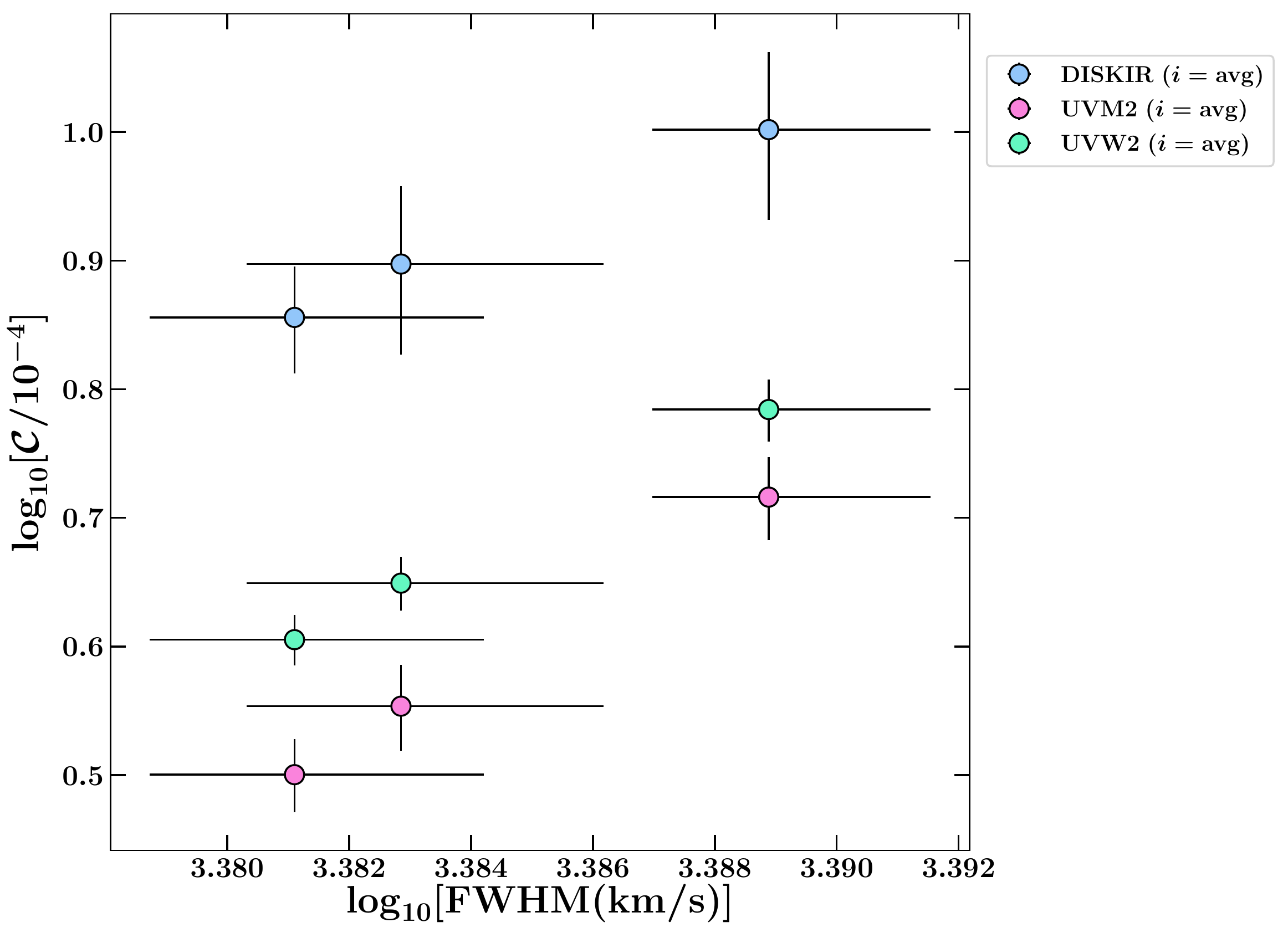}\hfill
    \includegraphics[width=1.0\linewidth,height=0.75\linewidth]{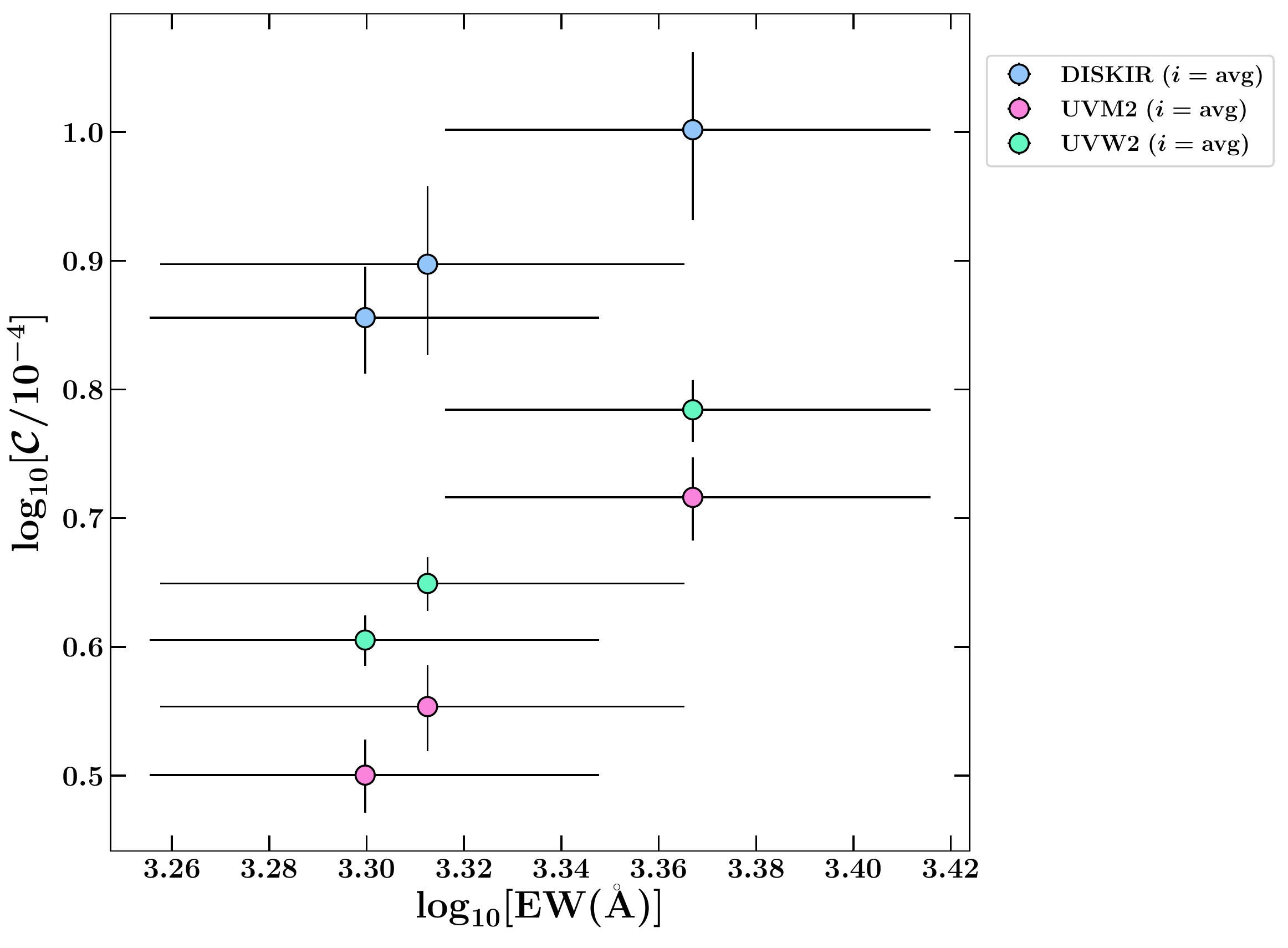}
    \caption{Correlation between the fraction of X-rays intercepted and reprocessed in the outer disc ($\cal C$) and \textit{(top)} FWHM (in km/s) and \textit{(bottom)} EW (in \AA) of the He {\sc ii} 4686 \AA\ disc emission line. Positive correlations are plotted using X-ray irradiation heating computed from the (i) broad-band SED fits (DISKIR), assuming an average over all inclination angles (blue circles), and (ii) light-curve data in both the UVW2 (green circles) and UVM2 (pink circles) UVOT filters.}
    \label{fig:corr_c_lines}
\end{figure}

\begin{figure}
    \center
    \includegraphics[width=1.0\linewidth,height=0.75\linewidth]{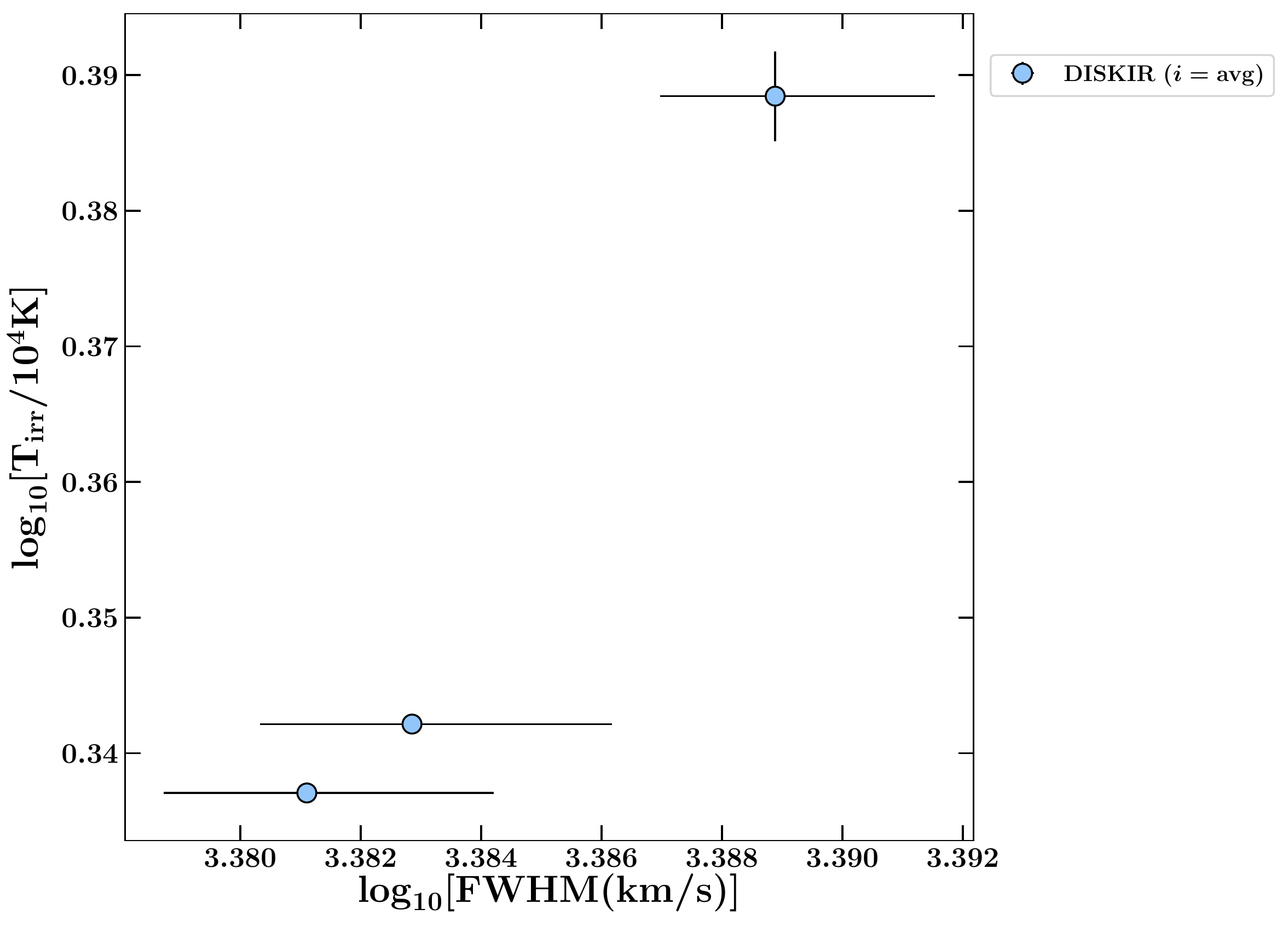}\hfill
    \includegraphics[width=1.0\linewidth,height=0.75\linewidth]{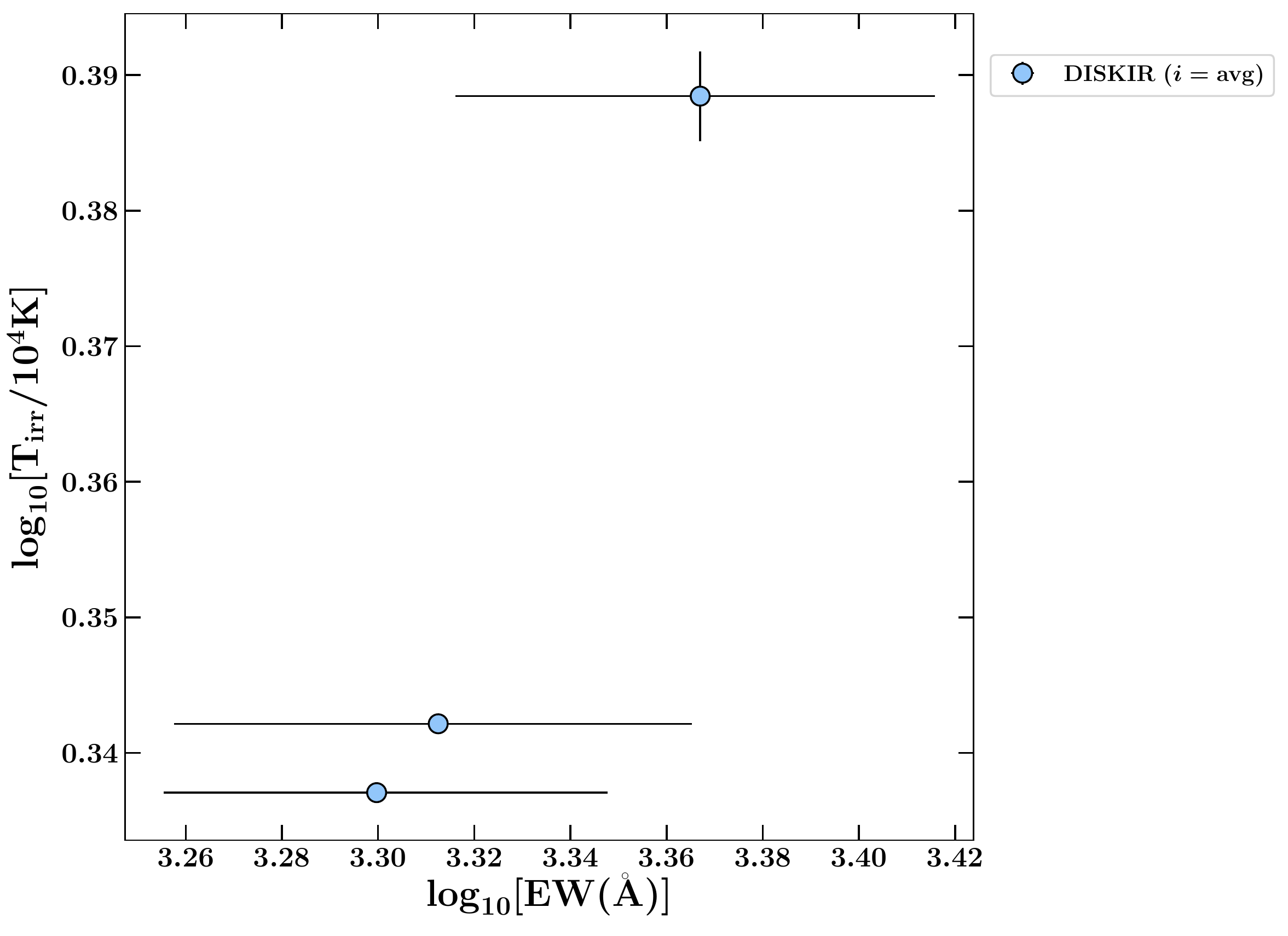}
    \caption{Correlation between the irradiation temperature in the outer disc region ($T_{\rm irr}$) and \textit{(top)} FWHM (in km/s) and \textit{(bottom)} EW (in \AA) of the He {\sc ii} 4686 \AA\ disc emission line. Positive correlations are plotted using X-ray irradiation heating computed from the broad-band SED fits (DISKIR), assuming an average over all inclination angles (blue circles).}
    \label{fig:corr_t_lines}
\end{figure}

 Figures \ref{fig:corr_c_lines} and \ref{fig:corr_t_lines} plots the correlations, found when making use of both (i) estimates of ${\cal C}_{\rm DISKIR}$, computed from the broad-band SED fits (see Section \ref{sec:broadbandspec}), as well as (ii) the $\cal C$ computed from a combination of bolometric and far-UV (UVM2, UVW2) light-curves (see Section \ref{sec:lc_method}), and (iii) $T_{\rm irr,DISKIR}$, computed from the broad-band SED fits (see Section \ref{sec:broadbandspec}). Unfortunately, $T_{\rm irr}$ is not well enough constrained using the light-curve method,
largely due to our lack of knowledge of the orbital parameters of the system. Thus, we are not able to find a correlation with the emission line properties using this data.

Using the Spearman rank-order test\footnote{\url{https://docs.scipy.org/doc/scipy/reference/generated/scipy.stats.spearmanr.html}} from the {\tt scipy} package
in {\sc python}, we test for a positive correlation existing in each data-set.
This algorithm calculates the Spearman correlation coefficient and its associated p-value. The correlation coefficent varies between $-1$ and $1$, with a value of zero implying no correlation exists. The p-value represents the probability of an uncorrelated system producing the same correlation as the one computed from the data. To take into account uncertainties in the data, we perform monte-carlo sampling (1000 samples) over the the errors in each parameter, and take the mean of the sampled distribution as the final result. 

For the FWHM vs. $\cal C$ correlation, the Spearman rank-order test results in  (correlation coefficient, p-value) of (0.97,0.036), (1.0,0.0), and (1.0,0.0), for the FWHM correlation with ${\cal C}_{\rm DISKIR, avg}$, ${\cal C}_{\rm UVW2,lc}$, and ${\cal C}_{\rm UVM2,lc}$, respectively. 
For the EW vs. $\cal C$ correlation, the Spearman rank-order test results in  (correlation coefficient, p-value) of (0.99,0.015), (0.98,0.027), and (0.98,0.020), for the EW correlation with ${\cal C}_{\rm DISKIR,avg}$, ${\cal C}_{\rm UVW2,lc}$, and ${\cal C}_{\rm UVM2,lc}$, respectively. 
For the FWHM vs. $T_{\rm irr,DISKIR}$, and EW vs. $T_{\rm irr,DISKIR}$, 
correlations, the Spearman rank-order test results in (correlation coefficient, p-value) of (1.0,0.0) and (0.98,0.027), respectively. Note that, for small data-sets like these ones, the stand-alone p-values are not particularly meaningful.

Changes in emission line profile shape (single-peaked vs double-peaked), as well emission line properties themselves (e.g., EW), during outburst have also been previously associated with spectral accretion state in BH-LMXBs (e.g., GROJ1655$-$40; \citealt{soria2000}). Unfortunately, all of our Gemini/GMOS epochs were taken when J0637 was in the soft accretion state and no significant changes to properties of the X-ray spectrum irradiating  the disc occurred during the 15-day time-period in which our Gemini/GMOS observations took place.  While we cannot test for correlations between the He {\sc ii} 4686 \AA\ emission line profile and spectral accretion state in our data, we do note that the double-peaked shape is a property that has been previously associated with soft-state BH-LMXBs  \citep{soria2000}.

These statistical results individually, combined with the fact that we observe statistical evidence for a positive correlation existing between the irradiation heating ($\cal C$ and $T_{\rm irr}$) and both the FWHM and EW of the He {\sc ii} 4686 \AA\ emission line, for irradiation properties computed using two completely independent methods (broad-band spectral fitting vs multi-wavelength time-series anlaysis), suggests that (i) the double-peaked emission line profiles in BH-LMXBs are likely the result of X-ray irradiation of the disc, and (ii) changes in physical properties of the irradiation-heating over an outburst can possibly be imprinted within the line profile properties themselves. Ultimately, more spectral data, taken over a larger range of time, flux, and spectral accretion state, during outburst, will allow us to put firmer constraints on the relationship between irradiation heating and disc emission line properties in BH-LMXBs.

\section{Summary}\label{sec:summary}

The double-peaked profiles of the H and He recombination lines observed in optical spectra can be used as powerful diagnostics for mapping the geometry and structure of the accretion discs in LMXB systems. In LMXBs, the detection of these emission lines during outburst is thought to be the result of X-ray irradiation heating the disc. This irradiation-heating plays a key role in regulating outburst cycles in these binary systems. By illuminating the disc surface, it controls both the overall evolution of an outburst from peak to quiescence, and the amount of mass able to be accreted during the outburst\citep{dubus1999,dubus2001}. 
However, despite decades of theoretical and observational efforts, the physical mechanism(s) driving X-ray irradiation remains largely unknown. 
The problem is three-fold. Neither (i) the disc radial profile, nor (ii) the geometry of the source heating such discs, nor (iii) the effect of the illuminating spectrum on the thermal properties of the disc, are well understood
 \citep{dubus1999,dubus2001,vrtilek1990,king1996}.
Answers to these open questions lie in understanding the evolution of the physical parameters of the irradiation source throughout an outburst cycle as functions of time and disc radius. These parameters include: the fraction of the X-ray flux intercepted and reprocessed in the outer disc ($\cal C$), the temperature at which the disc surface is heated ($T_{\rm irr}$), and the spectrum of the irradiation source \citep{tetarenko2018b,esin2000a}.

Accordingly, we have combined phase-resolved optical spectroscopy of new candidate BH-LMXB J0637, with multi-wavelength photometry, in an attempt to track and quantify how variations in irradiation-heating over an LMXB outburst affect the physical properties of the disc through its emission line profiles.
We were able to monitor this system (on daily-weekly timescales) through its 8 month long outburst with \textit{Swift}, obtaining time-series optical and UV photometry with UVOT, simultaneous with X-ray spectroscopic data from XRT. In addition, we were also able to obtain optical spectroscopic data from Gemini/GMOS during this time. These data are quasi-simultaneous over three individual days of the \textit{Swift} monitoring, covering the end of the first viscous decay stage and transition to the irradiation-controlled decay stage in the light-curve.

Using the \textit{Swift} UV, optical, and X-ray light-curves and spectra, we first derived the time-series evolution of key properties of the X-ray irradiation source heating the disc in J0637 over the entire outburst cycle using two independent observational methods. The first method involved fitting the broad-band (XRT+UVOT) SED with an irradiated disc model. The second method involved the recently developed empirical algorithm by \citet{tetarenko2020}, which makes use of multi-wavelength light-curves to build a three-dimensional X-ray irradiation profile (reprocessed fraction $\cal C$, irradiation temperature $T_{\rm irr}$, and size of the irradiated region of the disc $R_{\rm irr}$) of the disc. Second, we fit double-peaked Gaussian profiles to the strongest emission line, He {\sc ii} 4686 \AA, present in our daily-averaged Gemini/GMOS optical spectra. From here we were able to derive key properties defining these line profiles: FWHM, EW, DP seperation, and HWZI. Lastly, we built doppler tomograms, to analyze where in the disc the (irradiation-tracing) He {\sc ii} 4686 \AA\ emission was coming from.

Then, we looked for a correlation between the X-ray irradiation properties heating the disc in J0637 and the key parameters defining the detected He {\sc ii} 4686 \AA\ emission line profiles. Making use of the Spearman rank-order test, we find statistical evidence for positive correlations existing between $\cal C$ and both the FWHM and EW of the He {\sc ii} 4686 \AA\ line, and $T_{\rm irr}$ and both the FWHM and EW of the He {\sc ii} 4686 \AA\ line.
These positive correlations exist when making use of the X-ray irradiation properties derived from two independent observational methods, broad-band SED fitting and the light-curve comparision modelling of \citet{tetarenko2020}.
These results provide both evidence for He recombination line profiles in LMXBs being caused by irradiation of the disc, and demonstrate that changes in physical properties of the irradiation-heating of the disc over an LMXB outburst can be imprinted within the He line profile themselves.

The pilot-program presented here is the first crucial step toward our overall goal of developing a method to use phase-resolved spectroscopy to effectively probe the geometry and structure of the gas making up an outbursting LMXB accretion disc. However, what this pilot-study lacks is high-quality, sequential spectral data-sets, taken over a larger range of flux, and during different stages of the outburst decay, where the irradiating spectrum heating the disc changes significantly in shape. Future data will not only allow us to place firm constraints on the relationship we have found between irradiation-heating and disc emission line properties in LMXBs, but will also allow us to use multi-epoch tomography on higher time/wavelength resolution data, delivering a more coherent picture of accretion disc dynamics that one-off maps produced here do not \citep{marsh2001,marsh2005}.

\section*{Acknowledgements}
The authors would like to thank the anonymous referee for their insightful comments that improved this manuscript.
BET would like to thank (i) J. Turner and the Gemini Staff for helpful advice regarding the Gemini/GMOS data reduction, (ii) G. Dubus for helpful discussions and suggestions that improved the manuscript, and (iii) J.V. Hernandez Santisteban for use of his {\sc python} scripts to plot doppler tomograms.
BET acknowledges support from the University of Michigan through the McLaughlin Fellowship. This research is, in part, based on observations obtained at the international Gemini Observatory, a program of NSF’s NOIRLab, which is managed by the Association of Universities for Research in Astronomy (AURA) under a cooperative agreement with the National Science Foundation on behalf of the Gemini Observatory partnership: the National Science Foundation (United States), National Research Council (Canada), Agencia Nacional de Investigaci\'{o}n y Desarrollo (Chile), Ministerio de Ciencia, Tecnolog\'{i}a e Innovaci\'{o}n (Argentina), Minist\'{e}rio da Ci\^{e}ncia, Tecnologia, Inova\c{c}\~{o}es e Comunica\c{c}\~{o}es (Brazil), and Korea Astronomy and Space Science Institute (Republic of Korea). We acknowledge with thanks the variable star observations from the AAVSO International Database contributed by observers worldwide and used in this research. The Australia Telescope Compact Array (ATCA) is part of the Australia Telescope National Facility which is funded by the Australian Government for operation as a National Facility managed by CSIRO. We acknowledge the Gomeroi people as the traditional owners of the ATCA Observatory site.
This research has also made use of: (i) data, software, and/or web tools obtained from the High Energy Astrophysics Science Archive Research Center (HEASARC), a service of the Astrophysics Science Division at NASA Goddard Space Flight Center (GSFC) and of the Smithsonian Astrophysical Observatory’s High Energy Astrophysics Division, (ii) data supplied by the UK \textit{Swift} Science Data Centre at the University of Leicester, and (iii) NASA’s Astrophysics Data System (ADS).

\section*{Data Availability}

The observational data presented in this work are available online in the HEASARC Archive (Swift ObsIds: 00012168001-00012172094;\url{https://heasarc.gsfc.nasa.gov/docs/archive.html}), Gemini Data Archive (Program ID: GS-2019B-Q-233, PI Tetarenko; \url{https://archive.gemini.edu/}), AAVSO International Database (\url{https://www.aavso.org}), and the Australia Telescope Online Archive (\url{https://atoa.atnf.csiro.au/query.jsp}).



\bibliographystyle{mnras}
\bibliography{J0637_refs.bib}



\appendix

\section{Gemini Data Details}\label{sec:appC}

\onecolumn{
\begin{longtable}{lccccc}
\caption{Gemini/GMOS Long-slit Data Details}\label{tab:gemtab}\\
		\hline
		Number&ObsID&Date&Time Start&Time End&Exposure \\
		&&&(UT)&(UT)&Time (s) \\
		\hline
\endhead
\endfoot
1 & 20191215S0088 & 2019-12-15 & 05:15:42.2334 & 05:25:42.2334 & 600\\[0.05cm]
2 & 20191215S0089 & 2019-12-15 & 05:26:34.1167 & 05:36:34.1167 & 600\\[0.05cm]
3 & 20191215S0090 & 2019-12-15 & 05:37:26.4334 & 05:47:26.4334 & 600\\[0.05cm]
4 & 20191215S0091 & 2019-12-15 & 05:48:19.4333 & 05:58:19.4334 & 600\\[0.05cm]
5 & 20191215S0092 & 2019-12-15 & 05:59:11.3000 & 06:09:11.3000 & 600\\[0.05cm]
6 & 20191215S0093 & 2019-12-15 & 06:10:03.1667 & 06:20:03.1667 & 600\\[0.05cm]
7 & 20191215S0094 & 2019-12-15 & 06:20:54.2167 & 06:30:54.2167 & 600\\[0.05cm]
8 & 20191215S0095 & 2019-12-15 & 06:31:46.2334 & 06:41:46.2334 & 600\\[0.05cm]
9 & 20191215S0096 & 2019-12-15 & 06:42:39.3833 & 06:52:39.3834 & 600\\[0.05cm]
10 & 20191215S0097 & 2019-12-15 & 06:53:30.2334 & 07:03:30.2334 & 600\\[0.05cm]
11 & 20191215S0098 & 2019-12-15 & 07:04:22.1334 & 07:14:22.1334 & 600\\[0.05cm]
12 & 20191220S0054 & 2019-12-20 & 04:09:12.3834 & 04:19:12.3834 & 600\\[0.05cm]
13 & 20191220S0055 & 2019-12-20 & 04:20:03.1500 & 04:30:03.1500 & 600\\[0.05cm]
14 & 20191220S0056 & 2019-12-20 & 04:30:54.1334 & 04:40:54.1334 & 600\\[0.05cm]
15 & 20191220S0057 & 2019-12-20 & 04:41:45.1500 & 04:51:45.1500 & 600\\[0.05cm]
16 & 20191220S0058 & 2019-12-20 & 04:52:36.1334 & 05:02:36.1334 & 600\\[0.05cm]
17 & 20191220S0059 & 2019-12-20 & 05:03:28.4667 & 05:13:28.4667 & 600\\[0.05cm]
18 & 20191220S0060 & 2019-12-20 & 05:14:20.1167 & 05:24:20.1167 & 600\\[0.05cm]
19 & 20191220S0061 & 2019-12-20 & 05:25:11.834 & 05:35:11.834 & 600\\[0.05cm]
20 & 20191220S0062 & 2019-12-20 & 05:36:02.834 & 05:46:02.834 & 600\\[0.05cm]
21 & 20191220S0063 & 2019-12-20 & 05:46:53.4000 & 05:56:53.4000 & 600\\[0.05cm]
22 & 20191220S0064 & 2019-12-20 & 05:57:45.1500 & 06:07:45.1500 & 600\\[0.05cm]
23 & 20191225S0311 & 2019-12-25 & 03:30:53.8167 & 03:40:53.8167 & 600\\[0.05cm]
24 & 20191225S0312 & 2019-12-25 & 03:41:44.9500 & 03:51:44.9500 & 600\\[0.05cm]
25 & 20191225S0313 & 2019-12-25 & 03:52:36.8167 & 04:02:36.8167 & 600\\[0.05cm]
26 & 20191225S0314 & 2019-12-25 & 04:03:28.9833 & 04:13:28.9834 & 600\\[0.05cm]
27 & 20191225S0315 & 2019-12-25 & 04:14:20.9667 & 04:24:20.9667 & 600\\[0.05cm]
28 & 20191225S0316 & 2019-12-25 & 04:25:12.8833 & 04:35:12.8834 & 600\\[0.05cm]
29 & 20191225S0317 & 2019-12-25 & 04:36:04.8167 & 04:46:04.8167 & 600\\[0.05cm]
30 & 20191225S0318 & 2019-12-25 & 04:46:57.1167 & 04:56:57.1167 & 600\\[0.05cm]
31 & 20191225S0319 & 2019-12-25 & 04:57:48.9667 & 05:07:48.9667 & 600\\[0.05cm]
32 & 20191225S0320 & 2019-12-25 & 05:08:40.9333 & 05:18:40.9334 & 600\\[0.05cm]
33 & 20191225S0321 & 2019-12-25 & 05:19:32.9500 & 05:29:32.9500 & 600\\[0.05cm]
34 & 20191225S0323 & 2019-12-25 & 05:31:52.9500 & 05:41:52.9500 & 600\\[0.05cm]
35 & 20191225S0324 & 2019-12-25 & 05:42:44.9333 & 05:52:44.9334 & 600\\[0.05cm]
36 & 20191225S0325 & 2019-12-25 & 05:53:34.9833 & 06:03:34.9834 & 600\\[0.05cm]
37 & 20191225S0326 & 2019-12-25 & 06:04:28.9000 & 06:14:28.9000 & 600\\[0.05cm]
38 & 20191225S0327 & 2019-12-25 & 06:15:21.834 & 06:25:21.834 & 600\\[0.05cm]
39 & 20191225S0328 & 2019-12-25 & 06:26:11.9667 & 06:36:11.9667 & 600\\[0.05cm]
40 & 20191225S0329 & 2019-12-25 & 06:37:04.8333 & 06:47:04.8334 & 600\\[0.05cm]
41 & 20191225S0330 & 2019-12-25 & 06:47:56.9500 & 06:57:56.9500 & 600\\[0.05cm]
42 & 20191225S0331 & 2019-12-25 & 06:58:48.9500 & 07:08:48.9500 & 600\\[0.05cm]
    \hline
\end{longtable}
}

\clearpage

\section{Detailed \textit{Swift} Spectral Fit Data}\label{sec:appB}

\onecolumn{
\begin{longtable}{lcccccccc}
\caption{X-ray (XRT) Spectral Fits for MAXI\,J0637$-$430}\label{tab:xrayfitstab}\\
		\hline
		ObsID$^a$&MJD&model fit$^b$&$F_X$&$T_{\rm in}$&$\gamma$&$N_{\rm pl}$&reduced&dof \\
		&(days)&&($\times10^{-10} \,\rm{ erg \,cm^{-2} \,s^{-1}}$)&(keV)&&($\times10^{-4}$)&$\chi^2_\nu$& \\
		\hline
\endhead
\endfoot
12168001a & 58790.5828 & 0 & ${90.3}^{+5.0}_{-4.8}$ & ${0.681}^{+0.039}_{-0.037}$ & $\cdots$ & $\cdots$ & 0.98 & 110\\[0.05cm]
    12172001b & 58794.0326 & 0 & $12.20 \pm 0.13$ & $0.6786 \pm 0.0073$ & $\cdots$ & $\cdots$ & 1.3 & 360\\[0.05cm]
    12172002 & 58795.9662 & 2 & ${85.44}^{+0.77}_{-0.76}$ & ${0.782}^{+0.012}_{-0.011}$ & ${1.76}^{+0.13}_{-0.17}$ & ${2.10}^{+0.61}_{-0.63}$ & 1.4 & 310\\[0.05cm]
    12172003 & 58796.7424 & 2 & $57.53 \pm 0.34$ & ${0.7337}^{+0.0077}_{-0.0075}$ & ${1.959}^{+0.058}_{-0.062}$ & $2.70 \pm 0.33$ & 1.3 & 380\\[0.05cm]
    12172004a & 58797.8193 & 2 & ${65.39}^{+0.40}_{-0.41}$ & ${0.7180}^{+0.0078}_{-0.0076}$ & ${1.878}^{+0.065}_{-0.073}$ & ${2.29}^{+0.34}_{-0.33}$ & 1.1 & 370\\[0.05cm]
    12172004b & 58797.8815 & 2 & $62.14 \pm 0.40$ & ${0.7211}^{+0.0081}_{-0.0079}$ & ${1.959}^{+0.073}_{-0.080}$ & $2.37 \pm 0.36$ & 1.1 & 390\\[0.05cm]
    12172005 & 58798.6026 & 0 & ${59.97}^{+0.66}_{-0.65}$ & ${0.718}^{+0.012}_{-0.011}$ & ${1.79}^{+0.19}_{-0.33}$ & ${1.21}^{+0.53}_{-0.58}$ & 1.2 & 290\\[0.05cm]
    12172006a & 58799.6076 & 2 & $66.01 \pm 0.38$ & ${0.6713}^{+0.0064}_{-0.0063}$ & ${2.139}^{+0.071}_{-0.075}$ & $1.59 \pm 0.21$ & 1.2 & 370\\[0.05cm]
    12172006b & 58799.6768 & 2 & ${58.56}^{+0.34}_{-0.33}$ & ${0.6786}^{+0.0065}_{-0.0063}$ & ${1.92}^{+0.09}_{-0.10}$ & $1.33 \pm 0.24$ & 1.0 & 390\\[0.05cm]
    12172007a & 58800.3939 & 0 & $73.0 \pm 3.8$ & ${0.623}^{+0.040}_{-0.038}$ & $\cdots$ & $\cdots$ & 1.1 & 73\\[0.05cm]
    12172007b & 58800.4601 & 2 & $65.72 \pm 0.52$ & ${0.6273}^{+0.0090}_{-0.0087}$ & ${2.18}^{+0.14}_{-0.16}$ & $1.00 \pm 0.23$ & 1.1 & 390\\[0.05cm]
    12172008a & 58801.4022 & 2 & $39.93 \pm 0.29$ & ${0.6414}^{+0.0078}_{-0.0075}$ & ${2.17}^{+0.13}_{-0.15}$ & ${0.89}^{+0.20}_{-0.21}$ & 1.2 & 310\\[0.05cm]
    12172008b & 58801.9278 & 2 & $52.12 \pm 0.26$ & ${0.6664}^{+0.0054}_{-0.0053}$ & $2.314 \pm 0.073$ & $1.22 \pm 0.14$ & 1.4 & 320\\[0.05cm]
    12172010a & 58804.8447 & 2 & $41.20 \pm 0.21$ & ${0.6543}^{+0.0061}_{-0.0060}$ & ${2.177}^{+0.056}_{-0.058}$ & ${1.80}^{+0.19}_{-0.18}$ & 1.2 & 400\\[0.05cm]
    12172010b & 58804.9950 & 2 & $49.54 \pm 0.39$ & ${0.6376}^{+0.0097}_{-0.0094}$ & ${2.12}^{+0.09}_{-0.10}$ & ${1.71}^{+0.30}_{-0.29}$ & 1.2 & 410\\[0.05cm]
    12172011 & 58805.9082 & 2 & $34.39 \pm 0.25$ & ${0.5791}^{+0.0091}_{-0.0087}$ & ${2.116}^{+0.051}_{-0.055}$ & ${2.30}^{+0.31}_{-0.29}$ & 1.1 & 450\\[0.05cm]
    12172012 & 58806.3129 & 2 & ${31.75}^{+0.32}_{-0.31}$ & $0.596 \pm 0.012$ & ${2.149}^{+0.077}_{-0.082}$ & ${2.23}^{+0.42}_{-0.38}$ & 1.1 & 370\\[0.05cm]
    12172013 & 58807.7646 & 2 & $44.49 \pm 0.20$ & ${0.6221}^{+0.0052}_{-0.0051}$ & ${2.081}^{+0.050}_{-0.053}$ & $1.63 \pm 0.16$ & 1.6 & 520\\[0.05cm]
    12172014 & 58808.8953 & 2 & $41.74 \pm 0.18$ & ${0.6175}^{+0.0048}_{-0.0047}$ & ${2.196}^{+0.060}_{-0.063}$ & $1.22 \pm 0.13$ & 1.3 & 490\\[0.05cm]
    12172015 & 58809.2947 & 2 & $41.35 \pm 0.28$ & ${0.5910}^{+0.0073}_{-0.0070}$ & $2.27 \pm 0.10$ & $1.01 \pm 0.17$ & 1.3 & 400\\[0.05cm]
    12172016 & 58810.1059 & 0 & $35.32 \pm 0.28$ & $0.6284 \pm 0.0047$ & $\cdots$ & $\cdots$ & 1.3 & 280\\[0.05cm]
    12172018 & 58812.814 & 0 & $27.11 \pm 0.13$ & $0.5461 \pm 0.0029$ & $\cdots$ & $\cdots$ & 1.9 & 380\\[0.05cm]
    12172019 & 58813.2278 & 0 & $33.52 \pm 0.24$ & $0.5894 \pm 0.0044$ & $\cdots$ & $\cdots$ & 1.3 & 340\\[0.05cm]
    12172020 & 58814.1491 & 2 & $32.21 \pm 0.19$ & ${0.5768}^{+0.0060}_{-0.0061}$ & ${2.73}^{+0.18}_{-0.15}$ & ${0.616}^{+0.099}_{-0.094}$ & 1.3 & 370\\[0.05cm]
    12172021a & 58818.1966 & 0 & ${27.14}^{+0.20}_{-0.19}$ & $0.5475 \pm 0.0042$ & $\cdots$ & $\cdots$ & 1.4 & 320\\[0.05cm]
    12172021b & 58818.4035 & 2 & $36.29 \pm 0.26$ & $0.5281 \pm 0.0063$ & ${2.85}^{+0.20}_{-0.17}$ & ${0.418}^{+0.085}_{-0.081}$ & 1.4 & 320\\[0.05cm]
    12172021c & 58818.2615 & 0 & $19.55 \pm 0.14$ & $0.5343 \pm 0.0040$ & $\cdots$ & $\cdots$ & 1.4 & 310\\[0.05cm]
    12172022a & 58819.0045 & 2 & $27.30 \pm 0.19$ & ${0.5400}^{+0.0062}_{-0.0063}$ & ${2.93}^{+0.21}_{-0.17}$ & ${0.487}^{+0.091}_{-0.086}$ & 1.2 & 330\\[0.05cm]
    12172022b & 58819.0672 & 2 & $23.67 \pm 0.14$ & ${0.5422}^{+0.0054}_{-0.0055}$ & ${2.92}^{+0.16}_{-0.13}$ & ${0.573}^{+0.081}_{-0.078}$ & 1.5 & 350\\[0.05cm]
    12172023 & 58820.8515 & 2 & $25.88 \pm 0.12$ & $0.5364 \pm 0.0043$ & ${2.850}^{+0.109}_{-0.097}$ & ${0.559}^{+0.061}_{-0.059}$ & 1.4 & 390\\[0.05cm]
    12172024 & 58821.1917 & 2 & $24.39 \pm 0.15$ & $0.5112 \pm 0.0058$ & ${2.75}^{+0.14}_{-0.13}$ & ${0.460}^{+0.074}_{-0.071}$ & 1.1 & 330\\[0.05cm]
    12172025 & 58822.6442 & 2 & $22.78 \pm 0.18$ & ${0.5215}^{+0.0073}_{-0.0074}$ & ${2.85}^{+0.22}_{-0.18}$ & ${0.467}^{+0.096}_{-0.090}$ & 1.1 & 300\\[0.05cm]
    12172026 & 58823.9753 & 2 & $18.76 \pm 0.10$ & $0.5236 \pm 0.0050$ & ${0.8}^{+2.1}_{--1.9}$ & ${0.583}^{+0.071}_{-0.068}$ & 1.4 & 360\\[0.05cm]
    12172027 & 58824.8343 & 2 & $22.36 \pm 0.11$ & $0.4955 \pm 0.0044$ & ${2.729}^{+0.077}_{-0.076}$ & ${0.523}^{+0.059}_{-0.057}$ & 1.4 & 370\\[0.05cm]
    12172028 & 58825.9005 & 2 & $21.83 \pm 0.16$ & $0.5128 \pm 0.0064$ & ${2.81}^{+0.16}_{-0.15}$ & ${0.457}^{+0.083}_{-0.079}$ & 1.1 & 320\\[0.05cm]
    12172029 & 58826.6273 & 2 & $21.68 \pm 0.11$ & ${0.5046}^{+0.4955}_{-0.0042}$ & ${2.861}^{+0.092}_{-0.087}$ & ${0.474}^{+0.055}_{-0.053}$ & 1.4 & 360\\[0.05cm]
    12172030 & 58827.0259 & 2 & $20.82 \pm 0.11$ & $0.5079 \pm 0.0046$ & ${2.849}^{+0.098}_{-0.092}$ & ${0.516}^{+0.060}_{-0.058}$ & 1.4 & 360\\[0.05cm]
    12172031 & 58828.6267 & 2 & $18.52 \pm 0.21$ & $0.503 \pm 0.010$ & ${2.92}^{+0.31}_{-0.24}$ & ${0.43}^{+0.12}_{-0.11}$ & 1.2 & 260\\[0.05cm]
    12172032 & 58829.5478 & 2 & $19.21 \pm 0.12$ & ${0.4831}^{+0.0055}_{-0.0054}$ & ${2.81}^{+0.12}_{-0.11}$ & ${0.413}^{+0.064}_{-0.062}$ & 1.1 & 320\\[0.05cm]
    12172033a & 58830.6098 & 2 & $17.06 \pm 0.12$ & $0.4726 \pm 0.0058$ & ${3.07}^{+0.14}_{-0.13}$ & ${0.403}^{+0.062}_{-0.058}$ & 1.3 & 300\\[0.05cm]
    12172033b & 58830.6859 & 2 & $18.82 \pm 0.15$ & $0.4734 \pm 0.0067$ & ${2.95}^{+0.14}_{-0.13}$ & ${0.421}^{+0.075}_{-0.070}$ & 1.3 & 280\\[0.05cm]
    12172034a & 58831.5451 & 2 & $17.87 \pm 0.12$ & $0.4728 \pm 0.0054$ & ${3.10}^{+0.11}_{-0.10}$ & ${0.470}^{+0.061}_{-0.058}$ & 1.3 & 310\\[0.05cm]
    12172034b & 58831.6057 & 2 & $18.60 \pm 0.13$ & ${0.4711}^{+0.0058}_{-0.0057}$ & ${2.88}^{+0.11}_{-0.10}$ & ${0.436}^{+0.066}_{-0.063}$ & 1.2 & 310\\[0.05cm]
    12172035a & 58832.5348 & 2 & $16.96 \pm 0.11$ & ${0.4626}^{+0.0055}_{-0.0054}$ & $2.94 \pm 0.11$ & ${0.376}^{+0.058}_{-0.055}$ & 1.2 & 300\\[0.05cm]
    12172035b & 58832.6011 & 2 & $17.64 \pm 0.11$ & ${0.4704}^{+0.0056}_{-0.0055}$ & $2.86 \pm 0.11$ & ${0.413}^{+0.061}_{-0.059}$ & 1.2 & 310\\[0.05cm]
    12172036 & 58833.8724 & 2 & ${16.839}^{+0.096}_{-0.095}$ & $0.4733 \pm 0.0046$ & ${3.12}^{+0.12}_{-0.11}$ & ${0.372}^{+0.048}_{-0.046}$ & 1.2 & 320\\[0.05cm]
    12172037 & 58834.599 & 2 & $16.26 \pm 0.11$ & $0.4668 \pm 0.0057$ & ${3.14}^{+0.14}_{-0.12}$ & ${0.395}^{+0.061}_{-0.057}$ & 1.1 & 300\\[0.05cm]
    12172038a & 58835.5906 & 2 & ${16.34}^{+0.13}_{-0.12}$ & ${0.4639}^{+0.0064}_{-0.0063}$ & ${2.93}^{+0.13}_{-0.12}$ & ${0.416}^{+0.069}_{-0.065}$ & 1.2 & 290\\[0.05cm]
    12172038b & 58835.6554 & 2 & $17.13 \pm 0.14$ & ${0.4821}^{+0.0071}_{-0.0072}$ & ${3.04}^{+0.21}_{-0.17}$ & ${0.410}^{+0.078}_{-0.073}$ & 1.0 & 280\\[0.05cm]
    12172039 & 58836.7168 & 2 & $14.909 \pm 0.087$ & $0.4452 \pm 0.0047$ & ${2.944}^{+0.085}_{-0.084}$ & ${0.369}^{+0.047}_{-0.045}$ & 1.4 & 320\\[0.05cm]
    12172040 & 58837.6459 & 2 & $14.920 \pm 0.080$ & $0.4458 \pm 0.0043$ & ${2.910}^{+0.080}_{-0.079}$ & ${0.367}^{+0.044}_{-0.042}$ & 1.4 & 330\\[0.05cm]
    12172041 & 58839.6383 & 2 & $13.53 \pm 0.15$ & ${0.4282}^{+0.0091}_{-0.0089}$ & ${3.03}^{+0.17}_{-0.16}$ & ${0.345}^{+0.084}_{-0.076}$ & 1.2 & 230\\[0.05cm]
    12172042 & 58840.6992 & 2 & $13.305 \pm 0.081$ & ${0.4305}^{+0.0051}_{-0.0050}$ & $2.920 \pm 0.076$ & ${0.414}^{+0.052}_{-0.049}$ & 1.4 & 310\\[0.05cm]
    12172043a & 58841.6469 & 2 & $13.04 \pm 0.12$ & ${0.4340}^{+0.0075}_{-0.0074}$ & ${2.98}^{+0.20}_{-0.19}$ & ${0.269}^{+0.062}_{-0.058}$ & 1.1 & 250\\[0.05cm]
    12172043b & 58841.7734 & 2 & $11.147 \pm 0.088$ & ${0.4286}^{+0.0063}_{-0.0062}$ & ${3.02}^{+0.12}_{-0.11}$ & ${0.334}^{+0.057}_{-0.053}$ & 1.6 & 270\\[0.05cm]
    12172044a & 58842.6917 & 2 & $12.56 \pm 0.10$ & ${0.4166}^{+0.0063}_{-0.0060}$ & ${2.81}^{+0.13}_{-0.14}$ & ${0.266}^{+0.056}_{-0.053}$ & 1.3 & 270\\[0.05cm]
    12172044b & 58842.7578& 2 & $12.59 \pm 0.11$ & ${0.4218}^{+0.0071}_{-0.0069}$ & ${2.88}^{+0.14}_{-0.15}$ & ${0.282}^{+0.062}_{-0.058}$ & 1.2 & 260\\[0.05cm]
    12172045a & 58843.699 & 2 & $17.72 \pm 0.14$ & ${0.4123}^{+0.0066}_{-0.0064}$ & $2.994 \pm 0.093$ & ${0.376}^{+0.062}_{-0.057}$ & 1.4 & 270\\[0.05cm]
    12172045b & 58843.7645 & 2 & $13.77 \pm 0.14$ & ${0.4148}^{+0.0085}_{-0.0082}$ & $2.98 \pm 0.12$ & ${0.377}^{+0.081}_{-0.073}$ & 1.2 & 240\\[0.05cm]
    12172046a & 58844.289 & 2 & $11.58 \pm 0.10$ & ${0.4307}^{+0.0073}_{-0.0071}$ & $2.90 \pm 0.12$ & ${0.366}^{+0.072}_{-0.067}$ & 1.2 & 260\\[0.05cm]
    12172046b & 58844.3522 & 2 & $12.16 \pm 0.11$ & ${0.4319}^{+0.0071}_{-0.0069}$ & ${2.80}^{+0.14}_{-0.15}$ & ${0.304}^{+0.068}_{-0.065}$ & 1.1 & 260\\[0.05cm]
    12172047 & 58845.8187 & 2 & $11.422 \pm 0.090$ & ${0.4031}^{+0.0063}_{-0.0061}$ & ${2.977}^{+0.093}_{-0.094}$ & ${0.337}^{+0.056}_{-0.052}$ & 1.1 & 270\\[0.05cm]
    12172048a & 58846.8847 & 2 & $11.13 \pm 0.10$ & ${0.4050}^{+0.0070}_{-0.0067}$ & ${2.83}^{+0.15}_{-0.16}$ & ${0.232}^{+0.057}_{-0.054}$ & 1.3 & 240\\[0.05cm]
    12172048b & 58846.9539 & 2 & $10.34 \pm 0.23$ & ${0.401}^{+0.018}_{-0.016}$ & ${2.94}^{+0.28}_{-0.30}$ & ${0.32}^{+0.17}_{-0.14}$ & 0.94 & 160\\[0.05cm]
    12172049a & 58847.7398 & 2 & ${11.30}^{+0.11}_{-0.12}$ & ${0.3915}^{+0.0050}_{-0.0041}$ & ${2.06}^{+0.26}_{-0.27}$ & ${0.054}^{+0.029}_{-0.020}$ & 1.6 & 250\\[0.05cm]
    12172049b & 58847.8106 & 2 & ${10.682}^{+0.097}_{-0.095}$ & $0.3984 \pm 0.0035$ & ${1.80}^{+0.25}_{-0.40}$ & ${0.040}^{+0.021}_{-0.019}$ & 1.5 & 270\\[0.05cm]
    12172050 & 58848.883 & 2 & $9.72 \pm 0.19$ & ${0.398}^{+0.015}_{-0.013}$ & ${2.88}^{+0.27}_{-0.32}$ & ${0.26}^{+0.13}_{-0.11}$ & 0.97 & 170\\[0.05cm]
    12172051a & 58849.7313 & 2 & $10.466 \pm 0.089$ & ${0.3912}^{+0.0062}_{-0.0059}$ & ${2.86}^{+0.11}_{-0.12}$ & ${0.246}^{+0.051}_{-0.047}$ & 1.3 & 260\\[0.05cm]
    12172051b & 58849.8012& 2 & $10.114 \pm 0.093$ & ${0.3882}^{+0.0067}_{-0.0064}$ & ${2.87}^{+0.13}_{-0.14}$ & ${0.221}^{+0.051}_{-0.047}$ & 1.2 & 240\\[0.05cm]
    12172052 & 58850.2593 & 2 & $9.986 \pm 0.090$ & ${0.4058}^{+0.0067}_{-0.0064}$ & ${2.85}^{+0.15}_{-0.16}$ & ${0.222}^{+0.054}_{-0.051}$ & 1.3 & 250\\[0.05cm]
    12172053 & 58851.254 & 2 & $10.149 \pm 0.091$ & ${0.3926}^{+0.0064}_{-0.0062}$ & ${3.09}^{+0.14}_{-0.13}$ & ${0.218}^{+0.045}_{-0.042}$ & 1.3 & 240\\[0.05cm]
    12172054 & 58852.4495 & 2 & $9.672 \pm 0.089$ & ${0.3886}^{+0.0065}_{-0.0063}$ & ${2.99}^{+0.15}_{-0.16}$ & ${0.191}^{+0.046}_{-0.043}$ & 1.1 & 240\\[0.05cm]
    12172055 & 58853.3848 & 2 & $8.99 \pm 0.11$ & ${0.3779}^{+0.0081}_{-0.0075}$ & ${2.86}^{+0.18}_{-0.21}$ & ${0.175}^{+0.057}_{-0.053}$ & 1.1 & 220\\[0.05cm]
    12172056 & 58854.315 & 2 & $8.987 \pm 0.068$ & ${0.3736}^{+0.0053}_{-0.0050}$ & ${2.90}^{+0.10}_{-0.11}$ & ${0.192}^{+0.037}_{-0.034}$ & 1.4 & 260\\[0.05cm]
    12172057 & 58855.643 & 2 & ${9.166}^{+0.081}_{-0.082}$ & ${0.3967}^{+0.0071}_{-0.0069}$ & $3.07 \pm 0.11$ & ${0.329}^{+0.059}_{-0.054}$ & 1.2 & 250\\[0.05cm]
    12172059a & 58857.0981 & 2 & $8.310 \pm 0.084$ & ${0.3535}^{+0.0070}_{-0.0065}$ & ${2.88}^{+0.13}_{-0.15}$ & ${0.175}^{+0.046}_{-0.041}$ & 1.4 & 220\\[0.05cm]
    12172059b & 58857.1742& 2 & $8.761 \pm 0.088$ & ${0.3755}^{+0.0071}_{-0.0067}$ & ${2.87}^{+0.13}_{-0.15}$ & ${0.208}^{+0.053}_{-0.048}$ & 1.2 & 230\\[0.05cm]
    12172060a & 58858.426 & 2 & $8.868 \pm 0.089$ & ${0.3513}^{+0.0066}_{-0.0060}$ & ${2.56}^{+0.10}_{-0.12}$ & ${0.238}^{+0.060}_{-0.051}$ & 1.2 & 250\\[0.05cm]
    12172060b & 58858.4916 & 2 & $9.32 \pm 0.10$ & ${0.3714}^{+0.0085}_{-0.0078}$ & ${2.66}^{+0.11}_{-0.13}$ & ${0.308}^{+0.083}_{-0.071}$ & 1.3 & 240\\[0.05cm]
    12172064 & 58864.6897 & 2 & $3.30 \pm 0.15$ & $0.174 \pm 0.014$ & $2.01 \pm 0.20$ & ${0.0182}^{+0.0070}_{-0.0050}$ & 1.1 & 120\\[0.05cm]
    12172066 & 58866.6131 & 2 & ${2.901}^{+0.098}_{-0.096}$ & ${0.1648}^{+0.0097}_{-0.0096}$ & $1.95 \pm 0.13$ & ${0.0133}^{+0.0037}_{-0.0029}$ & 1.0 & 160\\[0.05cm]
    12172067a & 58869.4598 & 2 & ${1.505}^{+0.096}_{-0.093}$ & ${0.124}^{+0.016}_{-0.015}$ & $1.71 \pm 0.13$ & ${0.0046}^{+0.0044}_{-0.0024}$ & 0.78 & 97\\[0.05cm]
    12172067b & 58869.2 & 2 & ${1.370}^{+0.063}_{-0.061}$ & ${0.132}^{+0.014}_{-0.013}$ & $1.81 \pm 0.11$ & ${0.0067}^{+0.0043}_{-0.0028}$ & 0.95 & 150\\[0.05cm]
    12172071 & 58873.9762 & 1 & $0.703 \pm 0.030$ & $\cdots$ & ${2.066}^{+0.065}_{-0.063}$ & $\cdots$ & 1.1 & 280\\[0.05cm]
    12172072 & 58874.5156 & 1 & ${0.606}^{+0.047}_{-0.046}$ & $\cdots$ & ${2.02}^{+0.12}_{-0.11}$ & $\cdots$ & 1.0 & 160\\[0.05cm]
    12172073 & 58875.4391 & 1 & ${0.642}^{+0.040}_{-0.039}$ & $\cdots$ & ${1.929}^{+0.085}_{-0.082}$ & $\cdots$ & 1.0 & 210\\[0.05cm]
    12172074 & 58876.5592 & 1 & ${0.480}^{+0.036}_{-0.035}$ & $\cdots$ & ${1.804}^{+0.098}_{-0.095}$ & $\cdots$ & 0.95 & 190\\[0.05cm]
    12172075 & 58877.0235 & 1 & ${0.485}^{+0.039}_{-0.038}$ & $\cdots$ & ${1.93}^{+0.11}_{-0.10}$ & $\cdots$ & 0.86 & 140\\[0.05cm]
    12172076 & 58878.1523 & 1 & ${0.402}^{+0.040}_{-0.038}$ & $\cdots$ & $1.74 \pm 0.12$ & $\cdots$ & 0.91 & 140\\[0.05cm]
    12172077 & 58879.2234 & 1 & ${0.380}^{+0.028}_{-0.027}$ & $\cdots$ & ${1.689}^{+0.090}_{-0.088}$ & $\cdots$ & 1.1 & 200\\[0.05cm]
    12172080 & 58885.7255 & 1 & ${0.228}^{+0.032}_{-0.033}$ & $\cdots$ & ${1.51}^{+0.20}_{-0.13}$ & $\cdots$ & 0.80 & 60\\[0.05cm]
    12172081 & 58886.1173 & 1 & ${0.263}^{+0.086}_{-0.064}$ & $\cdots$ & ${1.38}^{+0.45}_{-0.15}$ & $\cdots$ & 0.58 & 20\\[0.05cm]
    12172082 & 58887.1160 & 1 & ${0.190}^{+0.031}_{-0.022}$ & $\cdots$ & ${1.62}^{+0.29}_{-0.12}$ & $\cdots$ & 0.81 & 46\\[0.05cm]
    12172083 & 58888.8541 & 1 & ${0.216}^{+0.028}_{-0.025}$ & $\cdots$ & ${1.46}^{+0.25}_{-0.11}$ & $\cdots$ & 0.73 & 50\\[0.05cm]
    12172084 & 58889.5696 & 1 & ${0.180}^{+0.063}_{-0.039}$ & $\cdots$ & ${1.94}^{+0.76}_{-0.54}$ & $\cdots$ & 1.0 & 23\\[0.05cm]
    12172085 & 58890.4982 & 1 & ${0.227}^{+0.035}_{-0.032}$ & $\cdots$ & ${1.51}^{+0.30}_{-0.13}$ & $\cdots$ & 0.69 & 46\\[0.05cm]
    12172086 & 58891.1628 & 1 & ${0.093}^{+0.017}_{-0.014}$ & $\cdots$ & ${1.44}^{+0.25}_{-0.14}$ & $\cdots$ & 0.88 & 44\\[0.05cm]
    12172087 & 58892.2259 & 1 & ${0.168}^{+0.024}_{-0.022}$ & $\cdots$ & ${1.49}^{+0.29}_{-0.18}$ & $\cdots$ & 0.97 & 65\\[0.05cm]
    12172088 & 58893.1549 & 1 & ${0.127}^{+0.016}_{-0.015}$ & $\cdots$ & ${1.66}^{+0.30}_{-0.15}$ & $\cdots$ & 0.82 & 71\\[0.05cm]
    12172089 & 58894.2224 & 1 & $0.153 \pm 0.016$ & $\cdots$ & ${1.64}^{+0.23}_{-0.22}$ & $\cdots$ & 0.77 & 90\\[0.05cm]
    12172090 & 58896.1467 & 1 & ${0.150}^{+0.023}_{-0.022}$ & $\cdots$ & ${1.64}^{+0.34}_{-0.19}$ & $\cdots$ & 0.76 & 52\\[0.05cm]
    12172091 & 58902.8604 & 1 & ${0.103}^{+0.020}_{-0.017}$ & $\cdots$ & ${1.61}^{+0.27}_{-0.18}$ & $\cdots$ & 0.82 & 41\\[0.05cm]
    12172092 & 58909.8978 & 1 & ${0.086}^{+0.046}_{-0.020}$ & $\cdots$ & ${2.49}^{+0.76}_{-0.59}$ & $\cdots$ & 0.85 & 28\\[0.05cm]
    12172093 & 58916.871 & 1 & ${0.083}^{+0.042}_{-0.031}$ & $\cdots$ & ${1.36}^{+0.52}_{-0.49}$ & $\cdots$ & 0.52 & 7.0\\[0.05cm]
    12172094 & 58923.9088 & 1 & ${0.044}^{+0.011}_{-0.010}$ & $\cdots$ & ${1.85}^{+0.35}_{-0.32}$ & $\cdots$ & 0.59 & 16\\[0.05cm]
    \hline
    \multicolumn{9}{p{0.89\columnwidth}}{\hangindent=1ex $^a$ObsIDs containing letters are observations that have been split into multiple GTIs.} \\
    \multicolumn{9}{p{0.89\columnwidth}}{\hangindent=1ex $^b$ {\tt tbabs*cflux*diskbb} (model 0), {\tt tbabs*cflux*powerlaw} (model 1), {\tt tbabs*cflux*(diskbb+powerlaw)} (model 2). Note: for use of the {\tt cflux} {\sc xspec} model component to obtain a flux estimate, either the {\tt diskbb} normalization (for models 0 and 2) or {\tt powerlaw} normalization (for model 1) must be frozen at 1.0.}
\end{longtable}
}

\onecolumn{
\begin{longtable}{lcccccccc}
\caption{Broad-band (XRT+UVOT) Spectral Fits for MAXI\,J0637$-$430}\label{tab:broadfitstab}\\
		\hline
		ObsID$^a$&MJD&$T_{\rm in}$&$f_{\rm out}$&logrout&$N_{\rm disc}$&$L_{\rm c}/L_{\rm d}$$^b$&reduced&dof \\
		&(days)&(keV)&&&&&$\chi^2$& \\
		\hline
\endhead
\endfoot
12172001 & 58794.0326 & ${0.71166}^{+0.00585}_{-0.00599}$ & ${377.9}^{+34.9}_{-35.4}$ & ${4.5428}^{+0.0248}_{-0.0211}$ & ${233.72}^{+8.94}_{-8.19}$ & 0.0 & 1.81 & 364\\[0.05cm]
12172002 & 58795.9662 & $0.7667 \pm 0.0143$ & ${24.46}^{+4.16}_{-3.22}$ & ${4.1689}^{+0.0455}_{-0.0467}$ & ${1144.2}^{+69.4}_{-64.1}$ & ${0.3820}^{+0.0710}_{-0.0690}$ & 1.44 & 319\\[0.05cm]
12172003 & 58796.7424 & ${0.70719}^{+0.00911}_{-0.00913}$ & ${79.3}^{+13.2}_{-10.6}$ & $4.0988 \pm 0.0297$ & ${1034.2}^{+44.6}_{-42.4}$ & ${0.4641}^{+0.0464}_{-0.0454}$ & 1.34 & 384\\[0.05cm]
12172004a & 58797.8193 & ${0.69294}^{+0.00943}_{-0.00944}$ & ${65.65}^{+10.23}_{-8.33}$ & ${4.0330}^{+0.0281}_{-0.0282}$ & ${1268.8}^{+57.2}_{-54.0}$ & ${0.5015}^{+0.0490}_{-0.0481}$ & 1.09 & 371\\[0.05cm]
12172004b & 58797.8815 & $0.6807 \pm 0.0105$ & ${71.99}^{+11.94}_{-9.54}$ & ${4.0230}^{+0.0291}_{-0.0294}$ & ${1286.6}^{+66.4}_{-62.2}$ & ${0.4319}^{+0.0499}_{-0.0489}$ & 1.14 & 393\\[0.05cm]
12172005 & 58798.6026 & ${0.7084}^{+0.0165}_{-0.0162}$ & ${81.8}^{+35.2}_{-22.0}$ & ${4.0379}^{+0.0523}_{-0.0517}$ & ${1106.2}^{+86.7}_{-82.5}$ & ${0.3242}^{+0.0781}_{-0.0765}$ & 1.26 & 242\\[0.05cm]
12172006a & 58799.6076 & ${0.69404}^{+0.00577}_{-0.00573}$ & ${65.14}^{+8.93}_{-7.45}$ & ${4.0288}^{+0.0263}_{-0.0264}$ & ${1475.6}^{+42.7}_{-41.5}$ & 0.1 & 1.41 & 323\\[0.05cm]
12172006b & 58799.6076 & ${0.66826}^{+0.00770}_{-0.00769}$ & ${72.43}^{+10.72}_{-8.73}$ & ${4.0326}^{+0.0275}_{-0.0278}$ & ${1379.2}^{+53.8}_{-51.4}$ & ${0.3493}^{+0.0403}_{-0.0398}$ & 1.02 & 378\\[0.05cm]
12172007a & 58800.3939 & ${0.6378}^{+0.0419}_{-0.0394}$ & ${85.3}^{+31.2}_{-19.4}$ & ${3.8392}^{+0.0741}_{-0.0759}$ & ${2371}^{+681.4}_{-529}$ & 0.1 & 1.14 & 76.0\\[0.05cm]
12172007b & 58800.4601& ${0.64098}^{+0.00486}_{-0.00481}$ & ${93.3}^{+27.4}_{-18.8}$ & ${3.8899}^{+0.0465}_{-0.0471}$ & ${2002.4}^{+64.7}_{-64.2}$ & ${0.2044}^{+0.0460}_{-0.0456}$ & 1.12 & 394\\[0.05cm]
12172008a & 58801.4022 & ${0.64959}^{+0.00528}_{-0.00530}$ & ${129.0}^{+19.2}_{-15.6}$ & ${4.0559}^{+0.0261}_{-0.0264}$ & ${1122.8}^{+35.6}_{-34.5}$ & 0.1 & 1.24 & 251\\[0.05cm]
12172008b & 58801.9278 & $0.67458 \pm 0.00320$ & ${94.4}^{+13.5}_{-11.0}$ & ${4.0294}^{+0.0254}_{-0.0258}$ & ${1287.8}^{+23.9}_{-23.8}$ & 0.1 & 1.38 & 331\\[0.05cm]
12172009 & 58803.7124 & ${0.63314}^{+0.00716}_{-0.00717}$ & ${41.35}^{+7.08}_{-5.57}$ & ${3.8671}^{+0.0312}_{-0.0319}$ & ${2649.5}^{+100.7}_{-95.8}$ & ${0.4099}^{+0.0350}_{-0.0345}$ & 1.23 & 355\\[0.05cm]
12172010a & 58804.8447 & ${0.61965}^{+0.00762}_{-0.00765}$ & ${90.3}^{+14.6}_{-11.6}$ & ${4.0170}^{+0.0292}_{-0.0297}$ & ${1299.5}^{+55.4}_{-52.5}$ & ${0.4035}^{+0.0369}_{-0.0364}$ & 1.01 & 335\\[0.05cm]
12172010b & 58804.9950 & ${0.61693}^{+0.00908}_{-0.00901}$ & ${71.58}^{+11.66}_{-9.25}$ & ${3.9731}^{+0.0301}_{-0.0308}$ & ${1623.4}^{+84.4}_{-79.8}$ & ${0.4009}^{+0.0546}_{-0.0537}$ & 1.21 & 416\\[0.05cm]
12172011 & 58805.9082 & ${0.52217}^{+0.00777}_{-0.00772}$ & ${68.66}^{+10.99}_{-8.61}$ & ${3.9478}^{+0.0324}_{-0.0333}$ & ${1971.6}^{+105.8}_{-99.9}$ & ${0.8793}^{+0.0634}_{-0.0622}$ & 1.11 & 448\\[0.05cm]
12172012 & 58806.3129 & ${0.5517}^{+0.0110}_{-0.0109}$ & ${109.1}^{+24.3}_{-17.7}$ & ${3.9575}^{+0.0368}_{-0.0375}$ & ${1497}^{+110.1}_{-103}$ & ${0.6917}^{+0.0764}_{-0.0744}$ & 1.09 & 374\\[0.05cm]
12172013 & 58807.7646 & ${0.59949}^{+0.00499}_{-0.00495}$ & ${67.67}^{+10.79}_{-8.59}$ & ${3.9775}^{+0.0305}_{-0.0310}$ & ${1617.4}^{+48.1}_{-47.0}$ & ${0.4559}^{+0.0302}_{-0.0300}$ & 1.61 & 525\\[0.05cm]
12172014 & 58808.8953 & ${0.60779}^{+0.00452}_{-0.00449}$ & ${73.77}^{+12.09}_{-9.61}$ & ${3.9671}^{+0.0310}_{-0.0314}$ & ${1519.1}^{+41.1}_{-40.2}$ & ${0.2808}^{+0.0249}_{-0.0247}$ & 1.31 & 494\\[0.05cm]
12172015 & 58809.2947 & ${0.58267}^{+0.00654}_{-0.00648}$ & ${66.79}^{+11.55}_{-8.98}$ & ${3.9212}^{+0.0338}_{-0.0348}$ & ${1824.5}^{+74.6}_{-71.6}$ & ${0.2525}^{+0.0362}_{-0.0359}$ & 1.31 & 407\\[0.05cm]
12172016 & 58810.1059 & $0.6168 \pm 0.0110$ & ${121.3}^{+33.0}_{-23.7}$ & ${3.9481}^{+0.0443}_{-0.0437}$ & ${1239.5}^{+78.8}_{-73.3}$ & ${0.1357}^{+0.0470}_{-0.0463}$ & 1.05 & 212\\[0.05cm]
12172018 & 58812.814 & ${0.55017}^{+0.00296}_{-0.00295}$ & ${131.5}^{+26.3}_{-20.0}$ & ${3.8824}^{+0.0331}_{-0.0336}$ & ${1627.2}^{+39.8}_{-39.7}$ & 0.1 & 1.32 & 381\\[0.05cm]
12172019 & 58813.2278 & ${0.59218}^{+0.00445}_{-0.00443}$ & ${80.4}^{+20.5}_{-14.5}$ & ${3.9621}^{+0.0505}_{-0.0516}$ & ${1482.3}^{+47.4}_{-46.2}$ & 0.1 & 1.31 & 343\\[0.05cm]
12172020 & 58814.1491 & ${0.57378}^{+0.00342}_{-0.00340}$ & ${107.5}^{+24.5}_{-18.1}$ & ${3.8993}^{+0.0395}_{-0.0401}$ & ${1594.0}^{+41.9}_{-41.5}$ & 0.1 & 1.26 & 378\\[0.05cm]
12172021a & 58818.1966 & ${0.55016}^{+0.00419}_{-0.00416}$ & ${102.1}^{+19.2}_{-14.7}$ & ${3.8674}^{+0.0335}_{-0.0343}$ & ${1639.7}^{+53.8}_{-52.3}$ & 0.1 & 1.38 & 320\\[0.05cm]
12172021b & 58818.4035 & ${0.52378}^{+0.00379}_{-0.00377}$ & ${76.5}^{+15.7}_{-11.7}$ & ${3.7525}^{+0.0353}_{-0.0367}$ & ${2690.3}^{+84.1}_{-81.7}$ & 0.1 & 1.34 & 325\\[0.05cm]
12172021c & 58818.2615 & ${0.55356}^{+0.00413}_{-0.00412}$ & ${162.5}^{+31.9}_{-24.1}$ & ${3.9342}^{+0.0335}_{-0.0342}$ & ${1167.9}^{+38.8}_{-37.9}$ & 0.0 & 1.18 & 317\\[0.05cm]
12172022a & 58819.0045 & ${0.53266}^{+0.00383}_{-0.00381}$ & ${85.6}^{+14.8}_{-11.4}$ & ${3.8642}^{+0.0334}_{-0.0345}$ & ${1875.2}^{+57.9}_{-56.3}$ & 0.1 & 1.17 & 333\\[0.05cm]
12172022b & 58819.0672 & ${0.53396}^{+0.00323}_{-0.00322}$ & ${102.1}^{+18.1}_{-13.8}$ & ${3.8949}^{+0.0332}_{-0.0345}$ & ${1598.4}^{+42.1}_{-41.3}$ & 0.1 & 1.46 & 352\\[0.05cm]
12172023 & 58820.8515 & ${0.53043}^{+0.00249}_{-0.00248}$ & ${94.0}^{+19.7}_{-14.4}$ & ${3.8490}^{+0.0374}_{-0.0391}$ & ${1801.5}^{+37.6}_{-37.5}$ & 0.1 & 1.40 & 392\\[0.05cm]
12172024 & 58821.1917 & ${0.50991}^{+0.00471}_{-0.00467}$ & ${89.1}^{+16.2}_{-12.3}$ & ${3.8558}^{+0.0355}_{-0.0367}$ & ${2015.5}^{+72.9}_{-70.5}$ & 0.1 & 1.10 & 340\\[0.05cm]
12172025 & 58822.6442 & ${0.51740}^{+0.00420}_{-0.00417}$ & ${111.9}^{+39.6}_{-26.0}$ & ${3.8450}^{+0.0511}_{-0.0516}$ & ${1760.3}^{+64.8}_{-64.2}$ & 0.1 & 1.04 & 305\\[0.05cm]
12172026 & 58823.9753 & ${0.51808}^{+0.00292}_{-0.00291}$ & ${106.5}^{+19.9}_{-14.8}$ & ${3.9301}^{+0.0372}_{-0.0390}$ & ${1438.0}^{+35.8}_{-35.3}$ & 0.1 & 1.40 & 365\\[0.05cm]
12172027 & 58824.8343 & ${0.48960}^{+0.00345}_{-0.00343}$ & ${63.19}^{+12.38}_{-8.96}$ & ${3.8806}^{+0.0488}_{-0.0527}$ & ${2186.7}^{+60.9}_{-59.4}$ & 0.1 & 1.37 & 372\\[0.05cm]
12172028 & 58825.9005 & ${0.51036}^{+0.00374}_{-0.00372}$ & ${122.6}^{+36.5}_{-24.8}$ & ${3.7904}^{+0.0454}_{-0.0466}$ & ${1783.7}^{+59.6}_{-59.2}$ & 0.1 & 1.12 & 321\\[0.05cm]
12172029 & 58826.6273 & ${0.49987}^{+0.00254}_{-0.00253}$ & ${90.2}^{+17.6}_{-13.1}$ & ${3.8473}^{+0.0381}_{-0.0398}$ & ${1953.9}^{+44.2}_{-43.5}$ & 0.1 & 1.34 & 365\\[0.05cm]
12172030 & 58827.0259 & ${0.50336}^{+0.00268}_{-0.00267}$ & ${111.5}^{+26.9}_{-19.0}$ & ${3.8123}^{+0.0406}_{-0.0426}$ & ${1807.7}^{+44.2}_{-44.1}$ & 0.1 & 1.32 & 365\\[0.05cm]
12172031 & 58828.6267 & ${0.49698}^{+0.00560}_{-0.00558}$ & ${78.9}^{+24.0}_{-15.4}$ & ${3.9203}^{+0.0663}_{-0.0705}$ & ${1720.5}^{+85.4}_{-81.2}$ & 0.1 & 1.12 & 262\\[0.05cm]
12172032 & 58829.5478 & ${0.48073}^{+0.00431}_{-0.00427}$ & ${103.0}^{+27.3}_{-18.6}$ & ${3.7876}^{+0.0462}_{-0.0487}$ & ${2039.5}^{+74.8}_{-72.8}$ & 0.1 & 1.04 & 326\\[0.05cm]
12172033a & 58830.6098 & ${0.46345}^{+0.00340}_{-0.00337}$ & ${117.0}^{+26.0}_{-18.6}$ & ${3.7613}^{+0.0382}_{-0.0401}$ & ${2126.5}^{+70.6}_{-68.8}$ & 0.1 & 1.26 & 303\\[0.05cm]
12172033b & 58830.6859 & ${0.46735}^{+0.00391}_{-0.00390}$ & ${103.7}^{+22.3}_{-16.2}$ & ${3.7516}^{+0.0380}_{-0.0398}$ & ${2265.5}^{+85.1}_{-82.1}$ & 0.1 & 1.30 & 291\\[0.05cm]
12172034a & 58831.5451 & ${0.46035}^{+0.00317}_{-0.00316}$ & ${91.4}^{+16.6}_{-12.4}$ & ${3.7936}^{+0.0362}_{-0.0380}$ & ${2302.8}^{+70.9}_{-68.8}$ & 0.1 & 1.38 & 314\\[0.05cm]
12172034b & 58831.6057 & ${0.46701}^{+0.00331}_{-0.00330}$ & ${86.1}^{+14.8}_{-11.4}$ & ${3.8038}^{+0.0356}_{-0.0369}$ & ${2252.8}^{+70.9}_{-68.6}$ & 0.1 & 1.15 & 312\\[0.05cm]
12172035a & 58832.5348 & ${0.45805}^{+0.00323}_{-0.00322}$ & ${99.3}^{+19.6}_{-14.5}$ & ${3.7752}^{+0.0376}_{-0.0392}$ & ${2234.3}^{+71.0}_{-68.8}$ & 0.1 & 1.23 & 308\\[0.05cm]
12172035b & 58832.6011& ${0.46761}^{+0.00314}_{-0.00313}$ & ${96.1}^{+18.9}_{-14.0}$ & ${3.7869}^{+0.0374}_{-0.0391}$ & ${2120.5}^{+63.8}_{-62.1}$ & 0.1 & 1.21 & 317\\[0.05cm]
12172036 & 58833.8724 & ${0.46329}^{+0.00276}_{-0.00274}$ & ${96.6}^{+25.3}_{-17.2}$ & ${3.7535}^{+0.0462}_{-0.0490}$ & ${2117.1}^{+57.5}_{-56.7}$ & 0.1 & 1.31 & 327\\[0.05cm]
12172037 & 58834.599 & ${0.45508}^{+0.00337}_{-0.00336}$ & ${76.8}^{+19.4}_{-13.2}$ & ${3.7939}^{+0.0520}_{-0.0556}$ & ${2219.4}^{+73.8}_{-71.5}$ & 0.1 & 1.20 & 302\\[0.05cm]
12172038a & 58835.5906 & $0.45863 \pm 0.00360$ & ${84.4}^{+15.9}_{-11.9}$ & ${3.7985}^{+0.0394}_{-0.0412}$ & ${2144.9}^{+75.3}_{-72.6}$ & 0.1 & 1.22 & 292\\[0.05cm]
12172038b & 58835.6554 & ${0.47316}^{+0.00400}_{-0.00398}$ & ${80.4}^{+14.9}_{-11.1}$ & ${3.8192}^{+0.0392}_{-0.0409}$ & ${1969.8}^{+73.6}_{-70.8}$ & 0.1 & 1.01 & 286\\[0.05cm]
12172039 & 58836.7168 & ${0.44177}^{+0.00276}_{-0.00275}$ & ${144.0}^{+46.6}_{-29.3}$ & ${3.6867}^{+0.0459}_{-0.0496}$ & ${2272.4}^{+69.5}_{-71.0}$ & 0.1 & 1.38 & 322\\[0.05cm]
12172040 & 58837.6459 & $0.44314 \pm 0.00250$ & ${108.7}^{+23.0}_{-16.7}$ & ${3.7566}^{+0.0389}_{-0.0409}$ & ${2263.3}^{+59.0}_{-57.8}$ & 0.1 & 1.40 & 335\\[0.05cm]
12172041 & 58839.6383 & ${0.42725}^{+0.00810}_{-0.00997}$ & ${170.6}^{+12.2}_{-90.3}$ & ${3.619}^{+0.128}_{-0.154}$ & ${2482}^{+181.3}_{-658}$ & 0.1 & 1.22 & 235\\[0.05cm]
12172042 & 58840.6992 & ${0.42024}^{+0.00358}_{-0.00357}$ & ${121.1}^{+34.2}_{-22.5}$ & ${3.6938}^{+0.0459}_{-0.0492}$ & ${2484.5}^{+91.8}_{-89.2}$ & ${0.1590}^{+0.0183}_{-0.0182}$ & 1.52 & 315\\[0.05cm]
12172043a & 58841.6469 & ${0.43116}^{+0.00408}_{-0.00406}$ & ${115.5}^{+25.1}_{-18.1}$ & ${3.7353}^{+0.0404}_{-0.0422}$ & ${2243.2}^{+97.4}_{-93.4}$ & 0.1 & 1.12 & 257\\[0.05cm]
12172043b & 58841.7734 & ${0.42350}^{+0.00355}_{-0.00354}$ & ${142.9}^{+34.0}_{-23.7}$ & ${3.7449}^{+0.0413}_{-0.0437}$ & ${2050.3}^{+80.6}_{-78.0}$ & 0.1 & 1.54 & 276\\[0.05cm]
12172044a & 58842.6917 & ${0.42046}^{+0.00353}_{-0.00350}$ & ${140.6}^{+37.6}_{-25.7}$ & ${3.6752}^{+0.0425}_{-0.0447}$ & ${2384.6}^{+94.4}_{-91.8}$ & 0.1 & 1.29 & 275\\[0.05cm]
12172044b & 58842.7578 & ${0.42311}^{+0.00392}_{-0.00391}$ & ${141.2}^{+38.3}_{-26.0}$ & ${3.6794}^{+0.0429}_{-0.0451}$ & ${2327.8}^{+101.3}_{-97.8}$ & 0.1 & 1.14 & 262\\[0.05cm]
12172045a & 58843.699 & ${0.40695}^{+0.00354}_{-0.00357}$ & ${66.54}^{+13.16}_{-9.79}$ & ${3.6517}^{+0.0412}_{-0.0430}$ & ${3940}^{+159.1}_{-151}$ & 0.1 & 1.57 & 273\\[0.05cm]
12172045b & 58843.7645 & ${0.41042}^{+0.00458}_{-0.00456}$ & ${89.1}^{+17.2}_{-12.8}$ & ${3.7159}^{+0.0409}_{-0.0426}$ & ${2931}^{+150.8}_{-143}$ & 0.1 & 1.33 & 244\\[0.05cm]
12172046a & 58844.289 & $0.42936 \pm 0.00400$ & ${115.9}^{+24.9}_{-18.0}$ & ${3.7737}^{+0.0422}_{-0.0442}$ & ${2009.0}^{+86.3}_{-82.3}$ & 0.1 & 1.23 & 268\\[0.05cm]
12172046b & 58844.3522 & ${0.43443}^{+0.00399}_{-0.00397}$ & ${108.4}^{+22.6}_{-16.4}$ & ${3.7771}^{+0.0416}_{-0.0435}$ & ${2013.6}^{+84.3}_{-80.8}$ & 0.1 & 1.09 & 268\\[0.05cm]
12172047 & 58845.8187 & $0.40029 \pm 0.00337$ & ${103.5}^{+27.1}_{-18.1}$ & ${3.7250}^{+0.0517}_{-0.0552}$ & ${2710}^{+107.3}_{-103}$ & 0.1 & 1.26 & 272\\[0.05cm]
12172048a & 58846.8847 & ${0.40890}^{+0.00394}_{-0.00392}$ & ${114.7}^{+26.8}_{-19.3}$ & ${3.7236}^{+0.0444}_{-0.0459}$ & ${2411}^{+108.1}_{-104}$ & 0.1 & 1.26 & 249\\[0.05cm]
12172048b & 58846.9539 & ${0.40036}^{+0.00958}_{-0.00945}$ & ${123.1}^{+28.7}_{-20.7}$ & ${3.7226}^{+0.0489}_{-0.0503}$ & ${2438}^{+277.9}_{-248}$ & 0.1 & 1.00 & 168\\[0.05cm]
12172049a & 58847.7398 & ${0.39681}^{+0.00555}_{-0.00548}$ & ${98.7}^{+21.0}_{-15.5}$ & ${3.7010}^{+0.0427}_{-0.0443}$ & ${2830}^{+171.3}_{-161}$ & ${0.1255}^{+0.0274}_{-0.0272}$ & 1.38 & 236\\[0.05cm]
12172049b & 58847.8106 & ${0.40229}^{+0.00477}_{-0.00473}$ & ${105.9}^{+22.4}_{-16.6}$ & ${3.7271}^{+0.0421}_{-0.0435}$ & ${2501}^{+127.3}_{-121}$ & ${0.1340}^{+0.0244}_{-0.0243}$ & 1.03 & 256\\[0.05cm]
12172050 & 58848.883 & ${0.39490}^{+0.01026}_{-0.00998}$ & ${142.3}^{+251.5}_{-64.7}$ & ${3.693}^{+0.124}_{-0.138}$ & ${2401}^{+283.8}_{-287}$ & ${0.1476}^{+0.0541}_{-0.0534}$ & 0.970 & 174\\[0.05cm]
12172052 & 58850.2593 & ${0.40587}^{+0.00483}_{-0.00478}$ & ${121.7}^{+25.9}_{-18.9}$ & ${3.7388}^{+0.0418}_{-0.0434}$ & ${2220}^{+112.8}_{-107}$ & ${0.1250}^{+0.0249}_{-0.0247}$ & 1.31 & 251\\[0.05cm]
12172053 & 58851.254 & ${0.40075}^{+0.00531}_{-0.00524}$ & ${152.7}^{+-27.5}_{-27.5}$ & ${3.6968}^{+0.0450}_{-0.0473}$ & ${2165}^{+124.8}_{-119}$ & ${0.0193}^{+0.0704}_{--0.0977}$ & 0.144 & 1.20\\[0.05cm]
12172054 & 58852.4495 & ${0.37949}^{+0.00482}_{-0.00477}$ & ${150.5}^{+42.5}_{-28.3}$ & ${3.6319}^{+0.0463}_{-0.0490}$ & ${2767}^{+157.6}_{-150}$ & ${0.1057}^{+0.0233}_{-0.0231}$ & 1.31 & 236\\[0.05cm]
12172055 & 58853.3848 & ${0.37969}^{+0.00570}_{-0.00562}$ & ${119.6}^{+26.1}_{-18.9}$ & ${3.6892}^{+0.0446}_{-0.0464}$ & ${2696}^{+177.1}_{-166}$ & ${0.1276}^{+0.0302}_{-0.0300}$ & 1.06 & 223\\[0.05cm]
12172056 & 58854.315 & ${0.37282}^{+0.00361}_{-0.00358}$ & ${112.9}^{+25.2}_{-17.9}$ & ${3.6856}^{+0.0451}_{-0.0474}$ & ${2910}^{+124.3}_{-119}$ & ${0.1437}^{+0.0190}_{-0.0188}$ & 1.41 & 267\\[0.05cm]
12172057 & 58855.643 & ${0.38526}^{+0.00458}_{-0.00454}$ & ${104.6}^{+24.0}_{-16.9}$ & ${3.7094}^{+0.0473}_{-0.0499}$ & ${2550}^{+131.9}_{-125}$ & ${0.1456}^{+0.0238}_{-0.0236}$ & 1.30 & 255\\[0.05cm]
12172058 & 58856.5709 & ${0.36603}^{+0.00317}_{-0.00316}$ & ${89.6}^{+17.1}_{-12.7}$ & ${3.6687}^{+0.0427}_{-0.0446}$ & ${3465}^{+133.8}_{-129}$ & ${0.1214}^{+0.0155}_{-0.0154}$ & 1.46 & 269\\[0.05cm]
12172059a & 58857.0981 & ${0.35184}^{+0.00466}_{-0.00461}$ & ${116.7}^{+25.8}_{-18.6}$ & ${3.6204}^{+0.0435}_{-0.0455}$ & ${3446}^{+203.8}_{-192}$ & ${0.1793}^{+0.0266}_{-0.0264}$ & 1.36 & 229\\[0.05cm]
12172059b & 58857.1742 & ${0.37271}^{+0.00505}_{-0.00500}$ & ${114.9}^{+25.2}_{-18.2}$ & ${3.6651}^{+0.0434}_{-0.0452}$ & ${2809}^{+166.1}_{-156}$ & ${0.1705}^{+0.0291}_{-0.0288}$ & 1.19 & 236\\[0.05cm]
12172060a & 58858.426 & ${0.34663}^{+0.00488}_{-0.00482}$ & ${103.4}^{+24.7}_{-17.4}$ & ${3.6072}^{+0.0467}_{-0.0487}$ & ${3589}^{+218.8}_{-206}$ & ${0.3843}^{+0.0348}_{-0.0344}$ & 1.29 & 251\\[0.05cm]
12172060b& 58858.4916 & ${0.36263}^{+0.00565}_{-0.00557}$ & ${102.9}^{+25.0}_{-17.5}$ & ${3.6349}^{+0.0472}_{-0.0493}$ & ${3140}^{+210.2}_{-197}$ & ${0.3391}^{+0.0377}_{-0.0373}$ & 1.30 & 249\\[0.05cm]
12172064 & 58864.6897 & ${0.15640}^{+0.00944}_{-0.00867}$ & ${159.2}^{+96.3}_{-47.2}$ & ${3.139}^{+0.107}_{-0.113}$ & ${29120}^{+9808}_{-7390}$ & ${1.801}^{+0.239}_{-0.203}$ & 1.09 & 128\\[0.05cm]
12172066 & 58866.6131 & ${0.14375}^{+0.00775}_{-0.00464}$ & ${221.0}^{+46.3}_{-73.4}$ & ${3.01}^{+0.10}_{-3.01}$ & ${36210}^{+6481}_{-8140}$ & ${1.888}^{+0.197}_{-0.186}$ & 1.07 & 170\\[0.05cm]
    \hline
    \multicolumn{9}{p{0.89\columnwidth}}{\hangindent=1ex $^a$ObsIDs containing letters are observations that have been split into multiple GTIs.}\\
    \multicolumn{9}{p{0.89\columnwidth}}{\hangindent=1ex $^b$When the $L_{\rm c}/L_{\rm d}$ parameter could not be contrained by the data, it was fixed at the default value of 0.1.}
\end{longtable}
}

\clearpage

\section{Example Broad-band Spectra}\label{sec:appA}

\begin{figure*}
  \includegraphics[width=0.45\linewidth,height=.4\linewidth]{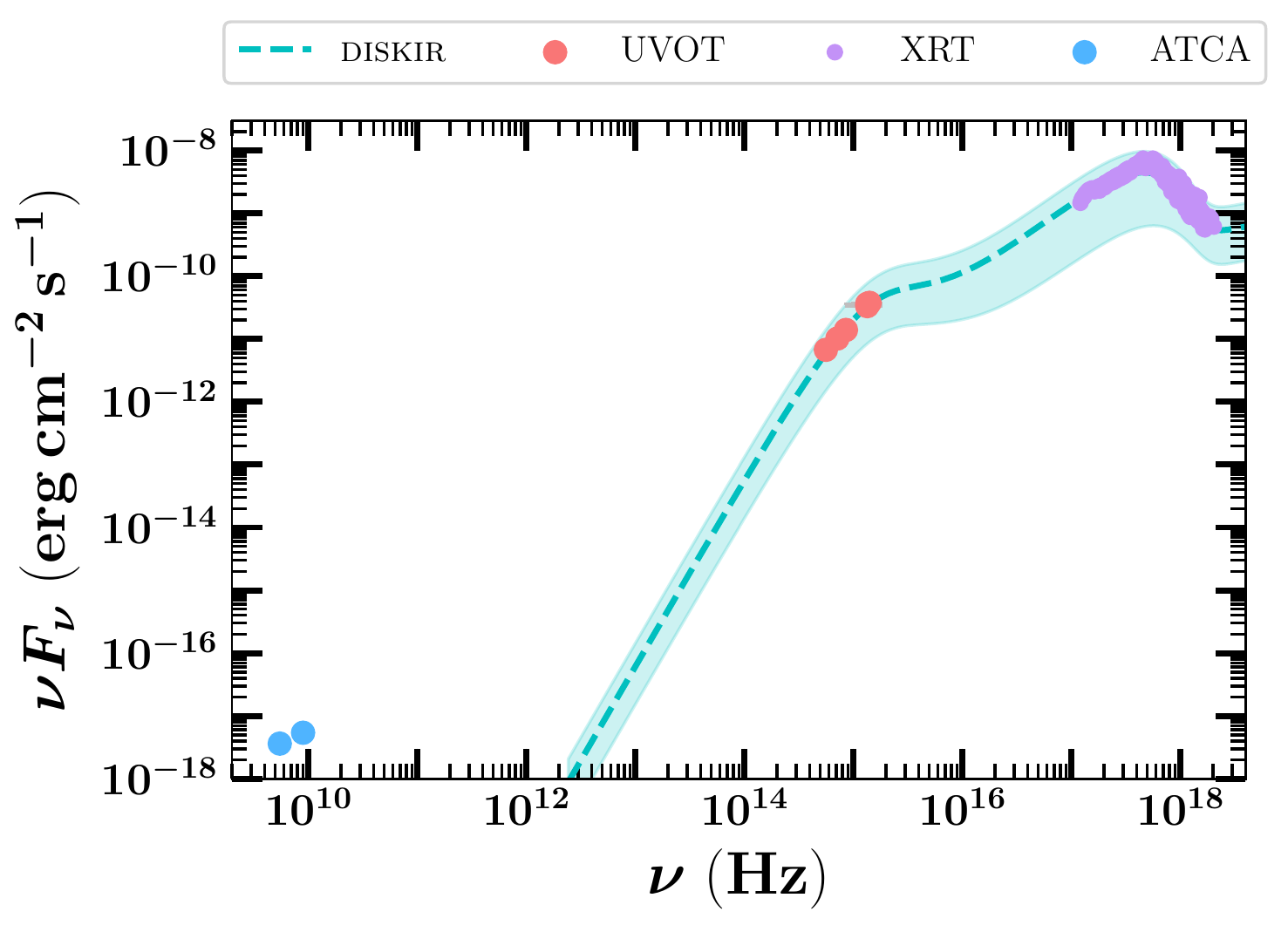}
  \includegraphics[width=0.45\linewidth,height=.4\linewidth]{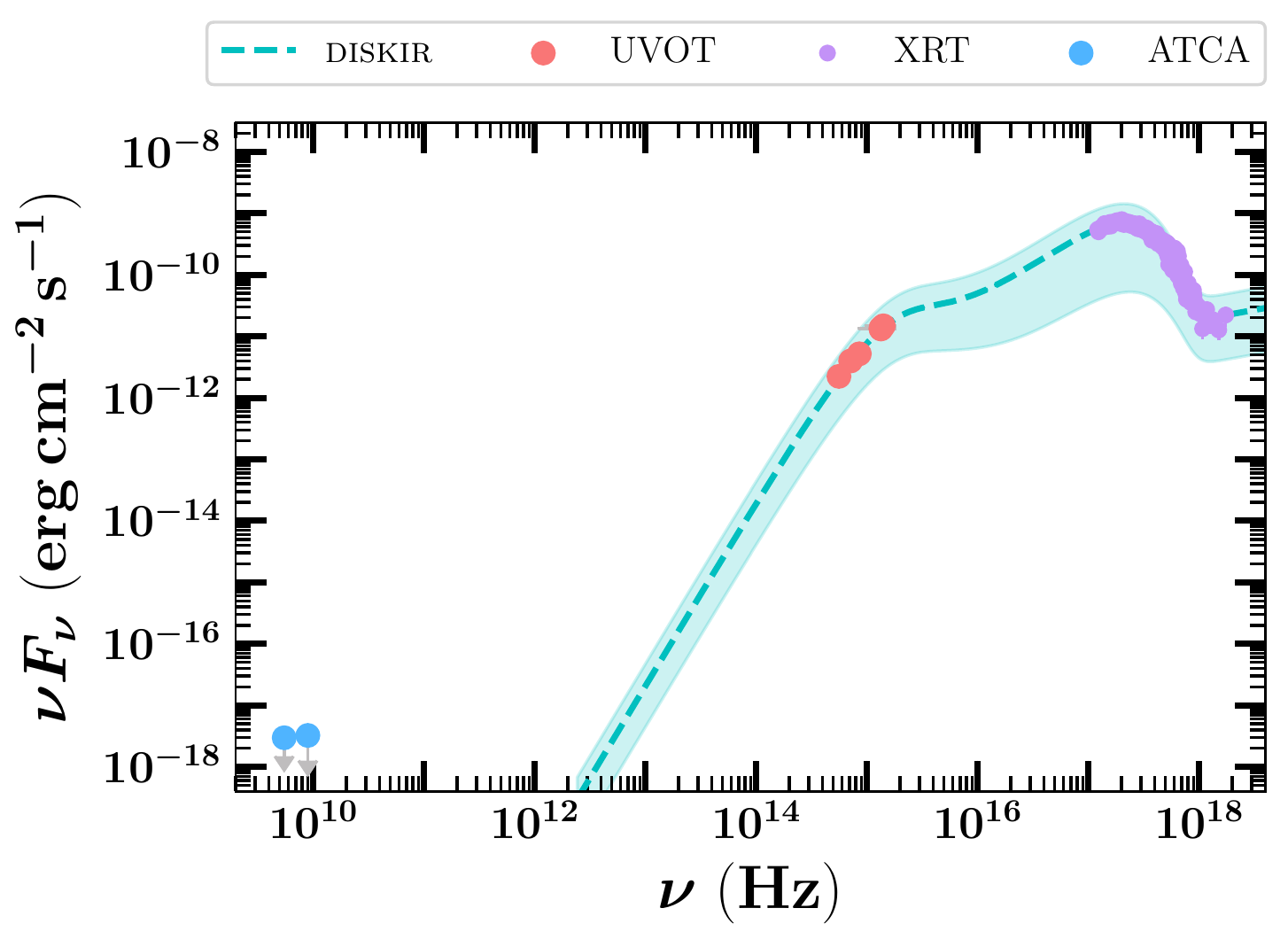}
  \includegraphics[width=0.45\linewidth,height=.4\linewidth]{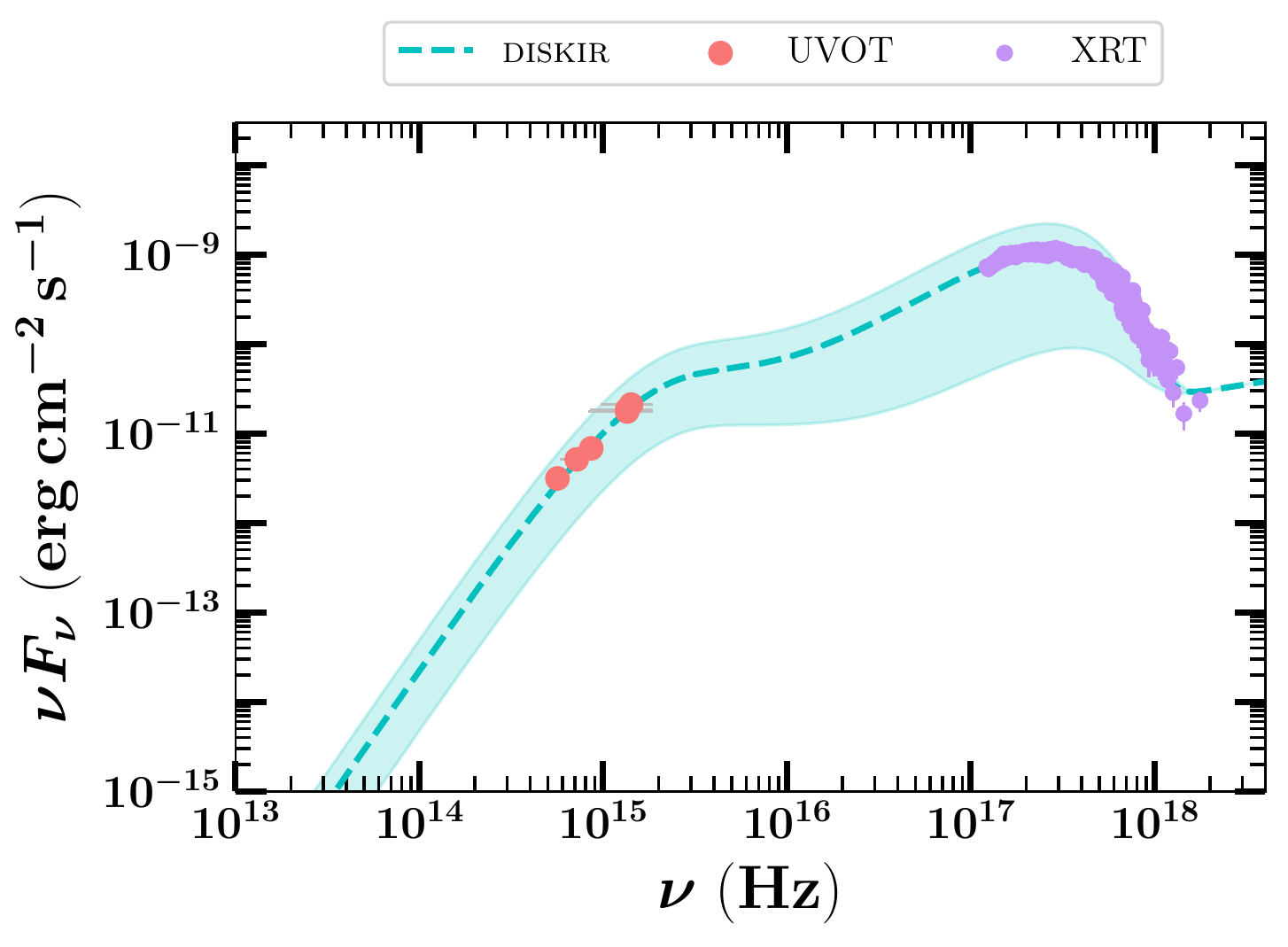}
  \includegraphics[width=0.45\linewidth,height=.4\linewidth]{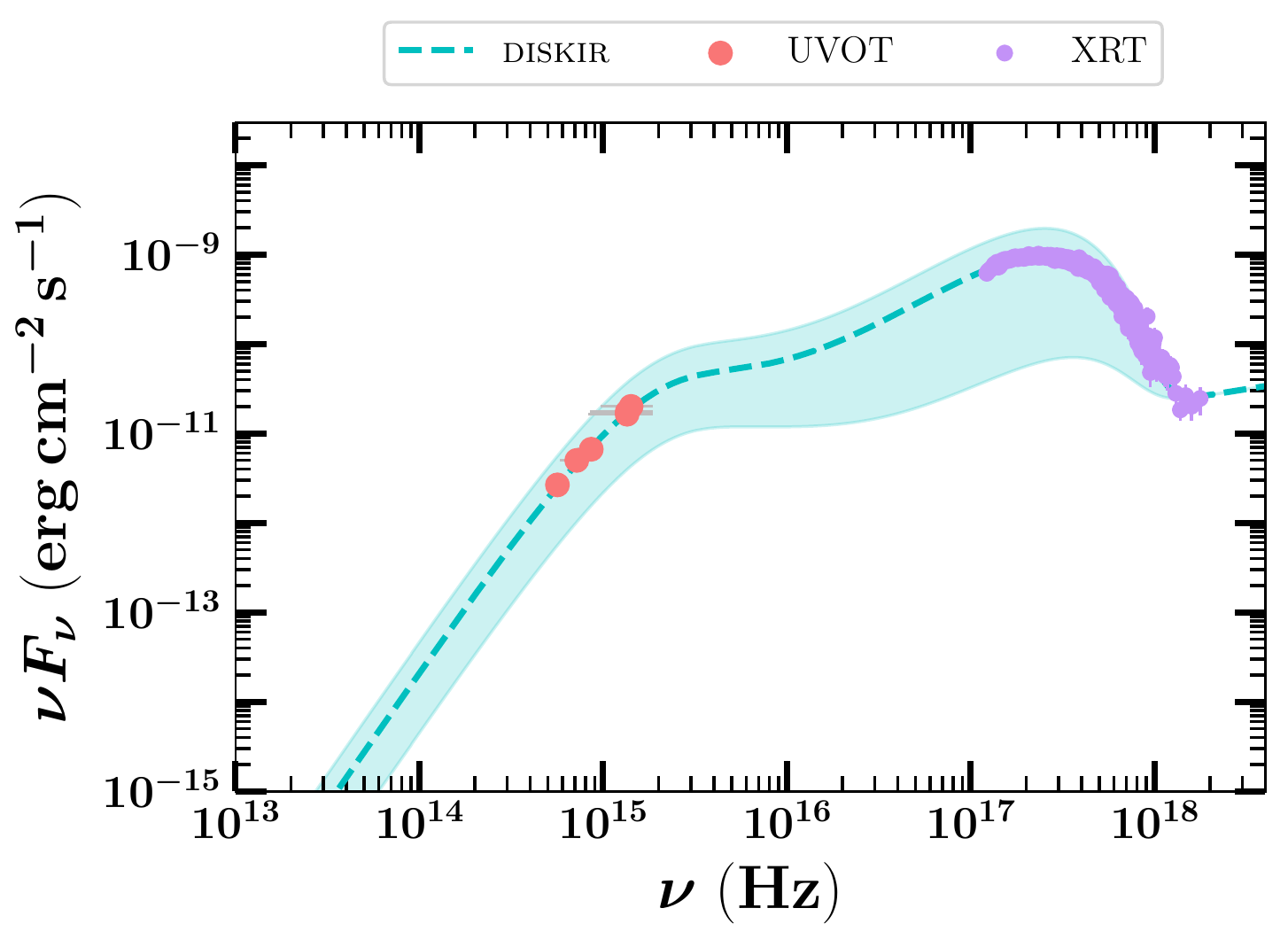}
  \includegraphics[width=0.45\linewidth,height=.4\linewidth]{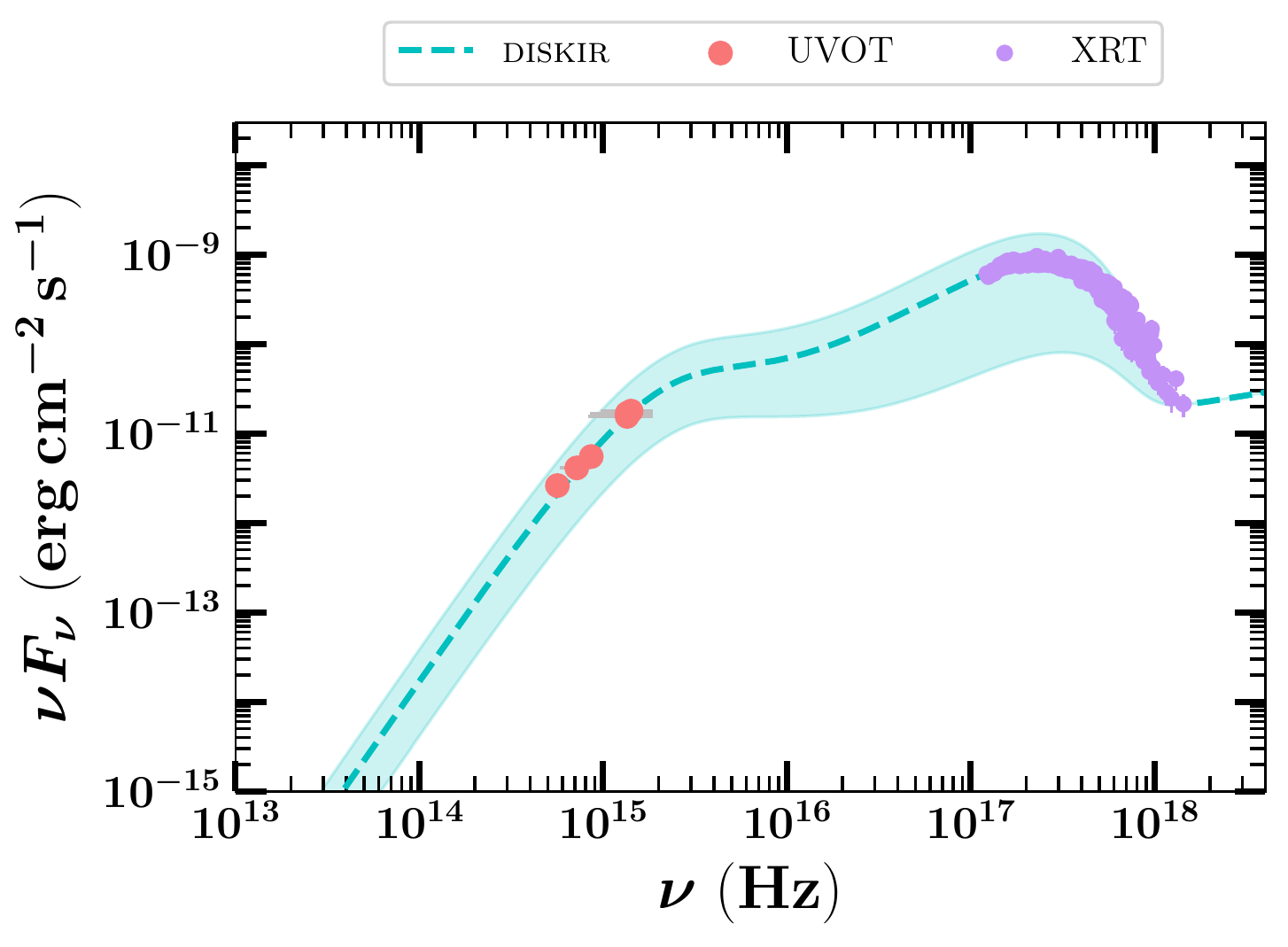}
  \caption{Example broad-band SEDs quasi-simultaneous (within at most 10hrs of each other) with the \textit{(top panel)} two epochs in which ATCA data were taken, and \textit{(middle and bottom panels)} and three epochs in which Gemini/GMOS data were taken. \textit{Swift}/XRT, \textit{Swift}/UVOT, and ATCA (5.5 and 9 GHZ) data are shown as the purple, red, blue circular data points, respectively. The best-fit {\tt diskir} model (dashed cyan line) and $1\sigma$ confidence interval on the fit (shaded cyan region) are also displayed in each SED.}
  \label{fig:ex_seds}
 \end{figure*}


\bsp	
\label{lastpage}
\end{document}